%% file: main.tex
\documentclass[3p,times]{elsarticle} 
\makeatletter
\def\ps@pprintTitle{%
  \let\@oddhead\@empty
  \let\@evenhead\@empty
  \def\@oddfoot{%
    \makebox[\textwidth][s]{
      \hfill
      \makebox[0pt][c]{\thepage}
      \hfill
      \small{\textit{\today}}
    }%
  }%
  \let\@evenfoot\@oddfoot}
\makeatother

\input{headings}

\biboptions{authoryear}
\journal{TBD}

\usepackage{setspace}
\usepackage{booktabs}
\usepackage{multirow}
\usepackage{lipsum}
\usepackage{makecell}
\usepackage{ulem}
\usepackage{tabularx}
\usepackage{booktabs}
\usepackage{array}

\usepackage{endnotes}

\begin{document}

\begin{frontmatter}

\title{Expected shortfall estimation and capital coverage\\ under overlapping scenarios}



\author[label1]{Martin Aichele}
\author[label2]{Igor Cialenco}
\author[label3]{Damian Jelito\corref{cor1}}\ead{damian.jelito@uj.edu.pl}
\author[label3]{Marcin Pitera}
\address[label1]{European Central Bank, Sonnemannstra\ss e 22, 60314 Frankfurt am Main, Germany} 
\address[label2]{School of Computing, Illinois Institute of Technology, 10 W 32nd Str, Building REC, Room 125B, Chicago, IL 60616, USA}
\address[label3]{Institute of Mathematics, Jagiellonian University, S. {\L}ojasiewicza 6, 30-348 Krak{\'o}w, Poland}
   \cortext[cor1]{Corresponding author}

\begin{abstract}
Expected shortfall (ES) is routinely estimated from a short sample of profit-and-loss (P\&L) scenarios built from overlapping multi-day returns, while the non-parametric estimators applied to such data are designed under independent sampling. We study how this mismatch affects estimation accuracy and capital coverage, and how it interacts with serial dependence and heavy tails in the underlying P\&L. Working within the coherent risk estimation framework, we compare six order-statistic-based ES estimators on fit, bias, and capital-coverage metrics. Following the Fundamental Review of the Trading Book (FRTB), we use a one-year estimation window and ES at the $2.5\%$ confidence level as a benchmark setup. We combine simulations with market data on twelve risk factors and ninety-nine equity portfolios. We show that overlapping sampling does not merely add noise but shifts ES estimates systematically downward, thereby weakening capital coverage. Serial dependence acts in the same direction, and heavy tails amplify both effects. More conservative weighting mitigates the underestimation but does not remove it. On market data, the confidence level at which the estimated capital covers realized losses roughly doubles in median under overlapping sampling for every estimator, while the $\textrm{ES}_{2.5\%}$-to-$\textrm{VaR}_{1\%}$ ratio, which links the FRTB base market risk capital metric to its pre-FRTB predecessor, changes comparatively little. Overlapping sampling can thus substantially understate ES-based capital. Since many banks apply a small set of standard estimators in their FRTB regulatory or economic capital model in a historical simulation approach, this understatement is likely to be common for many institutions, and combining simulation-based assessment of estimation risk with coverage-based diagnostics offers a practical method to detect it in model validation and monitoring.
\end{abstract}




\end{frontmatter}



\section{Introduction}

Risk management plays a critical role in safeguarding the solvency of financial institutions and, in aggregate, the stability of the financial system. This role rests on the ability to reliably estimate the capital required to absorb losses arising from realized adverse economic scenarios. For market risk, capital requirements are quantified through risk measures, which are backed by a well-developed mathematical theory. The central goal of this theory is to determine what a `good risk measure' should look like, and how to quantify, represent, and manage risk, see \citet{ArtDelEbeHea1999} for the seminal paper on \textit{coherent risk measures}, or \citet{FolSch2016} for a comprehensive treatment. 

In risk measure theory, one assumes that the true probability law of the underlying profit-and-loss (P\&L), cash-flows, or exposure vectors is known, which, practically speaking, can never be satisfied. Hence, the estimation of risk measures has become a central topic in its own right, see \citet{McNeilFreEmb2015}. The need for efficient estimation methods is reinforced by regulatory frameworks such as the Basel III reforms' Fundamental Review of the Trading Book (FRTB), under which expected shortfall (ES) at the $2.5\%$ confidence level replaced the traditional value-at-risk (VaR) at the $1\%$ confidence level as the main Pillar~I market risk metric for banks using internal models, see \citet{Bas2019}. This regulatory shift reflects the emphasis placed after the 2008 financial crisis on capturing the severity of tail losses rather than relying solely on a single quantile. In practice, market risk is computed, or more precisely \textit{estimated}, using finite samples, e.g. based on annual stress period data for Pillar~I purposes, and thus ES estimation accuracy is not a peripheral statistical concern. In particular, while the underlying risk measure can be coherent and sound, inadequate estimation procedures can lead to systematically biased capital reserves that are fragile to the shape of the P\&L distribution, see \citet{Dan2008,KelRos2016,BarKolVan2023} and references therein.

In this paper we focus on ES estimation risk and the related capital coverage sufficiency. A wide range of ES estimation methodologies have been proposed, from historical simulation and parametric approaches to extreme-value-theory-based and fully stochastic techniques; we refer to the survey of \citet{NadZhaCha2014}, which catalogs more than 45 estimation methods, and to the more recent treatments in \citet{RigCer2015}, \citet{Che2008}, and \citet{RuiNie2023}. While estimation risk has received considerable attention, see e.g. \citet{ConDegSca2010,RiskLudkovski2018,FraHer2012,Frank2016,LauZah2017,BartlEckstein2024}, the overwhelming majority of the ES estimation literature studies estimators through a traditional large-sample lens: consistency, rate of convergence, and asymptotic normality, typically under the assumption that the P\&L is observed as an independent and identically distributed (i.i.d.) sequence. While this is a standard and mathematically convenient assumption, it sits uneasily with the practical way risk estimators are actually deployed. Namely, risk estimates are not computed in asymptotic large sample size regimes but rather using a specific functional and a \textit{fixed} sample size, and then recomputed on a periodic basis, see \citet[MAR~33.6]{Bas2019}.

A recent strand of the literature takes this practical reality seriously by considering the desired estimator properties directly within finite-sample environments, rather than recovering them asymptotically, see e.g. \citet{BarTan2023}, \citet{Gas2026}, and references therein. In particular, \citet{AicCiaJelPit2025} introduce the notion of \textit{coherent risk estimators} (CREs): finite-sample analogues of the monotonicity, cash-additivity, positive homogeneity, and subadditivity axioms of coherent risk measures, imposed directly at the level of the estimator acting on a sample of fixed size $n$, together with law-invariance and comonotonicity. They show that the class of comonotonic, law-invariant CREs is \textit{exactly} the class of $L$-estimators with non-negative, decreasing, and normalized weights, i.e. negatives of weighted averages of the ordered P\&Ls. Crucially, this characterization is distribution-free, in the sense that a given CRE satisfies these properties for \textit{any} underlying P\&L distribution. This finite-sample, distribution-free perspective is particularly well suited to the market risk regulatory setup, where the sample size is often fixed to the number of observations over one year, for example when a historical simulation approach is calibrated to a period of financial stress, cf.~\citet[MAR~33.6]{Bas2019} and \citet[Article~325bc]{EU_CRR}, rather than treated as a parameter that can be sent to infinity.

The statistical properties of such estimators are often studied under the assumption that the observations form an i.i.d.\ sample from a fixed P\&L law. This assumption need not hold for the historical scenarios used in practice. While dynamic risk forecasting methods explicitly model features such as volatility clustering and serial dependence, see \citet{BarGiaVos1999,SoYu2006,Dan2011,RuiNie2023}, their application to individual return series should be distinguished from the scenario-based implementations involving simultaneous movements of multiple risk factors. In the latter, the same historical risk-factor scenarios are applied jointly to revalue the current trading book and aggregate trade-level P\&Ls before applying the risk estimator. Fitting a GARCH-type model to the aggregate P\&L series of the trading book does not, by itself, provide this joint scenario structure. A dynamic alternative serving the same operational purpose would also need to address joint modeling of a high-dimensional risk-factor set, consistent portfolio aggregation, and the associated calibration and validation requirements. The practical relevance of the historical-simulation framework is reflected in the EBA's 2024 benchmarking exercise, which reports that historical simulation was used by 71\% of banks reporting their VaR modeling approach \citep[para.~136]{EBA2025}. However, this construction does not ensure that the P\&L sample supplied to the estimator is independent. Two distinct sources of dependence are relevant. First, financial return series are themselves serially dependent -- volatility clustering, mean reversion, and structural breaks are the rule rather than the exception, so that even the ``raw'' historical P\&L series is not i.i.d.\ to begin with. Second, and specific to the Pillar~I market risk regulatory capital design, a base liquidity horizon of 10 days is set, and the standard device for producing a 10-day P\&L sample from daily data is the \textit{rolling-window overlapping scenario construction}: the 10-day P\&Ls are derived from the overlapping 10-day returns by rolling the single-observation window day-by-day, so that the consecutive observation windows overlap by nine out of ten days, cf.\ \citet[MAR~33.4]{Bas2019} and \citep[para.~97]{EGIM}. This construction induces serial dependence across observations \textit{even if} the underlying daily series are themselves i.i.d. The nominal sample size, say $n=250$ for one year, therefore vastly overstates the actual information content of the data, a point we return to and quantify directly in our numerical study, see also \citet{HanHod1980,RicSmi1991,HarBro2009,Lea1983,EmbKluMik1997}.

Despite this, the very same closed-form weighting schemes designed for i.i.d. samples are applied, unmodified, to overlapping, serially dependent samples as standard market practice for historical-simulation $\var$ and ES, see \citet[Annex~I]{EBA2023RTS05} for an ES estimation analysis of an exemplary catalog of non-parametric ES plug-in estimators. Direct application of these procedures can lead to significant errors in the estimated capital. The literature documents that overlapping samples bias standard variance and quantile estimators \citep{SumNelHanGuo2009}, and serial dependence leads to underestimation of risk-based capital \citep{DanZho2016}, see also \citet{Bak2006} and \citet{KalOko2011}, where the behavior of $L$-statistics under dependent sampling is studied. The question addressed here is not whether an alternative conditional forecasting model can perform better, but whether these procedures, as used for capital calculation, remain reliable when their inputs are overlapping and serially dependent. A further complication is that the distribution-free axiomatic properties of CREs do not imply distribution-free statistical performance. Even under i.i.d.\ sampling, the shape of the P\&L distribution, especially its tails, can materially affect estimation accuracy and capital coverage at a fixed, relatively small sample size. This concern is also reflected in the literature on capital adequacy under heavy-tailed models, see \citet{BarTan2023,PitSch2022}.

The aim of this study is to investigate, jointly and quantitatively, the impact of these three factors (the overlapping sampling scheme, serial dependence in the underlying P\&L, and the shape of the  P\&L distribution) on the accuracy and capital adequacy of ES estimation under a calibration motivated by the FRTB Pillar~I market risk framework, that is, for the sample size $n=250$, corresponding approximately to a one-year calibration window, and for the fixed confidence threshold $\alpha=2.5\%$. Our study is performed  within the CRE framework of \citet{AicCiaJelPit2025}, applied to six representative ES estimators derived from practice-relevant, order-statistic-based plug-in estimators considered in \citep[Annex~I]{EBA2023RTS05}, together with a $\var$ benchmark at the 1\% confidence level. The analysis is structured around nine metrics grouped into three economically meaningful families: \textit{Fit} (how tightly the estimate concentrates around the true ES), \textit{Bias} (whether the estimate is correctly centered, statistically and in a risk-management sense), and \textit{Safety} (how reliably the resulting capital reserve actually covers the realized risk). Building on existing analyses of overlapping risk estimation and coherent estimator design, we quantify how a fixed historical sample translates into estimation bias and secured-position risk across distributional and volatility regimes, using error metrics measured relative to the known ES in the spirit of \citet[Annex~I]{EBA2023RTS05}, and examine which simulation designs reproduce the loss of capital coverage observed on market data. Our aim is to complement the existing, often fragmented, risk estimation analyses, and also to provide more direct and targeted comment on capital coverage sufficiency, especially when market risk capital is evaluated. We also illustrate how model monitoring and model validation practices could be enhanced to account for the VaR-to-ES migration initiated by the FRTB reforms.

Our empirical investigation is organized along three axes. First, and running throughout the entire study, we cross every model and every estimator with two sampling schemes: non-overlapping and overlapping. Second, within this design we consider a comprehensive simulation study spanning a wide range of data-generating models: static, i.i.d.\ distributions of varying tail heaviness (Gaussian, Student-$t$, generalized Pareto, and normal-inverse-Gaussian laws with parameters compatible with daily trading-book P\&Ls), as well as dynamic, serially correlated GARCH-type models under both Gaussian and Student-$t$ innovations. Thus, the two sources of dependence motivated above, the sampling scheme and the serial dependence of the P\&L itself, can be examined both separately and jointly. Third, recognizing that synthetic data can only go so far in assessing a capital framework, we complement the simulation study with an extensive market-data analysis: twelve risk-factor series spanning interest rates, credit, equities, FX, and commodities, together with ninety-nine diversified equity portfolios from the Fama--French data library \citep{FamFre2026}, evaluated under both sampling schemes.

Our main findings can be summarized as follows. \textbf{First}, we document that moving from non-overlapping to overlapping sampling is associated with \textit{a regime shift, not merely a variance penalty}. Overlapping samples do more than shrink the effective sample size, they are associated with a systematic, directional bias that pushes estimated capital \textit{downward}, an adverse direction for capital adequacy, with a broad deterioration across fit, bias, and safety metrics. \textbf{Second}, we show that within the CREs we consider there is no uniformly best choice. Under non-overlapping sampling, shifting weight toward the most extreme order statistics improves bias and safety at a visible cost in symmetric fit. However, once overlapping and serial dependence are both present, symmetric fit is poor for every estimator, and more conservative weighting becomes weakly preferable, mainly raising the level of conservatism. \textbf{Third}, the market-data analysis corroborates the simulation evidence directly. Across our twelve risk-factor series and, separately, across our ninety-nine equity portfolios, the confidence level at which the estimated capital actually covers subsequent realized losses, ideally equal to the $2.5\%$ target, roughly doubles in median under the overlapping scheme for every ES estimator, while the realized $\ES_{2.5\%}/\var_{1\%}$ ratio moves comparatively little. This indicates that the ratio diagnostic alone is insufficient to assess ES estimation quality and should be read alongside coverage-based measures. \textbf{Fourth}, and more broadly, these results indicate that the use of historical-simulation ES estimators within the FRTB framework carries a substantial risk of understating capital, underscoring the need for enhanced estimator-specific validation and monitoring, and ultimately for a dedicated theory of risk estimation under overlapping, dependent sampling schemes. 

Our analysis offers a practical way to assess estimation risk and capital coverage in ES models used by many financial institutions. Regulatory capital formulas add some generic safeguards on top of the estimates, for example through averaging of the capital charge, calibration to a period of stress, or a multiplication factor linked to backtesting and qualitative add-ons~\cite[MAR~33.42]{Bas2019}. These safeguards, however, do not remove the need to control the accuracy of the underlying estimate itself. Since many banks apply historical simulation with a small set of standard order-statistic estimators, the understatement we document is likely to be common for many institutions, which links estimation risk at the level of a single model to the resilience of the banking sector. Our analysis therefore supports the initial and periodic internal model validation requirements of the FRTB framework; see \cite[MAR~30.17]{Bas2019}, \cite[Art.~325bj]{EU_CRR}, and \cite[Section~4.3]{EGIM}. 

Furthermore, while we focus on $\ES_{2.5\%}$ and $n=250$, motivated by the FRTB Pillar~I market risk framework, the underlying estimation problem is not specific to this regulatory calibration. Overlapping historical scenarios introduce dependence that needs to be accounted for when assessing estimation accuracy and capital coverage. The assessment framework considered here is therefore also relevant to other applications, including internal market risk monitoring and capital adequacy assessments, whenever risk is estimated from finite samples of overlapping scenarios.

The paper is organized as follows. Section~\ref{sec:coh.risk} recalls the notion of CREs and their $L$-estimator representation, and fixes the family of six ES estimators, together with a $\var$ benchmark, on which the study is built. Section~\ref{S:assessment} introduces the family of assessment metrics, grouped in three classes, the Fit, Bias, and Safety. Section~\ref{S:numerical.study} is devoted to the simulation study, contrasting non-overlapping and overlapping sampling across static and dynamic data-generating models and examining the resulting fit-bias-safety trade-off, capital adequacy, and the comparison between ES and $\var$ estimation risk. Section~\ref{S:market_data} turns to the market data, assessing capital coverage, comparing simulated and realized estimation risk, and examining the realized $\mathrm{ES}_{2.5\%}/\mathrm{VaR}_{1\%}$ ratio. Section~\ref{sec:concluding-rems} concludes and discusses implications for the supervisory validation of ES-based capital frameworks. Detailed numerical results are collected in the  Appendices.

\section{Coherent risk estimation and ES estimator selection}\label{sec:coh.risk}
On a theoretical level, the expected shortfall (ES) at confidence level $\alpha\in (0,1)$ is a risk measure given by
\begin{equation}\label{eq:ES}
 \textstyle \textrm{ES}_\alpha(X):= \frac{1}{\alpha} \int_0^\alpha \var_\beta(X)\dif \beta, \quad X\in\cX,
\end{equation}
where $\cX$ is a set of random variables corresponding, for example,  to the future financial position P\&Ls  and $\var_\beta(X):= \inf \set{m\in \mathbb{R} \mid \bP(X+m<0)\leq \beta}$, $X\in\cX$, is the value-at-risk (VaR) at confidence level $\beta\in (0,1)$. The risk measure defined in \eqref{eq:ES} is known to be a monetary risk measure that is coherent, law-invariant and comonotonic, see~\cite{FolSch2016} for details. For continuous random variables, \eqref{eq:ES} could be seen as the tail conditional expectation over VaR due to representation
\begin{equation}\label{eq:E2}
\textrm{ES}_\alpha(X)=\bE[-X|X\leq-\var_\alpha(X)].
\end{equation}

Both $\var$ and $\ES$ are standard measures used to quantify market risk, and the pair
$(\var_{1\%},\ES_{2.5\%})$ is the one of particular regulatory interest that is the focus of this paper. Namely, $\var_{1\%}$ is the
Pillar~I capital metric under the Basel~II IMA market risk framework, while $\ES_{2.5\%}$ is its  Basel III (FRTB) counterpart; note that $\var_{1\%}$ is retained for backtesting purposes under Basel III, 
cf.~\cite[MAR 32]{Bas2019}. 
The two reference confidence levels are deliberately matched. For a standard normal $X$ one has
$\var_{1\%}(X)=2.326$ and $\ES_{2.5\%}(X)=2.336$, so the two metrics secure essentially the
same capital in the Gaussian benchmark and their realized ratio is a natural diagnostic;
see~\cite{KelRos2016} and~\cite[Section 4.3.2]{EGIM}. Accordingly, throughout this paper
$\var_{1\%}$ is used as a benchmark that isolates $\ES_{2.5\%}$ estimation risk features,
not as a competing metric.

In addition to the key normative difference between ES and VaR linked to the lack of coherence of VaR (outside the elliptical distribution class), these risk measures follow different risk measurement philosophies. By~\eqref{eq:ES},
$\ES_\alpha$ measures the expected tail risk below the $\alpha$-quantile, whereas $\var_\alpha$
reads a single point of the distribution, see \cite{Emmeretal2015} for a comparison of the two from a practitioner’s perspective. Consequently, we need the information about the whole tail of $X\in\cX$ to calculate ES, which, as we later show, has a profound impact on ES estimation quality, especially when confronted with VaR estimation quality. 

Let us now introduce a class of estimators that are suitable for estimating coherent risk measures, and then propose a set of representative ES estimators.

\subsection{Coherent risk estimators}\label{S:coherent.risk.estimators}
The monetary risk measure axioms, together with law-invariance and comonotonicity, can be transferred to estimation level following the framework introduced in \cite{AicCiaJelPit2025}. From now on, we fix sample size $n\in\bN$ and use $\bfx\in \bR^n$ to denote a generic sample from the underlying position $X\in\cX$.

\begin{definition}[Coherent risk estimator]\label{def:CRE}
A function $\hat\rho_n\colon \bR^n\to \bR$ is a {\it coherent risk estimator} (CRE) if it satisfies 
\begin{enumerate}[(E1)]
\item {\it Monotonicity}, for any $\bfx,\bfx'\in\bR^n$ such that\footnote{For vector order  $\bfx\leq \bfy$ we use the componentwise comparison $x_i\leq y_i$, for  $i=1,2,\ldots,n$.}   $\bfx\geq\bfx'$, we have $\hat\rho_n(\bfx)\leq\hat\rho_n(\bfx')$;
\item {\it Cash additivity}, for any $\bfx\in\bR^n$ and $m\in\bR$, we have $\hat\rho_n(\bfx+m)=\hat\rho_n(\bfx)-m$;
\item {\it Positive homogeneity}, for any $\bfx\in\bR^n$ and $\lambda\geq 0$, we have  $\hat\rho_n(\lambda\bfx)=\lambda\hat\rho_n(\bfx)$;
\item {\it Subadditivity}, for any $\bfx,\bfx'\in\bR^n$, we have $\hat\rho_n(\bfx+\bfx')\leq\hat\rho_n(\bfx)+\hat\rho_n(\bfx')$.
\end{enumerate}
Moreover, we say that $\hat\rho_n$ is law-invariant and comonotonic if it satisfies
\begin{enumerate}[(E1)]
\item[(E5)] {\it Law-invariance}, for any $\bfx\in\bR^n$, we have $\hat\rho_n(\bfx)=\hat\rho_n(s(\bfx))$, where $s(\mathbf{x}):=(x_{1:n},\ldots, x_{n:n})$ is the ordered sample, that is,  $x_{i:n}$ is the $i$th smallest element of $(x_1,\ldots, x_n)$, $i=1,\ldots, n$. 
\item[(E6)] {\it Comonotonicity}, for any comonotonic vectors $\bfx\in\bR^n$ and $\mathbf{y}\in \mathbb{R}^n$, we have $\hat\rho_n(\mathbf{x}+\mathbf{y}) = \hat\rho_n(\mathbf{x})+\hat\rho_n(\mathbf{y})$.
\end{enumerate}
\end{definition}

Note that properties (E1)-(E6) are required on estimation level, that is, for any input sample, and they are not automatically satisfied by any ES risk estimator. In fact, the class of estimators which satisfies properties (E1)-(E6) has been recently characterized in \cite{AicCiaJelPit2025} via the robust representation.

\begin{theorem}[Robust  representation of comonotonic and law-invariant  CREs]\label{th:comonotonic_CRE}
    A risk estimator $\hat\rho_n\colon \bR^n\to \bR$ is a comonotonic law-invariant CRE if and only if there exists a unique $a=(a_1, \ldots, a_n)\in \cS_n$ satisfying
\begin{equation}\label{eq:comonotonic_representation}
   \textstyle \hat\rho_n(\bfx)= \langle a,-s(\bfx)\rangle = -\sum_{i=1}^n a_i x_{i:n}, \quad \mathbf{x}\in \mathbb{R}^n,
\end{equation}
where $\cS_n:=\{a\in\bR_+^{n}\colon \sum_{i=1}^{n}a_i=1\,\,\textrm{and}\,\, a_1\geq a_2\geq \ldots \geq a_n\}$.
\end{theorem}
In a nutshell, Theorem~\ref{th:comonotonic_CRE} says that the class of $L$-estimators with decreasing weight inputs that sum to one applied to samples of size $n$ is the class of coherent risk estimators that are additionally law-invariant and comonotonic. Notably, Theorem~\ref{th:comonotonic_CRE} shows that popular semi-parametric estimators based on Extreme Value Theory and GPD tail fits, which are not $L$-estimators, are not comonotonic law-invariant CREs; in fact, such estimators may fail subadditivity and hence coherence altogether, see \cite[Section~6.2]{AicCiaJelPit2025}. We refer to \cite{AicCiaJelPit2025} for more details and the discussion on how this interacts with other estimation methods.  We emphasize, as already mentioned in the introduction, that the main goal of this manuscript is to study the impact of applying estimators, such as \textit{comonotonic law-invariant} CREs, designed for i.i.d. samples, to non-i.i.d. samples.

\subsection{Expected shortfall estimator selection}\label{S:es.estimator.selection}
Having introduced coherent risk estimators, we now fix a concrete, practice-relevant family of ES estimators on which the numerical study is built. In many market risk applications, ES is estimated non-parametrically from the sorted P\&L sample, and supervisory texts typically prescribe desirable properties of such estimators rather than a single closed formula. A representative catalog of the estimators is collected in~\cite[Annex~I]{EBA2023RTS05}; we adopt the six estimators reported there, labeled \#1--\#6 in Table~\ref{T:estimators}, and study how their order-statistic weighting affects estimation quality. 

\begin{table}[htp!]
    \centering
    \scalebox{0.7}{
    \begin{tabular}{cp{7.2cm}p{13.5cm}}
Id & Estimator  & Comment\\\toprule
\#1 & $\widehat\ES^{1}_{\alpha,n}(\bfx):=\frac{-1}{\alpha n}\left(\sum_{i=1}^{\lfloor\alpha n\rfloor} x_{i:n}+(\alpha n-\lfloor\alpha n\rfloor) x_{(\lfloor\alpha n\rfloor+1):n}\right)$  & Non-parametric ES plug-in estimator for the empirical sample quantile.
\\\midrule
\#2 & $\widehat\ES_{\alpha,n}^2(\bfx):=\frac{-1}{\lfloor\alpha n\rfloor}\sum_{i=1}^{\lfloor\alpha n\rfloor} x_{i:n}$  & Average tail loss ES estimator based on \eqref{eq:E2} and sample conditional mean.\\\midrule
\#3 & $\widehat\ES_{\alpha,n}^{3}(\bfx):=\frac{-1}{M_6}\left(\frac{3}{2} x_{1:n}+\sum_{i=2}^{M_6} x_{i:n}\right)$ & ES plug-in integral estimator for Type 6 sample quantile, based on~\eqref{eq:ES} and flat extrapolation in which the integral in~\eqref{eq:ES} is restricted to $[0, M_6/(n+1)]$. For details, see \cite[Annex I]{EBA2023RTS05}.\\\midrule
\#4 & \vtop{$\begin{aligned}[t] \widehat\ES_{\alpha,n}^4(\bfx) & \textstyle :=\frac{-1}{\alpha(n+1)}\Big(\frac{3}{2} x_{1:n}+\sum_{i=2}^{M_6-1} x_{i:n}+\frac{1+2 R_6-R_6^2}{2} x_{M _6:n}\\ &\textstyle \phantom{:=\,\,}+\frac{R_6^2}{2} x_{(M_6+1):n}\Big) \end{aligned}$}
 & {ES plug-in integral estimator for Type 6 sample quantile, based on~\eqref{eq:ES} and flat extrapolation. For details, see \cite[Annex I]{EBA2023RTS05} and \cite{HynFan1996}.}\\\midrule
\#5 & $\widehat\ES_{\alpha,n}^{5}(\bfx):=\frac{-1}{M_6}\left(\left(\frac{1}{2}+\frac{1}{1-\xi}\right) x_{1:n}+\sum_{i=2}^{M_6} x_{i:n}\right)$  &  {ES plug-in integral estimator for Type 6 sample quantile, based on~\eqref{eq:ES} and Pareto-type extrapolation in which the integral in~\eqref{eq:ES} is restricted to $[0, M_6/(n+1)]$. For details, see \cite[Annex I]{EBA2023RTS05}.} \\\midrule
\#6 &
$\begin{aligned}[t]
\widehat\ES_{\alpha,n}^6(\bfx) &\textstyle :=\frac{-1}{\alpha(n+1)}\Big(\left(\frac{1}{2}+\frac{1}{1-\xi}\right) x_{1:n}+\sum_{i=2}^{M_6-1} x_{i:n}\\
& \textstyle\phantom{:=\,\,} +\frac{1+2 R_6-R_6^2}{2} x_{M_6:n}+\frac{R_6^2}{2} x_{(M_6+1):n}\Big)
\end{aligned}$ & 
{ES plug-in integral estimator for Type 6 sample quantile, based on~\eqref{eq:ES} and Pareto-type extrapolation. For details, see \cite[Annex I]{EBA2023RTS05} and \cite{McN1999}.}\\\midrule
\gray{\#VaR} &
$\gray{\begin{aligned}[t]
\widehat\var_{\alpha,n}(\bfx) &\textstyle :=-\left((1-R_6) x_{M_6:n}+R_6 x_{(M_6+1):n}\right)
\end{aligned}}$ & 
{\,\gray{Benchmark VaR plug-in estimator for the empirical sample quantile (Type 6 in the notation of~\cite{HynFan1996}), cf.~\cite[para.~100]{EGIM}.
}}
    \end{tabular}}
    \caption{The table presents six different non-parametric ES estimators that are considered in the comparative analysis. The estimators are stated as reported in \cite[Annex I]{EBA2023RTS05}; the order statistics weights of \#3, \#5, and \#6 do not sum to one and are normalized later. For brevity, we set $M_6:=\lfloor \alpha(n+1)\rfloor$ and $R_6:=\alpha(n+1)-\lfloor\alpha(n+1)\rfloor$, and consider estimators \#5 and \#6 with a fixed tail shape $\xi:=1/3$. The grayed row defines a benchmark VaR estimator; note that we typically set $\alpha=2.5\%$ for ES and $\alpha=1\%$ for VaR which results in different values of $R_6$.}
    \label{T:estimators}
\end{table}

The catalog spans the main non-parametric methodologies: the empirical plug-in (\#1) and the sample-conditional-mean estimator (\#2), together with Type 6 sample-quantile plug-ins using either flat (\#3--\#4) or Pareto-type (\#5--\#6) tail extrapolation. The extrapolation-based estimators place progressively more mass on the most extreme order statistics to account for unobserved tail losses, with the Pareto variants governed by a shape parameter~$\xi$; see~\cite{McN1999}.

First, we check if the considered estimators belong to the class of estimators introduced in Section~\ref{S:coherent.risk.estimators}. As ES is a comonotonic and law-invariant risk measure, one would expect that its estimators satisfy the same estimator-level axioms. Recalling Theorem~\ref{th:comonotonic_CRE}, instead of directly checking the axioms from Definition~\ref{def:CRE}, it is sufficient to check if the estimators admit the robust representation \eqref{eq:comonotonic_representation}. All six estimators presented in Table~\ref{T:estimators} are $L$-estimators with decreasing and non-negative weights. Furthermore, by direct calculation, we get that the weights assigned to order statistics for estimators \#1, \#2, and \#4 always sum to one, so these estimators are law-invariant and comonotonic CREs. On the other hand, the weights in estimators \#3, \#5, and \#6 do not always sum to one; these estimators either normalize the integral by the floored value $M_6$ rather than by $\alpha(n+1)$ (\#3, \#5), or load the smallest order statistic with the Pareto-type coefficient $\frac{1}{2}+\frac{1}{1-\xi}$ in place of the flat coefficient $\frac{3}{2}$ used by \#3 and \#4 -- in both cases the weights sum to more than one and cash additivity (E2) fails at estimator level. Consequently, to align these estimators with CRE theory, we normalize weight vectors of \#3, \#5, and \#6  by their total mass. The renormalization retains the relative magnitude, preserving the conservative emphasis on the most extreme losses while ensuring unit sum of weights. Also, note that the weights might be read as various discrete approximations of the ES risk spectrum, which provide an alternative renormalization justification, see \cite[Example 4.11]{AicCiaJelPit2025} and \cite{Ace2002}. For $n=250$ and $\alpha=2.5\%$ the unadjusted weights of \#3, \#5, and \#6 sum to $1.083$, $1.167$, and $1.080$, so after rescaling the values are decreased by $7.7\%$, $14.3\%$, and $7.4\%$, respectively. After this adjustment, all six estimators are comonotonic law-invariant CREs, which follows directly from Theorem~\ref{th:comonotonic_CRE}.  The \#1-\#6 weight vectors for $n=250$ and $\alpha=2.5\%$ that will be used in the numerical study are later reported in Table~\ref{T:weights_normalized}. Note that this normalization lowers the resulting estimate. The comparison carried out in this paper is therefore within the CRE class, and the reported metrics of the renormalized estimators are not directly comparable to those of \cite[Annex~I]{EBA2023RTS05}.

Second, we remark that while estimators \#1--\#6 are built on i.i.d. sample characteristics, they are routinely applied to overlapping samples, most notably to the 10-day overlapping P\&Ls of FRTB models, for which the i.i.d.\ property fails; cf.~\cite[MAR~33.4]{Bas2019} and~\cite[para.~97]{EGIM}. Overlapping scenarios raise the nominal sample size, but consecutive 10-day sums are autocorrelated, so the effective sample size (measured e.g. in terms of estimator variance) is usually considerably smaller than the typical reference size $n=250$, see Section~\ref{S:numerical.study} for details. This creates additional, non-negligible estimation risk, which is the main topic of this paper; see also \cite{SumNelHanGuo2009,AicCroReh2021,RuiNie2023,DanZho2016}. Of course, there are various ES estimators tailored to dependent or multi-period data available. For example, this includes estimators that incorporate volatility filtering \cite{BarGiaVos1999,McNeilFrey2000,KueMitPao2006}, quantile estimation for dependent sequences \cite{Lea1983,Dre2003,LiCleDro2023}, or direct modeling of the 10-day laws \cite{DanZho2016}; in each case an ES estimator follows by integrating the quantile estimator as in~\eqref{eq:ES}. We retain \#1--\#6 because they reflect the estimators effectively used in practice: market practice for VaR and ES alike is often historical simulation with fixed order-statistic weights applied unchanged to overlapping P\&Ls, and the i.i.d.-derived catalog of \cite[Annex~I]{EBA2023RTS05} is used in exactly this way, cf.~\cite{EBA2025}. This mismatch is the key motivation of our study, not an incidental simplification.

Third, we analyze the weight structure differences in the considered estimators. The weighting schemes differ markedly across estimators, especially in the boundary weights ($a_1$ and $a_6$ or $a_7$ for $n=250$). Such differences are unlikely to matter for moderately tailed distributions, but they can induce systematic bias under heavy tails, or when an exogenous shock injects extreme observations into an otherwise i.i.d.\ sample. To quantify these effects, and to assess estimator performance on both i.i.d.\ and non-i.i.d.\ data, we first introduce a panel of benchmark metrics and then carry out a numerical and market data study in the spirit of~\cite[Annex~I]{EBA2023RTS05}.

\section{Assessment framework}\label{S:assessment}

Having fixed the candidate ES estimators, we now specify how their performance is assessed. The structural differences between the weighting schemes of Table~\ref{T:estimators} translate into measurable differences in estimation characteristics, and the aim of this section is to make that translation precise.  We organize the diagnostics into three families, each answering a distinct question.  In Table~\ref{T:metrics.guide} we provide a reading guide together with metrics classification and the target value that signals best performance, while in Table~\ref{T:performance.metrics} we give the theoretical definition as well as the Monte Carlo implementation of each metric.

\begin{table}[htp!]
    \centering
    \footnotesize
    \setlength{\tabcolsep}{4pt}
    \renewcommand{\arraystretch}{1.15}
    \begin{tabular}{@{}l >{\raggedright\arraybackslash}p{0.20\textwidth} l >{\raggedright\arraybackslash}p{0.50\textwidth} c@{}}
    Type & Description & Metric & What it captures & Optimum \\
    \midrule
    \multirow{3}{*}{{\bf Fit}}
      & \multirow{3}{=}{How tightly the estimate concentrates around the true ES.}
      & ME        & Mean absolute deviation from true ES; outlier-robust dispersion. & $0$ \\
      & & SE      & Root mean squared error; dispersion that penalizes large errors. & $0$ \\
      & & DE      & SE restricted to under-estimations; dispersion of the unsafe side only. & $0$ \\
    \midrule
    \multirow{3}{*}{{\bf Bias}}
      & \multirow{3}{=}{Whether the estimate is correctly centered, in a statistical or a risk-management sense.}
      & SB        & Relative mean deviation; signed statistical (un)biasedness. & $0$ \\
      & & MB      & Relative median deviation; outlier-robust centering. & $0$ \\
      & & RB      & Residual risk of the position secured with the estimated reserve; (un)biasedness in the risk-management sense. & $0$ \\
    \midrule
    \multirow{3}{*}{{\bf Safety}}
      & \multirow{3}{=}{Reliability of capital adequacy: how often and how severely the estimate under-secures the position.}
      & UP$_5$    & Frequency of estimates falling below $95\%$ of the true ES. & $0$ \\
      & & UP$_{10}$ & Frequency of estimates falling below $90\%$ of the true ES. & $0$ \\
      & & CT      & Minimal ES confidence level at which the secured position is acceptable. & $\alpha$ \\
    \addlinespace
    \end{tabular}
    \caption{The top-level description of benchmark metrics; see Table~\ref{T:performance.metrics} for exact definition and full metric names. Unless stated otherwise, the Fit and Bias metrics are expressed relative to the true ES and reported in percent. The ``Optimum'' column reports the target value: dispersion (ME, SE, DE) and under-estimation frequencies (UP$_5$, UP$_{10}$) are best near $0$; the bias metrics (SB, MB, RB) are best near $0$ in signed terms, with the sign indicating over- ($+$) or under-estimation ($-$); CT is best near the reference level $\alpha$, that is, $\alpha=2.5\%$ for the ES estimators and $\alpha=1\%$ for the VaR benchmark.}
    \label{T:metrics.guide}
\end{table}

\begin{table}[htp!]
    \centering
\scalebox{0.80}{
    \begin{tabular}{l >{\raggedright\arraybackslash}p{4cm} l l}
Type &Metric &Theoretical formula & Implementation\\\toprule
&Mean Absolute Error & $\frac{\bE\left|\widehat\ES_{\alpha,n}(\bfX)-\ES_{\alpha}(X)\right|}{\ES_{\alpha}(X)}$ & $\textrm{ME}:=\frac{\tfrac{1}{K}\sum_{k=1}^{K}\left|\widehat\ES_{\alpha,n}(\bfx_k)-\ES_{\alpha}(X)\right|}{\ES_{\alpha}(X)}$ \\ 
{\bf Fit} &Root Mean Squared Error & $\frac{\sqrt{\bE\left[\left(\widehat\ES_{\alpha,n}(\bfX)-\ES_{\alpha}(X)\right)^2\right]}}{\ES_{\alpha}(X)}$ & $\textrm{SE}:=\frac{\sqrt{\tfrac{1}{K}\sum_{k=1}^{K}\left(\widehat\ES_{\alpha,n}(\bfx_k)-\ES_{\alpha}(X)\right)^2}}{\ES_{\alpha}(X)}$  \\
& Downside Root Mean Squared Error & $\frac{\sqrt{\bE\left[\left(\widehat\ES_{\alpha,n}(\bfX)-\ES_{\alpha}(X)\right)_-^2\right]}}{\ES_{\alpha}(X)}$ & $\textrm{DE}:=\frac{\sqrt{\tfrac{1}{K}\sum_{k=1}^{K}\left(\widehat\ES_{\alpha,n}(\bfx_k)-\ES_{\alpha}(X)\right)_-^2}}{\ES_{\alpha}(X)}$  \\\midrule
&Statistical Bias & $\frac{\bE[\widehat\ES_{\alpha,n}(\bfX)]-\ES_{\alpha}(X)}{\ES_{\alpha}(X)}$ & $\textrm{SB}:=\tfrac{1}{K}\sum_{k=1}^{K}\frac{\widehat\ES_{\alpha,n}(\bfx_k)}{\ES_{\alpha}(X)}-1$  \\
 {\bf Bias} &Median Bias & $\frac{\textrm{Med}[\widehat\ES_{\alpha,n}(\bfX)]-\ES_{\alpha}(X)}{\ES_{\alpha}(X)}$ & $\textrm{MB}:=\frac{\widehat{\textrm{Med}}\left[\widehat\ES_{\alpha,n}(\bfx_k)\right]}{\ES_{\alpha}(X)}-1$  \\
&Risk Bias & $-\frac{\ES_{\alpha}(X+\widehat\ES_{\alpha,n}(\bfX))}{\ES_{\alpha}(X)}$ & $\textrm{RB}:=-\frac{\widehat\ES^1_{\alpha,K}\left((\tilde x_{k}+\widehat\ES_{\alpha,n}(\bfx_k))_{k=1}^{K}\right)}{\ES_{\alpha}(X)}$  \\\midrule
&Underestimation Probability (5\%) & $\bP\left[\widehat\ES_{\alpha,n}(\bfX)<0.95\cdot \ES_{\alpha}(X)\right]$ & $\textrm{UP}_{5}:=\tfrac{1}{K}\sum_{k=1}^{K}\1_{\{\widehat\ES_{\alpha,n}(\bfx_k)<0.95\cdot \ES_{\alpha}(X)\}}$ \\
{\bf Safety} &Underestimation Probability (10\%) & $\bP\left[\widehat\ES_{\alpha,n}(\bfX)<0.90\cdot \ES_{\alpha}(X)\right]$ & $\textrm{UP}_{10}:=\tfrac{1}{K}\sum_{k=1}^{K}\1_{\{\widehat\ES_{\alpha,n}(\bfx_k)<0.90\cdot \ES_{\alpha}(X)\}}$ \\
&Safe Confidence Threshold & $\inf\{\beta\in (0,1): \ES_{\beta}(X+\widehat\ES_{\alpha,n}(\bfX))\leq 0\}$ & $\textrm{CT}:=\inf\left\{\beta\in (0,1): \widehat\ES^1_{\beta,K}\left((\tilde x_{k}+\widehat\ES_{\alpha,n}(\bfx_k))_{k=1}^{K}\right)\leq 0\right\}$ \\
    \end{tabular}}
  \caption{Benchmark metrics used in the comparative analysis, grouped into Fit, Bias, and Safety families; see Table~\ref{T:metrics.guide} for a reading guide. ME, SE, DE, SB, MB, and RB are expressed relative to the true ES and reported in percent; UP$_5$ and UP$_{10}$ are probabilities (in percent) and CT is a confidence threshold level. For each test distribution the outputs are computed by Monte Carlo as specified in the Implementation column; typically, we consider Monte Carlo size $K=10^6$, sample size $n=250$, and confidence threshold $\alpha=2.5\%$. The value $\tilde x_k$ denotes an independent out-of-sample observation from the same marginal law, not overlapping with the learning sample.
For the VaR benchmark, we use the same definitions with $\var$ and $\widehat\var$ in place of $\ES$ and $\widehat\ES$, and set the level to 1\%.
}
\label{T:performance.metrics}
\end{table}  

A few features of the chosen metrics are worth noting. The fit metrics are ordered, $\mathrm{ME}\le\mathrm{SE}$ and $\mathrm{DE}\le\mathrm{SE}$ (the first by Jensen's inequality, the second because the negative part is dominated pointwise), so SE is the dominant accuracy measure while DE isolates the under-estimation contribution to it. The bias family is signed: SB and MB capture statistical centering in the mean and median, while RB captures centering in the risk-management sense of~\cite{PitSch2016}; see also the discussion of risk fairness in~\cite{BieleckiEtAl2020}. Here, the secured position $X+\widehat\ES_{\alpha,n}(\bfX)$ should bear no residual risk: if the reserve is equal to the true risk then $\ES_{\alpha}(X+\ES_{\alpha}(X))=0$ by cash-additivity of $\ES_\alpha$, but estimation error leaves a residual $\ES_{\alpha}(X+\widehat\ES_{\alpha,n}(\bfX))$, positive when the position is under-estimated and negative when it is over-estimated, which RB normalizes. SB and MB carry the same sign convention one step earlier, at the level of the reserve itself rather than the residual risk it leaves behind: SB is negative when the reserve is too small on average, $\mathbb E[\widehat\ES_{\alpha,n}(\bfX)]<\ES_\alpha(X)$, and positive when it is too large on average; MB gives the same reading at the median rather than the mean, negative  when the reserve falls short of the true risk more often than not and positive when it exceeds it more often than not. RB, MB and SB vanish in the limit for consistent estimators, though not necessarily at the fixed, regulation-mandated sample size $n\in\bN$. Finally, RB and CT probe the same safety position: RB is its (normalized) residual ES at the reference level $\alpha$, while CT is the level at which that residual vanishes. This is why CT is compared with $\alpha$ rather than $0$.  We also refer to \cite{MolPit2017}, where the implementation of the CT metric in the context of backtesting is discussed in detail. For reading purposes we also adopt the traffic-light zones proposed in~\cite{MolPit2017}, which parallel the Basel VaR exception zones: the amber and red thresholds are $5\%$ and $10\%$ for the ES estimators at $\alpha=2.5\%$, and $2\%$ and $4\%$ for the VaR benchmark at $\alpha=1\%$. Both sets are the same multiples of the respective reference level, namely $2\alpha$ and $4\alpha$, so CT values become comparable across the two reference levels. We emphasize that these zones are a calibration proposed in~\cite{MolPit2017} rather than a supervisory requirement, and that they are calibrated to a backtesting sample of length $250$, so the classification should be treated as a reading aid rather than a formal test.


\section{Numerical study}\label{S:numerical.study}

In this numerical study we estimate $\ES_{2.5\%}$ and use $\var_{1\%}$ as a benchmark.\footnote{As mentioned earlier, for a standard normal $X$,  $\var_{1\%}(X)=2.326$ and $\ES_{2.5\%}(X)=2.336$, hence, the two benchmark confidence levels are matching the values.} Both metrics are computed at the learning-sample size $n=250$. As already mentioned, we use reference confidence levels and the one-year calibration windows of the FRTB internal-model approach for market-risk capital, cf.~\cite[MAR 33.6]{Bas2019}, the overall comparison design follows the framework of~\cite[Annex~I]{EBA2023RTS05}. We consider six $L$-estimators \#1--\#6 as ES estimators, presented in Table~\ref{T:estimators}, recalling that their weights are normalized, and the linear interpolation quantile estimator \#VaR as the benchmark VaR estimator. 

For convenience, we display in Table~\ref{T:weights_normalized} the corresponding order-statistic weights for $n=250$ and $\alpha=2.5\%$ for ES, and for $\alpha=1\%$ for $\var$. The weights for  ES estimators decrease across order statistics and sum to one, thus, in view of Theorem~\ref{th:comonotonic_CRE}, all six estimators \#1-\#6 are CREs.

\begin{table}[htp!]
    \centering
    \scalebox{0.8}{
    \begin{tabular}{ccccccccc}
        Estimator (Id) & $a_1$ & $a_2$ & $a_3$ & $a_4$ & $a_5$ & $a_6$ & $a_7$ & sum\\ \hline
        \#1 &  0.160 & 0.160 & 0.160 & 0.160 & 0.160 & 0.160 & 0.040& 1.000\\
        \#2 & 0.167 & 0.167 &  0.167 & 0.167 & 0.167 & 0.167 & 0.000 & 1.000 \\
        \#3 & 0.231 & 0.154 & 0.154 & 0.154 & 0.154 & 0.154 & 0.000& 1.000 \\
        \#4 & 0.239 & 0.159 & 0.159 & 0.159 & 0.159  & 0.117 & 0.006& 1.000 \\
        \#5 & 0.286 & 0.143 & 0.143 & 0.143 & 0.143 & 0.143 & 0.000 & 1.000 \\
        \#6 & 0.295 & 0.148 & 0.148 & 0.148 & 0.148 & 0.109 & 0.006&  1.000 \\\midrule
        \gray{\#VaR} & \gray{0.000} & \gray{0.490} & \gray{0.510} & \gray{0.000} & \gray{0.000} & \gray{0.000} & \gray{0.000}&  \gray{1.000}
    \end{tabular}}
    \caption{Order-statistic weights $(a_1,\ldots,a_7)$ of the ES estimators \#1--\#6 from Table~\ref{T:estimators}, for $\alpha=2.5\%$ and $n=250$; the remaining weights are zero. The weights decrease and sum to one for every ES estimator, consistent with the CRE representation~\eqref{eq:comonotonic_representation}. The \#VaR corresponds to $\alpha=1\%$ and $n=250$, and it is not a CRE, as the weights are not decreasing. Values are rounded to three decimals.}
    \label{T:weights_normalized}
\end{table}

We evaluate the estimators on two classes of data-generating models.

The \textit{first class is static}: the P\&L follows a process $Z_t=\varepsilon_t$ with i.i.d. innovations $\varepsilon_t$ from a fixed one-period law.   The set of considered laws consists of the standard normal law, the Student's $t$-law with $\nu=5$ degrees of freedom, six normal-inverse Gaussian (NIG) laws with
\[
(\alpha,\beta)\in\{(0.4,\pm0.14),\,(0.55,\pm0.3025),\,(0.4,\pm0.22)\},
\]
and two negative generalized Pareto distributions (GPD) with tail-shape parameter $\xi\in\{1/4,1/3\}$.  All distributions are standardized to zero mean and unit variance. Since all considered metrics are scale invariant, the scale normalization matters only for NIG,  where, following \cite[Annex~I]{EBA2023RTS05}, the parameters $(\alpha,\beta)$ refer to the standardized law, so that the scale and location parameters are fixed to $\delta=\tfrac{(\alpha^2-\beta^2)^{3/2}}{\alpha^2}$ and $\mu=-\delta\tfrac{\beta}{(\alpha^2-\beta^2)^{1/2}}$. The normal and Student~-~$t$ laws provide light- and moderate-tailed reference points. The NIG family spans a controlled grid of laws whose one-day parameters are compatible with values fitted to daily hypothetical trading-book P\&Ls over 2014--2022~\cite[Annex~I, p.~162]{EBA2023RTS05}, cf.~\cite{The2020}. As the NIG family is closed under convolution, the 10-day laws are again NIG, with closed-form skewness and kurtosis~\cite[Section~6.1]{ScottEtAl2011}, ranging from $(\pm0.946,3.628)$ through $(\pm1.360,4.505)$ to $(\pm1.870,8.517)$. The two GPD laws extend the design into the heavy-tailed regimes.

The \textit{second class is dynamic}: the P\&L follows a standard GARCH$(1,1)$ process $Z_t=\sigma_t\varepsilon_t$, $\sigma_t^2=\omega+\alpha Z_{t-1}^2+\beta\sigma_{t-1}^2$, with $t\in\bN$, i.i.d.\ innovations $\varepsilon_t$ that are either standard normal (GARCH-norm) or standardized Student-$t$ (GARCH-$t$) with $\nu=5$ degrees of freedom, and under a moderate-persistence parametrization $(\omega_1,\alpha_1,\beta_1)=(4\times10^{-6},0.17,0.80)$ (parameters fitted to the NASDAQ100 index in the period 2016--2021) and a near-integrated one $(\omega_2,\alpha_2,\beta_2) = (10^{-5},0.05,0.94)$ (parameters manually adjusted to reflect close-to-integrated regime), giving four dynamic models. In what follows we denote these models by GARCH-norm-$j$, GARCH-$t$-$j$, for $j=1,2$, correspondingly.  Furthermore, to increase sampling stability, noting an ergodic setup $\alpha+\beta<1$, we start simulation with long-run unconditional variance, i.e. we set $\sigma^2_0:=\omega/(1-\alpha-\beta)$. In each simulated path, we discard the first 500 observations as burn-in. The out-of-sample realizations $\tilde x_k$ used in RB and CT, and the samples used to compute the true risk values, come from separate independent paths with the same burn-in. All metrics therefore refer to the stationary marginal distribution. Unlike the static class, GARCH data are serially dependent through volatility clustering already in one-period form, which allows us to separate the effect of temporal dependence from that of the overlapping construction. We refer to \cite[Section II.4]{Ale2009} for details on GARCH modeling.

For each model we use two sampling schemes, reflecting the two standard ways of building P\&L series in market-risk practice. In the \textit{non-overlapping scheme} the sample $(X_i)_{i=1}^{250}$ is generated directly as $(Z_i)_{i=1}^{250}$, that is, i.i.d. $(Z_i)$'s for the corresponding static class  and a single trajectory $(Z_i)_{i=1}^{250}$ for the dynamic class. This encompasses the non-overlapping P\&Ls. In the \textit{overlapping scheme} each observation is a rolling sum of past ten consecutive values of the non-overlapping daily P\&L $Z_s$'s, namely, $X_i:=\sum_{j=0}^{9}Z_{i-j}$ with $(Z_i)_{i=-8}^{250}$ being  the underlying non-overlapping sample. This corresponds to the 10-day overlapping P\&Ls of the FRTB internal-model approach, cf.~\cite[MAR~33.4]{Bas2019} and~\cite[para.~97]{EGIM}. While the first setup might be seen as a 1-day setup and the second one as the 10-day setup, we do not rescale a 1-day ES to a 10-day horizon due to benchmark metric normalization. Namely, in each scheme, the estimation error for all benchmark metrics is measured against the true ES of the law that is actually sampled so that any time-period shift would cancel out. Still, under the overlapping scheme this target is the law of the 10-day aggregate, which might differ from the one-period law even for the static case -- it remains normal in the Gaussian case and NIG, with reduced relative skewness in the NIG case, and is computed numerically otherwise. We deliberately compare 1-day and 10-day samples, rather than drawing 250 i.i.d.\ observations from the 10-day aggregate law, because this matches the daily historical data available in practice. Each scheme is evaluated against the ES of its own marginal law, so scale effects cancel. However, the aggregation also changes the shape of the law. The comparison therefore mixes the effect of serial dependence with that of a target law closer to Gaussian, except the Gaussian case, where the 10-day law remains Gaussian and the effect of dependence is isolated. The market-data analysis in Section~\ref{S:market_data} complements this practice-oriented design, since the two effects cannot be separated on realized data.

As mentioned earlier, the estimators \#1--\#6 are built on the i.i.d. assumption underlying~\eqref{eq:ES}, in the sense that the estimators are based on integrals over quantile functions that are built using the i.i.d. assumption, so applying them to the serially dependent overlapping samples is a deliberate ``correct target, mis-specified estimator'' design. Note that the considered estimator construction approach is aligned with a typical market practice in which estimators built on i.i.d. samples are applied to market data, see Section~\ref{S:es.estimator.selection} for the discussion.

For every model, sampling scheme, and estimator we compute the nine benchmark metrics of Section~\ref{S:assessment}, namely ME, SE, DE for fit, SB, MB, RB for bias, and UP$_5$, UP$_{10}$, CT for safety, evaluated by Monte Carlo over $K=10^{6}$ independent learning samples; the true $\ES_{2.5\%}(X)$ and $\var_{1\%}(X)$ entering the metrics are taken in either closed form or from an independent large-sample Monte Carlo of the relevant 1-day or 10-day-aggregated law. The resulting tables span every model, scheme, and metric and are therefore long; we collect them in full in~\ref{S:tables} (Tables~\ref{tab:biases:table1}, \ref{tab:biases:table2}, and~\ref{tab:biases:garch}) while in the discussion that follows, we draw on excerpts and illustrations rather than reproducing the tables in the body. The discussion proceeds thematically, roughly split in subsections: we first contrast the non-overlapping i.i.d.\ benchmark with the overlapping and GARCH regimes and document the resulting shift in estimation error; we then examine how the order-statistic weighting traces out the fit--bias--safety trade-off across \#1--\#6; we turn to the capital adequacy of the secured positions; and we close with the ES-versus-VaR comparison, read as a diagnostic on ES-specific estimation rather than as a ranking.

\subsection{From non-overlapping to overlapping data: a regime shift} 

\begin{figure}[ht!]
    \centering
    \includegraphics[width=0.95\linewidth]{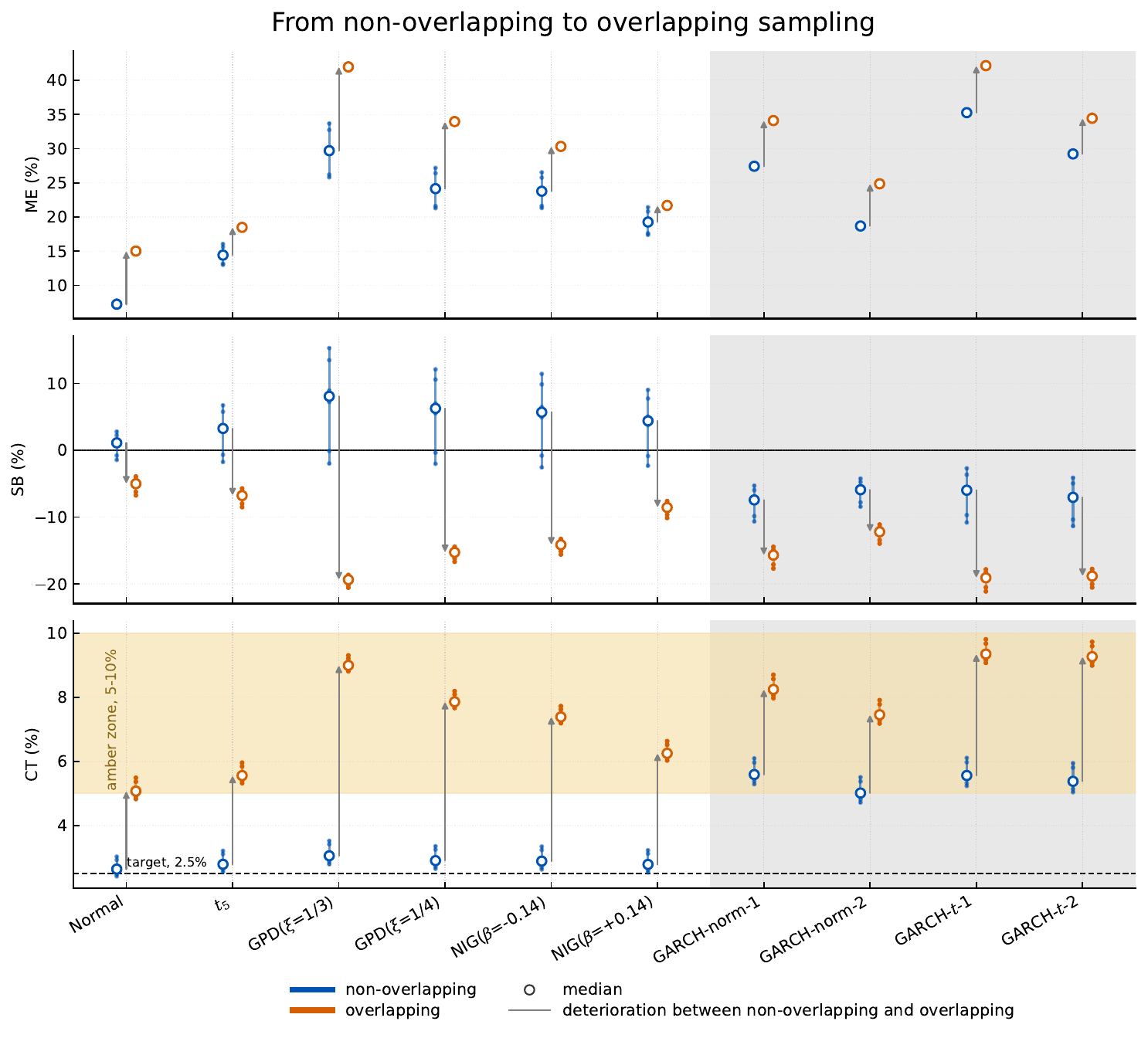}
    \caption{Metric values (range and median) across ES estimators \#1–\#6 for non-overlapping (blue) and overlapping (orange) schemes, by distribution (x-axis). ME (top), SB  (middle), and CT (bottom) correspond to the fit, bias, and safety groups, respectively; see Table~\ref{T:metrics.guide} for metrics' guide. For CT, the target value of 2.5\% is shown as a dashed line, with the amber zone (5–10\%) shaded in yellow. Vertical gray shading indicates dynamic GARCH models.   The overlapping sampling and serial dependence act in the same direction, jointly driving ES estimators toward systematic underestimation of risk.}
    \label{fig:non-verlap_overlap1}
\end{figure}

The numerical study reveals a significant performance decline between non-overlapping and overlapping sampling schemes, across all estimators, metrics, and classes of distributions. Figure~\ref{fig:non-verlap_overlap1} displays the range and median (circle markers) of estimators \#1–\#6 for three metrics, ME, SB and CT (one from each metric group), for representative distributions from the static and dynamic models. The non-overlapping scheme is shown in blue and the overlapping scheme in orange, with the arrow pointing towards deterioration of the corresponding metric. Other metrics exhibit similar trends; see also  Figure~\ref{fig:dumbell_grouped_all}, where we display the average of all metrics within the static and dynamic classes, respectively, for all estimators. Overall, our main conclusions are as follows.

We observe an across board \textbf{regime shift, not just a variance penalty}. Overlapping samples do more than inflate variance through a smaller effective sample size: they are associated with a directional bias, systematically underestimating risk across fit, bias, and safety metrics -- the most adverse direction for capital adequacy.

\textbf{Fit metrics deteriorate}, with median ME rising across the board from non-overlapping to overlapping scheme, Figure~\ref{fig:non-verlap_overlap1} (top). We also note that the dynamic class of distributions on average have worse fit metrics.  A similar pattern (not plotted) is observed for the SE metric, with the exception of two (out of six) NIG distributions, where the median SE is slightly smaller under the overlapping scheme. Figure~\ref{fig:dumbell_grouped_all} shows the average fit metrics (ME, SE, and DE) across models, where the deterioration from the non-overlapping to the overlapping sampling scheme is clearly apparent.

\textbf{Bias metrics reverse direction}, where for the stationary i.i.d. distributions median SB is near zero for normal or positive for others under non-overlapping data, it turns substantially negative under overlapping data, with values ranging from about $-5\%$ (Normal) to $-19\%$ (GPD $\xi=1/3$), see Figure~\ref{fig:non-verlap_overlap1} (middle). For the GARCH models, the bias becomes more negative for all estimators under both schemes, by roughly 7--11 percentage points for non-overlapping and 7--13 percentage points for overlapping samples, relative to the i.i.d.\ model with the same innovation family (Gaussian or Student-$t$); see Figure~\ref{fig:non-verlap_overlap1} (middle, gray shaded area). For MB and RB see Figure~\ref{fig:dumbell_grouped_all} (middle panels).

\textbf{Safety decreases}, with  UP$_5$, UP$_{10}$, and CT rising sharply.  Notably, median CT remains close to the target threshold of 2.5\% for most static models under the non-overlapping sampling scheme, with the GPD and left-skewed NIG distributions slightly elevated relative to the other static models; see Figure~\ref{fig:non-verlap_overlap1} (bottom). In contrast, the overlapping sampling scheme pushes all ES estimators towards the amber zone ($5$--$10\%$) of Section~\ref{S:assessment} across all models, with CT values clustering between roughly 5\% and 9.8\%. Moreover, for the dynamic GARCH-type models, even the non-overlapping scheme crosses into the amber zone, with CT already at 5--6\%. Underestimation probabilities routinely exceed 50\% and reach 57--79\% under the GARCH models; see Figure~\ref{fig:dumbell_grouped_all} (right).

We also note that the \textbf{serial dependence alone reproduces much of the deterioration}, without considering the overlapping scheme; Figure~\ref{fig:non-verlap_overlap1}, gray shaded region. The non-overlapping GARCH biases are more negative than the corresponding overlapping biases of the stationary light-tailed i.i.d. sequence. For example, estimator \#1 for GARCH-norm-1 under non-overlapping scheme has $\textrm{SB} =-10.7\%$, compared with $\textrm{SB} =-6.7\%$ for the normal distribution under the overlapping scheme. Deterioration of safety metrics is also apparent, as discussed above.    

The \textbf{effective sample size shrinks} sharply. Building 250 overlapping 10-day sums corresponds roughly to 25 independent 10-day blocks, so the nominal sample of $n=250$ carries far less information than its size suggests. Quantifying this precisely is hard, since existing results for L-statistics are asymptotic in nature \citep{Bak2006,KalOko2011}. Moreover, the effective sample size is distribution-dependent. The Gaussian case provides a clean, tractable indicator of this loss: matching the sampling variance of the empirical plug-in estimator \#1 for the overlapping sample of $n=250$ with that of an i.i.d. sample corresponds to only $n \approx 57$ independent observations, a more than fourfold loss of effective precision. This back of the envelope calculation, however, isolates only the variance penalty of overlap, and it says little about the changes in bias or shape.

Taken together, these results point to a compounded failure, not an isolated one. Overlapping sampling and serial dependence act in the same direction, jointly driving ES estimators toward systematic underestimation of risk. 
These findings motivate the next analysis, where we examine the trade-off between fit, bias, and safety across the different ES estimators.

\subsection{Fit, bias, and safety trade-offs}

\begin{figure}[ht!]
\centering
    \centering
    \includegraphics[width=.99\linewidth]{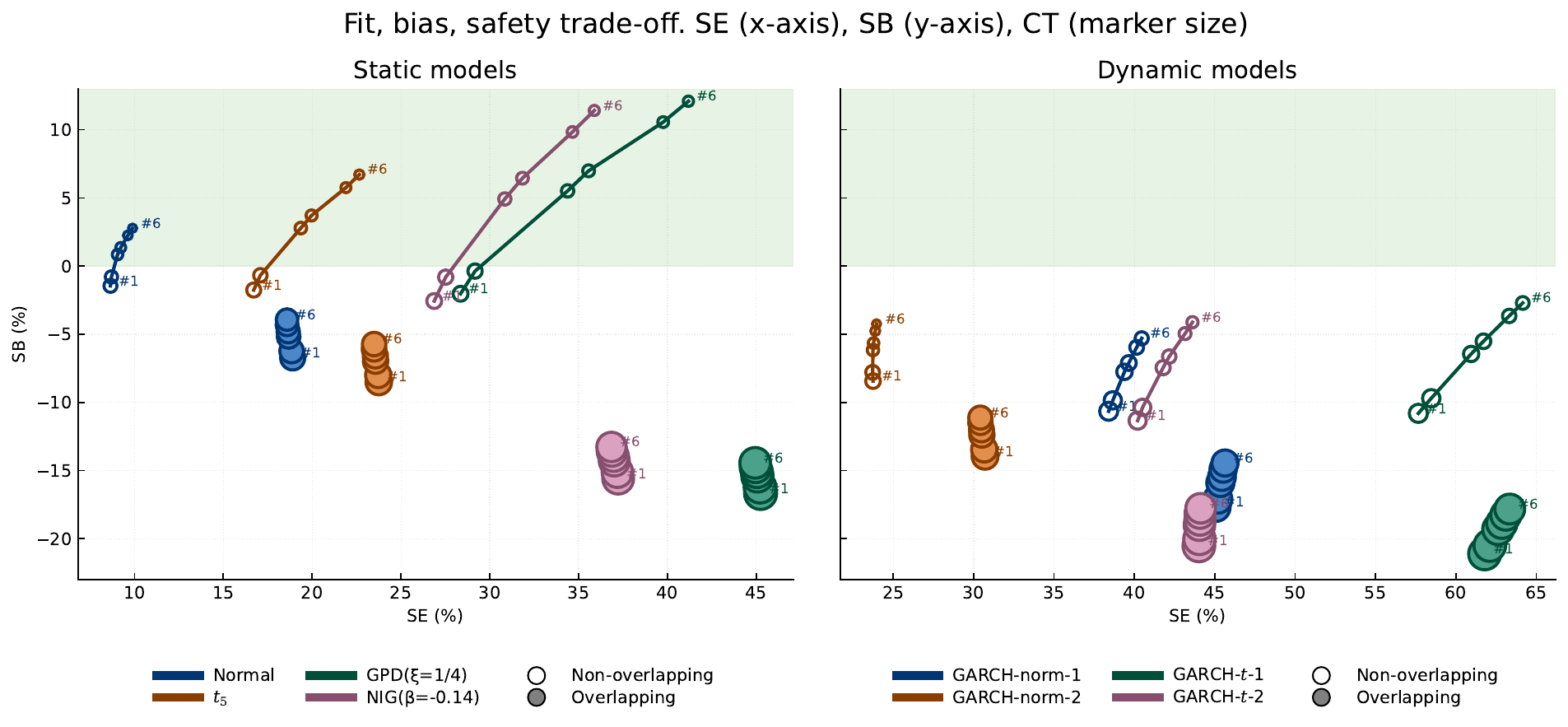}
\caption{ SE (x-axis), SB (y-axis), CT (size) metrics for static (left) and dynamic (right) series, and for non-overlapping (open circle) and overlapping (filled circles). Size of the markers corresponds to the CT value. There is no best CRE that would optimize across (symmetric) fit, bias and safety.} 
\label{fig:scater-fit-bias-tradeoff}
\end{figure}

Within the family of CREs, in particular those investigated herein, there is no universally superior estimator across all metrics. A particular choice of weights $(a_i)$ in \eqref{eq:comonotonic_representation} (see also Table~\ref{T:weights_normalized}) identifies a trade-off between fit, bias, and safety metrics. A snapshot of the interaction between the SE (fit) and SB (bias) metrics for the ES estimators is presented in Figure~\ref{fig:scater-fit-bias-tradeoff}. Open circles correspond to estimators applied to non-overlapping scheme, while filled circles to overlapping sampling scheme; the left panel is dedicated to static models, and the right panel to dynamic GARCH. The size  of the marker reflects the relative value of CT (safety). The light-green shaded upper  half plane corresponds to  risk overestimation. 

\textbf{The magnitude of the fit and bias trade-off depends on the sampling scheme.} Under non-overlapping scheme, the trade-off between fit and bias is clear, for both static and dynamic models: shifting the weights towards the first order statistic $a_1$ improves the bias (SB increases), while fit deteriorates (SE increases). Although, for dynamic GARCH model, even the most conservative ES estimator \#6 remains in the undesirable region $\textrm{SB}\leq 0$. For the overlapping scheme (filled circle markers), the trade-off between SE and SB vanishes especially for the static models (left panel). Moreover, for every static model under overlapping sampling, SE is almost unchanged while SB rises from \#1 to \#6. Notably, the CT (markers' relative size) also improves from \#1 to \#6, and as already mentioned earlier, it is significantly larger for overlapping schemes. Similar results hold true for ME (fit) and MB, RB (bias), as well as for the  UP$_5$, UP$_{10}$ (safety). However, the DE metric from the fit group, which in contrast to ME and SE is not symmetric, behaves similar to bias and safety metrics---decreases monotonically from estimator \#1 to \#6, across all distributions and sampling schemes, with larger gap for non-overlapping scheme; see also Figure~\ref{fig:dumbell_grouped_all}.  This is because conservative over-shooting contributes to symmetric errors ME and SE.  

Overall, as expected, \textbf{there is no best CRE that would optimize across (symmetric) fit, bias and safety}. Similar to the statistical variance-bias trade-off or the mean-variance trade-off in classical portfolio optimization, when designing an estimator one has to make choices that balance the statistical errors and capital adequacy. This choice can itself be cast as an optimization problem with additional constraints, reflecting regulatory or institutional priorities, that would map directly onto policy choices.

\subsection{Estimator performance and capital adequacy}

\begin{figure}[ht!]
    \centering
    \includegraphics[width=0.99\linewidth]{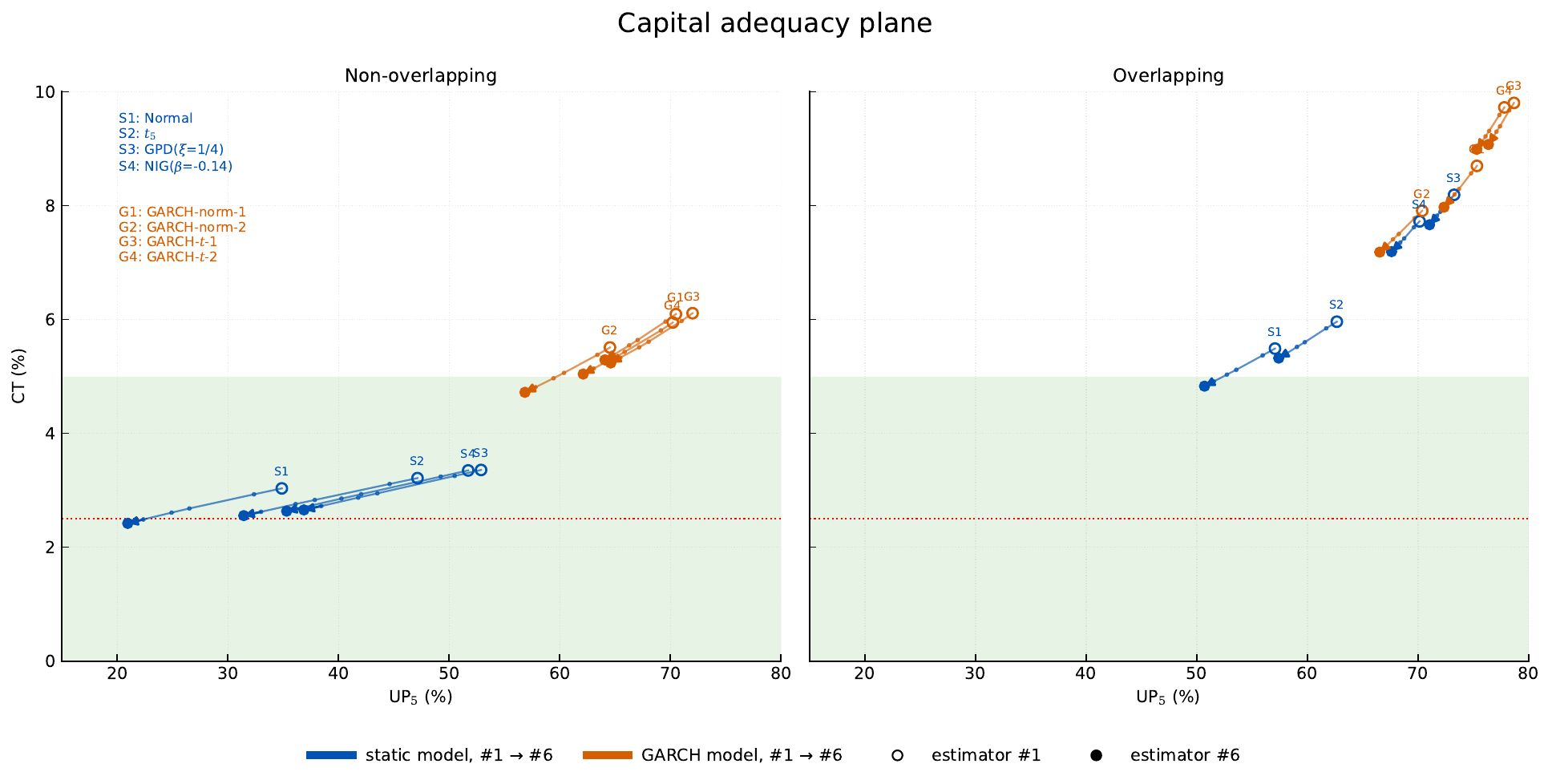}
    \caption{Safety metrics UP$_5$ vs CT for static (blue) and dynamic GARCH (orange), and for non-overlapping (left) and overlapping (right) sampling schemes. Horizontal dotted line corresponds to CT target value 2.5\%. Shaded green region reflects the capital adequacy safe zone CT $\leq 5\%$. The gradient arrow points from estimator \#1 to \#6. The apparent conservatism of the ES estimators is insufficient to ensure capital adequacy once heavy tails, serial dependence, and overlapping sampling are jointly present. }
    \label{fig:cap_adequacy1}
\end{figure}

We recall that, first and foremost, risk measures such as ES and VaR serve as tools to evaluate and determine the capital adequacy of a financial entity. The metrics in the safety group, and RB from the bias group, are designed to measure the degree to which an estimator's implied capital requirement satisfies this adequacy standard, rather than its accuracy in a purely statistical sense. We focus on CT vs UP$_5$, Figure~\ref{fig:cap_adequacy1}. For CT, a safety zone can be defined as violation rates below 5\%, cf.~Section~\ref{S:assessment}, shaded in green. In contrast, there is no clear reference value for the UP$_5$ and UP$_{10}$ metrics, and only the direction is clear with smaller values better. Accordingly, we use them in a comparative rather than absolute sense, ranking estimators relative to one another instead of against a fixed target.

The CT target  value of 2.5\% is marked by a horizontal red dotted line. Labels S1-S4 and blue color correspond to static models, and dynamic GARCH models G1-G4 are depicted in orange. The gradient arrow points from the least conservative estimator \#1 to most conservative \#6.  

\textbf{ES estimators perform well for static and non-overlapping schemes}, with more conservative estimators \#3-\#6 pushing performance deeper into the safe zone closer to the target CT value, even for heavier-tailed distributions. The UP$_5$ values are higher for heavier tailed static distributions. Serially correlated \textbf{GARCH models consistently remain in the adverse zone}, and only for non-overlapping data the more conservative estimators \#5 and \#6 for GARCH-norm-2 cross the CT safe threshold; Figure~\ref{fig:cap_adequacy1}, orange markers.

For all models the \textbf{shift in safety from non-overlapping to overlapping is significant.} On average across static distributions and estimators \#1--\#6, CT increases by roughly  138\% (from averages about 2.91\% to 6.95\%) and UP$_5$ by roughly  62\% (from averages  about 40.7\% to 65.8\%). For dynamic GARCH models, the CT increases by 59\%, and UP$_5$ by 13\%. Overall, the shift from non-overlapping to overlapping sampling raises CT by roughly 104.4\% and UP$_5$ by roughly 42.7\%, averaged across all models and estimators \#1–\#6. Putting more weight on lower order statistics in CRE mitigates some of the decline in safety, but only for light-tailed distributions and non-overlapping regime, and even then the estimator \#6 barely reaches the CT safety target of 2.5\%. The serial correlation in GARCH models amplifies the error further.

The risk bias metric RB shows that only the lightest-tailed distributions under non-overlapping sampling come close to a well-capitalized position (see Table~\ref{tab:biases:table1} and \ref{tab:biases:table2}). It turns positive for the Normal by estimator \#5, and stays  negative for the remaining static distributions across all estimators, whereas \textbf{under overlapping data RB remains deeply negative} across all estimators and all distributions;   about -12\% to -14\% for the Normal,  -39\% to -40\% for GPD($\xi=1/3$), comparable to some GARCH models. We also note that in the overlapping scheme GPD($\xi =1/4$) and NIG($\beta = -0.14$) are in the same region as GARCH models. 
Thus, in terms of capital adequacy, heavy tails combined with the overlapping scheme can have an effect similar to that of GARCH-type dynamics.

$\var$ exhibits the same pattern at materially lower levels. Overall, these results show that the apparent conservatism of the ES estimators is insufficient to ensure capital adequacy once heavy tails, serial dependence, and overlapping sampling are jointly present. This emphasizes the need for internal validation of CREs, particularly when the sampling scheme or dependence structure departs from the implicit i.i.d. assumptions under which the estimators were built in the first place. A natural next step is to assess the magnitude of this deterioration relative to some benchmarks, such as VaR estimators, which we do next.

\subsection{ES vs VaR}

\begin{figure}[ht!]
    \centering
    \includegraphics[width=0.95\linewidth]{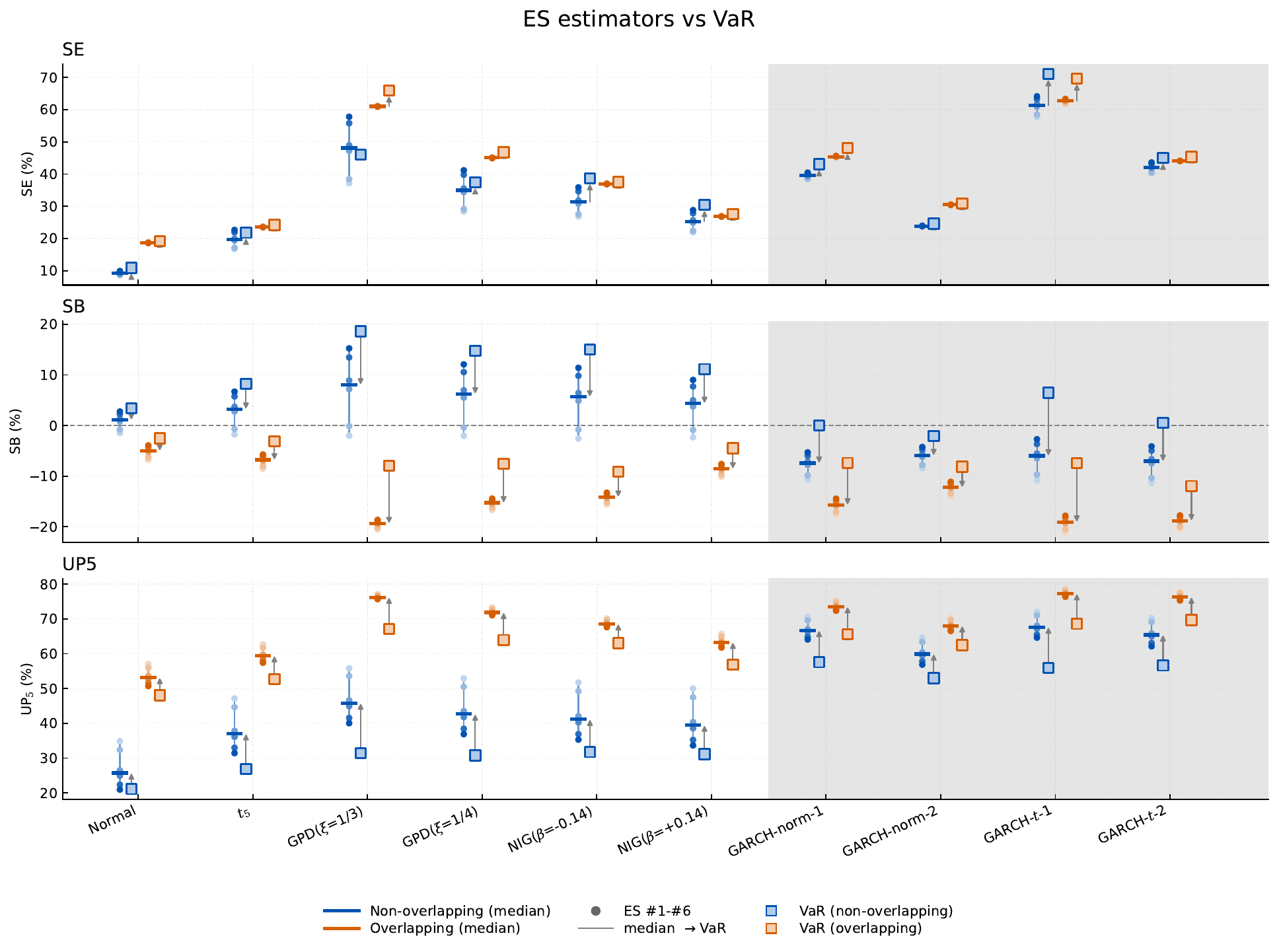}
    \caption{SE (top), SB (middle), and $\mathrm{UP}_5$ (bottom) values for the ES estimators \#1--\#6 (dot markers, colored by estimator's conservatism) and \#VaR (square markers), for selected models -- static models on the left, dynamic GARCH models on the right (shaded area). Blue corresponds to the non-overlapping scheme and orange to the overlapping scheme. The color intensity of the dot markers reflects the weight placed on the lower order statistics in the CRE (darker shades indicate more conservative estimators). Each distribution's short horizontal bar marks the median across estimators \#1--\#6, and the thin vertical line spans their min--max range. The gray arrow links the median ES and \#VaR, with the arrowhead landing at whichever is worse, indicating the direction and magnitude of deterioration.  
    While direct comparison between ES and VaR estimators is delicate, the results suggest a general deterioration in estimation quality for ES under overlapping sampling and serial dependence.}
    \label{fig:ES_vs_VAR}
\end{figure}

We recall that $\ES_{2.5\%}$ is the quantity of regulatory interest under the FRTB, while $\var_{1\%}$ is used mostly for historical reasons and as a reference point -- in particular because, for P\&L that follows a normal distribution, $\ES$ at significance level $2.5\%$ coincides with $\var$ at significance level $1\%$. Heavier-tailed distributions, and/or serially correlated models, may yield significantly different values. Thus, comparing $\ES$ and $\var$, and especially the performance of their respective estimators, is a delicate matter. A snapshot of three metrics, SE, SB, and UP$_5$ for ES and VaR is presented in Figure~\ref{fig:ES_vs_VAR}. The blue markers correspond to non-overlapping sampling, while orange markers correspond to overlapping sampling. The dot markers, colored by estimator conservatism, and square markers correspond to \#VaR, for selected models: static models on the left, and dynamic GARCH models on the right (shaded area). The color intensity of the dot markers reflects the weight placed on the lower-order statistics in the CRE (darker shades indicate more conservative estimators). The horizontal bar marks the median across estimators \#1--\#6, and the thin vertical line spans their min--max range. The gray arrow links the median ES and \#VaR, with the arrowhead landing at whichever is worse, indicating the direction and magnitude of deterioration.

The considered \textbf{ES estimators provide a better symmetric fit} than \#VaR. The median SE of ES estimators \#1-\#6 is lower than of \#VaR, with one exception for GPD($\xi=1/3$) in the non-overlapping case, Figure~\ref{fig:ES_vs_VAR} (top). Similarly, the median ME for ES is better than the ME for $\var$ for static non-overlapping, and remains roughly the same for dynamic models and the overlapping scheme; see Table~\ref{tab:biases:table1} and \ref{tab:biases:table2}. In contrast, DE is smaller for \#VaR across the board, with only exception for estimator \#6 for Normal and non-overlapping. 
At equivalent risk capital levels, our results challenge the common perception that ES estimators are inherently noisier than VaR ones, although this should not be generalized to broader claims, since other sensitivity measures continue to favor VaR, cf. \cite{YamYos2005,KelRos2016,DanZho2016}.

\textbf{Bias is directional and regime-dependent}. 
Under non-overlapping data and static models, the median SB metric for ES estimators (roughly corresponding to \#3-\#4) is  1.1\%–4.6\% for Normal, $t_5$, and right-skewed NIG, rising to 5.7\%–6.7\% for left-skewed NIG and to 6.3\%–8.1\% for GPD, while SB for every ES estimator turns materially negative once the sampling becomes overlapping, with magnitudes of roughly  3.9\%-20.6\% across estimators and models. For dynamic GARCH models, the median SB of ES estimators is already more negative under non-overlapping data, ranging from about $-6\%$ to $-7.5\%$ across the four specifications, and deteriorates further under overlapping to roughly $-12\%$ to $-19\%$. The \#VaR estimate is more conservative than the ES estimators throughout. Its SB sits further to the positive side under non-overlapping static data ($+3.4\%$ to $+18.6\%$, versus $+1.1\%$ to $+8.1\%$ for the ES median), but conservatism and accuracy point in opposite directions here: under non-overlapping static data, the ES median is in fact closer to zero, and hence less biased, than \#VaR. This relationship reverses once serial dependence enters the picture: under overlapping data and under GARCH dynamics, SB for \#VaR sits closer to zero than the ES median (except for GARCH-$t$-1 under non-overlapping data, where it overshoots), whose bias instead remains strongly negative. In particular, under non-overlapping GARCH data, SB for \#VaR ranges from mildly negative to positive ($-2.1\%$ to $+6.5\%$ across the four GARCH specifications), while SB for the ES estimators remains negatively biased throughout ($-5.9\%$ to  -7.5\% for the median). Estimating ES thus carries a larger directional under-estimation than \#VaR specifically in serially correlated samples, such as overlapping and dynamic GARCH,  which is exactly where FRTB compliance is most at stake.

\textbf{Tail sensitivity is intrinsic to ES}. By its very nature ES and its estimators are less stable than VaR at the same confidence level, since ES integrates over the whole left tail and a misspecified or under-sampled tail propagates into the estimates. In contrast, the single-quantile $\var$ estimator is relatively isolated. Consequently, in terms of underestimation, the \#VaR estimator is more stable across distributions with different tail characteristics, while the ES estimators react more strongly to tail heaviness. Under non-overlapping static data, the median $\textrm{UP}_5$ of the ES estimators ranges from 25.7\% (Normal) to 45.8\% (GPD($\xi=1/3$)), against 21.1\% to 33.0\% for \#VaR. The RB metric (see Table~\ref{tab:biases:table1} and \ref{tab:biases:table2}) shows even stronger tail stability of $\var$, staying between $+0.7\%$ and $+2.4\%$ for \#VaR, while the ES median falls from $-1.0\%$ to $-8.4\%$. Heavier tails thus push \#VaR towards over-estimation (its SB rises from $+3.4\%$ to $+18.6\%$) and the ES estimators towards under-estimation. This stability does not, however, extend to serial dependence. When moving to overlapping data, $\textrm{UP}_5$ of \#VaR rises at least as much as that of the ES median for all models except the Normal (e.g., by 35.7\% compared with 30.5\% for GPD($\xi=1/3$)), and its RB drops by a comparable or larger amount (from $+2.3\%$ to $-31.7\%$, compared with $-8.4\%$ to $-39.2\%$ for the ES median in the same case). Under overlapping data, \#VaR therefore remains less under-estimated than ES mainly because it starts from a more conservative level, not because it moves less. Both underestimation probability metrics $\textrm{UP}_5$ and $\textrm{UP}_{10}$ are higher for ES, with the only exception estimator \#6, Normal, non-overlapping. This greater sensitivity to tail heaviness is inherent to the tail-oriented philosophy of ES rather than a shortcoming of a particular estimator; it is the analytical price of measuring tail severity rather than a single quantile.

\textbf{ES estimation adequacy requires enhanced monitoring.} The increased tail sensitivity of ES results in the increased need of estimator stability monitoring and a backtesting framework that goes beyond regulatory VaR backtesting prescribed in FRTB \cite[MAR 32]{Bas2019}. In fact, ES specific supervisory mitigants are already put in place. This refers to both direct ES backtesting need, see e.g.~\cite[Art. 21(1)(b)]{EU2024_1085} and \cite[Section~4.3.2]{EGIM}, or a $\mathrm{ES}_{2.5\%}/\var_{1\%}$ ratio test, see \cite[Section~4.3.2]{EGIM}. 
Regarding the validation of ES forecasts, several ES backtests have also been proposed in the literature; cf.  \cite{Acerbi2014Risk,DuEsc2017,MolPit2017,KraLokMcN2018,BayDim2022}. That said, to the best of our knowledge there is currently no standard practice that would sufficiently quantify the aforementioned risks. To understand this better we take the ES/VaR ratio and safety metric CT, and study them on market data in the next section.

In summary, while direct comparison between ES and VaR estimators is delicate, the presented results show a general deterioration in estimation quality for ES under overlapping sampling and serial dependence, pointing to the need for more extensive internal validation and stricter risk management when ES is used under the updated framework. We next turn to market data to assess how ES estimators behave in a more practical environment.

\section{Market data analysis}\label{S:market_data}

Realized market P\&Ls exhibit a considerably richer structure than the canonical specifications of Section~\ref{S:numerical.study}, that is, their volatility dynamics are not confined to a stationary GARCH(1,1) with fixed parameters, and the tails, skewness, and dependence they display shift over time, cf. \citet{DanJamVal2016}. This motivates the analysis of this section, in which we assess the estimators directly on market data. Existing work on overlapping constructions has largely addressed the sampling standard error they induce and the design of multi-period forecasts; see~\cite{SumNelHanGuo2009,AicCroReh2021,RuiNie2023}. What these works leave open is the estimation risk that bears on the sufficiency of the capital number itself. In other words, whether the estimated reserve built by an i.i.d.-calibrated estimator from 250 overlapping scenarios actually secures the position it is meant to secure has received far less attention on realized data, despite its importance. To assess whether the conclusions of Section~\ref{S:numerical.study} carry over, we evaluate the estimators \#1--\#6 on two datasets, listed in Table~\ref{T:market.data}. The first dataset covers twelve risk-factor level series from five asset classes.  The second dataset represents composite (non risk-factor level) returns, and consists of the diversified equity portfolio returns of {\it 100 Portfolios Formed on Size and Book-to-Market} from the Fama--French dataset.

\begin{table}[htbp]
\centering
\scalebox{0.8}{
\begin{tabular}{|p{3cm}|p{6.5cm}|p{2.8cm}|p{3.5cm}|p{2.4cm}|}
\hline
\textbf{Category} & \textbf{Series / dataset} & \textbf{Source (ticker)} & \textbf{Time range} & \textbf{Data count} \\
\hline\hline
\multirow{5}{*}{Interest Rate}
  & SOFR        & NY FED      & 02.04.2018 - 31.12.2025 & 1936\\
  & US 1M Bill  & FRED        & 04.01.2010 - 31.12.2025 & 4173\\
  & US 2Y Note  & FRED        & 04.01.2010 - 31.12.2025 & 4173\\
  & US 10Y Note & FRED        & 04.01.2010 - 31.12.2025 & 4173\\
  & DE 10Y Bund & DB          & 04.01.2010 - 30.12.2025 & 4061\\
\hline
Credit
  & Xtrackers II iTraxx Crossover Short Daily Swap UCITS ETF 1C
                & YF (XTC5.MI) & 04.01.2010 - 30.12.2025 & 3078\\
\hline
\multirow{2}{*}{Equity}
  & S\&P 500 & YF (\textasciicircum GSPC) & 04.01.2010 - 30.12.2025 & 4023 \\
  & VIX      & YF (\textasciicircum VIX)  & 04.01.2010 - 30.12.2025 & 4023 \\
\hline
\multirow{2}{*}{FX}
  & EUR/USD Spot Rate & YF (EURUSD=X) & 01.01.2010 - 30.12.2025 & 4164 \\
  & USD/JPY Spot Rate & YF (JPY=X)    & 01.01.2010 - 30.12.2025 & 4164 \\
\hline
\multirow{2}{*}{Commodities}
  & Brent Oil Futures Price & YF (BZ=F) & 04.01.2010 - 30.12.2025 & 3992 \\
  & Gold Futures Price & YF (GC=F) & 04.01.2010 - 30.12.2025 & 4022 \\
\hline\hline
Portfolios (Equity) & 100 Portfolios Formed on Size and Book-to-Market & FF & 01.01.2010 - 31.12.2025 & $4024 \times 99$ \\
\hline
\end{tabular}
}
\caption{Market data used in the analysis. The first twelve rows are risk-factor level series; the last row is the portfolio level dataset. Quotes were sourced from public online data repositories linked to NY FED (Federal Reserve Bank of New York), FRED (Federal Reserve Economic Data), DB (Deutsche Bundesbank), YF (Yahoo Finance), and FF (Fama--French data library, see \cite{FamFre2026}). Following data quality checks, one Fama--French portfolio (ME9 BM10) was excluded due to incomplete records, so that data consists of 4024 daily relative returns for 99 portfolios.}\label{T:market.data}
\end{table}

To construct the \textit{market data set} used for estimation, we use the daily close price series $(S_i)_{i=0}^{n}$ of the risk factors listed in Table~\ref{T:market.data}, where the data count $n$ is reported in the last column. Following market practice, cf.~\cite[para.~87]{EGIM}, daily returns are taken to be \textit{absolute returns} $Z_i := S_i - S_{i-1}$ for the interest rate and credit factors, and \textit{relative returns} $Z_i := (S_i - S_{i-1})/S_{i-1}$ for the equity, FX, and commodity factors, $i = 1,\ldots,n$. The Fama--French series are already reported as daily relative returns and are used directly as $(Z_i)_{i=1}^{n}$. Consequently, in analogy with the two sampling schemes of Section~\ref{S:numerical.study}, we construct two time series from each market data set: \textit{non-overlapping 1-day returns} and \textit{overlapping 10-day returns}. In this \textit{non-overlapping scheme} we set $X_i := Z_i$. In the \textit{overlapping scheme} each observation is the 10-day return aggregated from the ten preceding daily returns, namely $X_i := \sum_{j=0}^{9} Z_{i-j} = S_i - S_{i-10}$ for absolute returns, and $X_i := \prod_{j=0}^{9}(1 + Z_{i-j}) - 1 = (S_i - S_{i-10})/S_{i-10}$ for relative returns. The former is precisely the rolling sum used for the simulated data in Section~\ref{S:numerical.study}, where the daily values play the role of log-returns and hence aggregate additively; for relative returns, summation is replaced by compounding. Finally, for each data set the estimators are computed at each point using the previous 250 data points.

\begin{figure}[htp!]
\centering
\begin{minipage}[c]{0.42\linewidth}
\centering
\footnotesize
\begin{tabular}{l|ccc}
\toprule
Indicator & $\hat\xi$ & Excess Kurtosis & Skewness\\ \midrule
SOFR & 0.97 & 423.56 & 1.44\\
US 1M Bill & 0.48 & 164.26 & 5.70\\
US 2Y Note & 0.40 & 13.31 & -0.77\\
US 10Y Note & 0.36 & 1.65 & -0.02\\
DE 10Y Bund & 0.37 & 3.81 & -0.08\\
Xtrackers & 0.52 & 16.38 & 0.14\\
S\&P500 & 0.42 & 6.45 & -0.43\\
VIX & 0.29 & 25.05 & 2.60\\
EUR-USD & 0.30 & 3.47 & 0.15\\
USD-JPY & 0.35 & 6.19 & 0.23\\
Brent & 0.33 & 18.44 & 0.40\\
Gold & 0.32 & 4.55 & 0.09\\\bottomrule
\end{tabular}
\end{minipage}\hfill
\begin{minipage}[c]{0.56\linewidth}
\centering
\includegraphics[width=\linewidth]{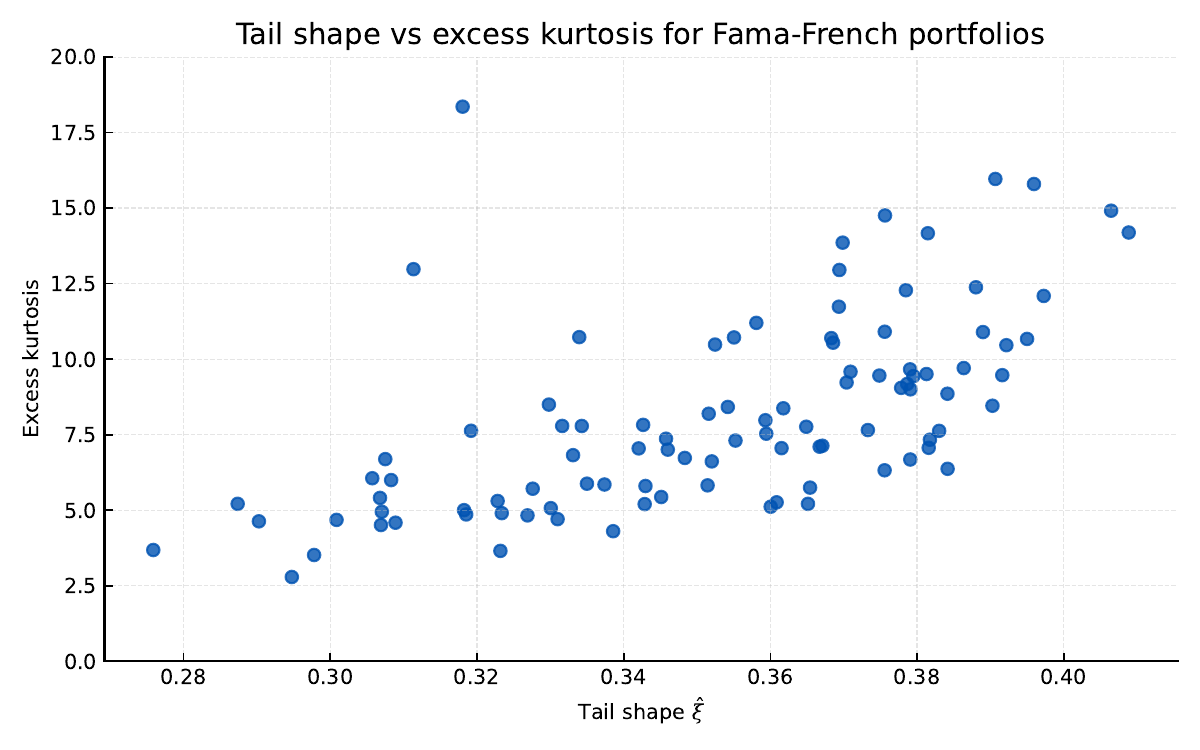}
\end{minipage}
\caption{Market data 1-day return tail heaviness diagnostics. Left: tail shape parameter $\hat\xi$, sample excess kurtosis, and sample skewness for the market data indicators. Right: tail shape parameter $\hat\xi$ against sample excess kurtosis for the daily returns of Fama--French portfolios. In both panels, the estimates of $\hat\xi$ are based on  Hill estimator and use the $5\%$ smallest observations.}
\label{F:xi_market}
\end{figure}

Before turning to risk estimation we briefly characterize the data. Figure~\ref{F:xi_market} reports selected 1-day tail characteristics  of $(Z_i)$: for the risk-factor series (left exhibit) the GPD tail-shape estimate $\hat\xi$, obtained from the Hill estimator applied to the $5\%$ smallest observations together with the sample kurtosis and skewness, and for the Fama--French portfolios (right panel) $\hat\xi$ against sample excess kurtosis, obtained with the same logic.

Most series display moderately heavy left tails, with $\hat\xi$ between roughly $0.3$ and $0.5$, that is, at and above the upper end of the range spanned by the GPD specifications ($\xi\in\{1/4,1/3\}$) and comparable to the GARCH-$t$ models, whose stationary returns are heavier-tailed than their innovations, so the simulated regimes of Section~\ref{S:numerical.study} provide natural reference points for the market results. The two short-rate series are the exception: their sample period combines a prolonged near-zero-rate regime with abrupt hiking cycles, giving $\hat\xi$ estimates of $0.97$ and $0.48$ and extreme kurtosis, so that any i.i.d.-based non-parametric estimator is severely mis-specified before the overlapping construction even enters. We retain these two series in the panel as a stress illustration and comment on them separately in Section~\ref{S:market.ct}. For the Fama-French long equity portfolios the skewness is not displayed; as expected, the returns are skewed to the left, with $90\%$ of the estimates being negative and a minimum of $-0.68$. As expected, the estimated tail shape and excess kurtosis for the Fama-French portfolios are roughly in line with those of the risk factors, spanning 0.25--0.41 and 2.5--20.0, respectively.

\begin{figure}[htp!]
\centering
\begin{minipage}[c]{0.49\linewidth}
\centering
\includegraphics[width=\linewidth]{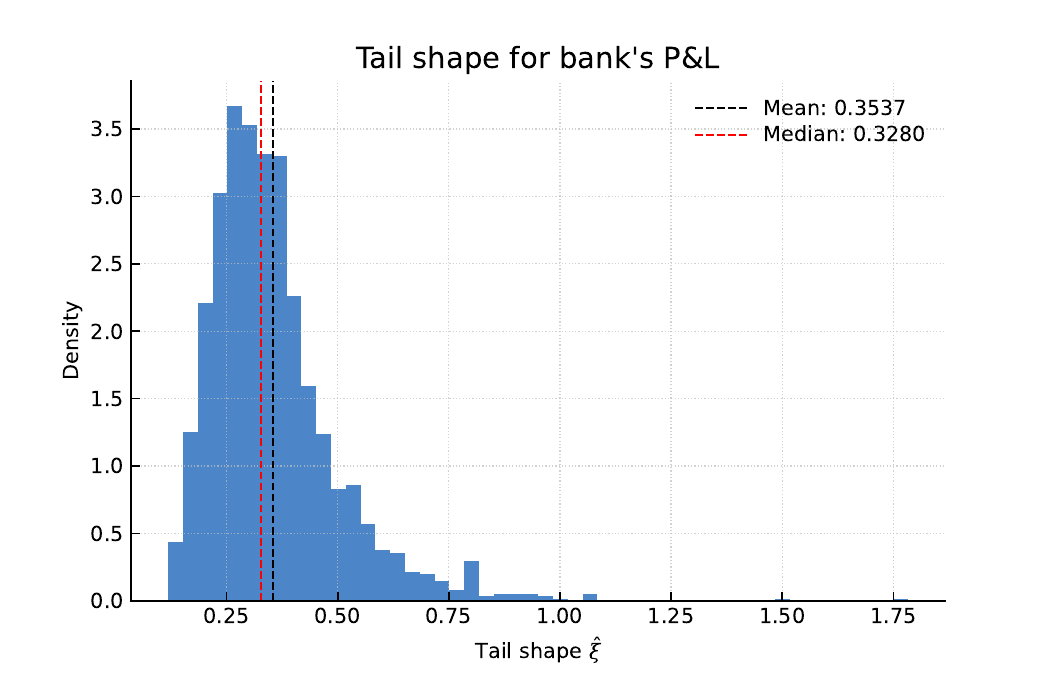}
\end{minipage}\hfill
\begin{minipage}[c]{0.49\linewidth}
\centering
\includegraphics[width=\linewidth]{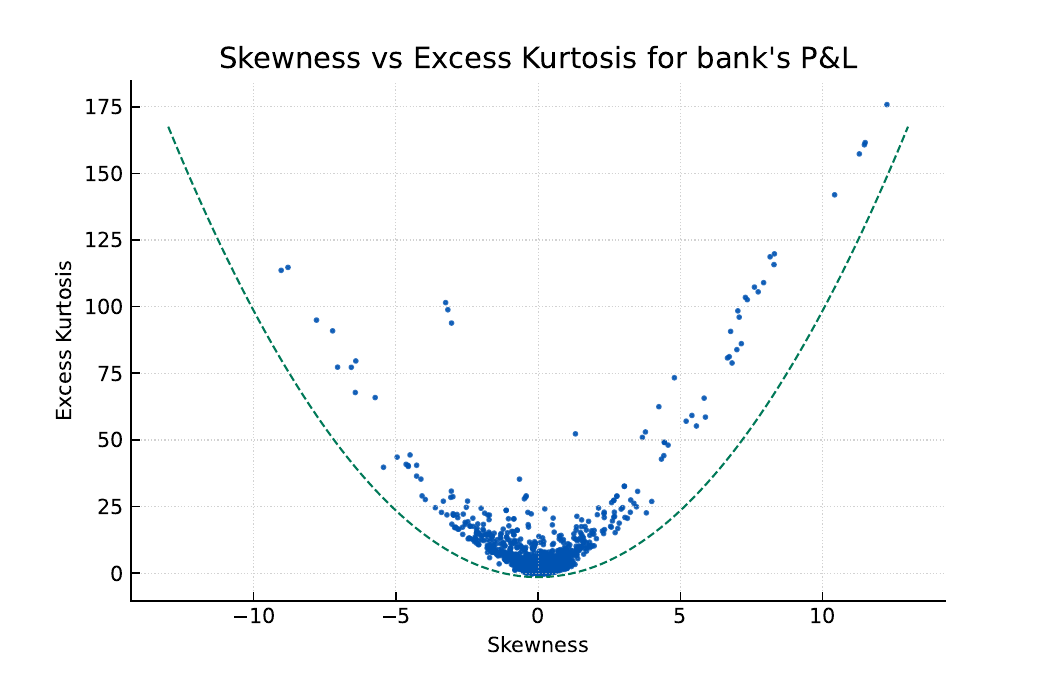}
\end{minipage}
\caption{Sample skewness and excess kurtosis diagnostics for quarterly shifted one-year moving blocks within ten years of 1-day P\&L data from European banks' regulatory trading books. Left: tail shape parameter $\hat\xi$ estimated with the Hill estimator applied to $5\%$ left tails (losses). The obtained $\hat\xi$ estimates indicate a heavy-tailed data structure that substantially departs from Gaussianity. Right: sample skewness and excess kurtosis for the considered data (sample moments as in \cite{JoaGil1998}); the dashed line marks the \cite{KlaassenEtAl2000, KlaassenVanEs2025} 
bound for unimodal distributions, $\gamma_2 \ge  \gamma_1^2 -\frac{186}{125}$ with $\gamma_1$ and $\gamma_2$ being skewness and excess kurtosis, respectively. Again, substantial deviations from Gaussianity are apparent.}
\label{F:PnL_Xi_Hill_estimates}
\end{figure}

To verify whether the considered market data tail characteristics are aligned with production-level portfolio P\&Ls, we also took anonymized standardized regulatory trading book buy-and-hold (hypothetical) 1-day P\&L data from 2014 to 2024 of large European banks with an internal model for market risk under the Basel~2.5 framework reported to the ECB as their supervisory authority and estimated their tail characteristics\footnote{We thank the ECB for providing the anonymized and standardized data from which the P\&L characteristics were extracted.}, see Figure~\ref{F:PnL_Xi_Hill_estimates}; here we used $n=252$ trading days as one year and a rolling window with quarterly ($63$ day) shifts, which resulted in 1849 samples. 
Because trading books hold both long and short positions, realized skewness takes both signs and excess kurtosis reaches high values, in part because P\&Ls load on interest rate factors subject to the same regime shifts visible in the short-rate series of Figure~\ref{F:xi_market}. The mean and median tail-shape estimates are close to $1/3$, which roughly aligns with the risk-factor panel and motivates our choices of the GPD parameter in the simulation study; note that $\xi=1/3$ is also the Pareto extrapolation parameter used in the construction of estimators \#5 and \#6, cf.~\cite[Annex~I, p.~158]{EBA2023RTS05}. Overall, the tail-shape, skewness, and kurtosis values observed on production P\&Ls are consistent with our datasets, which further highlights the importance of analyzing risk estimation under non-normal returns.

Unlike in the simulation study of Section~\ref{S:numerical.study}, the true ES is unknown in the market data setting, so we have to adapt the assessment framework of Section~\ref{S:assessment}. While most metrics introduced therein cannot be applied directly to market data, we use a market-data counterpart of CT based on rolling-window ES forecasts to assess capital coverage and compare simulated and market results (Sections~\ref{S:market.ct} and~\ref{S:sim.vs.market}). For completeness, we also analyze the ratio of estimated capital reserves, $\mathrm{ES}_{2.5\%}/\var_{1\%}$ (Section~\ref{S:market.ratio}). Whereas market data CT evaluates the reliability of ES forecasts against subsequent out-of-sample realizations, the ratio analysis provides a supporting consistency check between ES and VaR estimates. We include the ratio check in the market data analysis because, unlike in the simulation study, estimation errors cannot be assessed directly against known ES and VaR targets. We use the $0.9$ threshold for time-averaged 1-day ES/VaR ratios as a key diagnostic reference,  see \cite[Section~4.3.2, para.~76(c) and footnote~19]{EGIM}. Note that we focus on a long-run ratio distribution analysis rather than average annual mean ratios, as in \cite{EGIM}, so the threshold should be treated as a visualization tool rather than a statistical test failure indicator. Nevertheless, values below this threshold flag a low ES estimate relative to VaR that warrants investigation, although a ratio above it does not establish adequate capital coverage as, for instance, similar proportional underestimation of ES and VaR may leave the ratio largely unchanged. Examining the ratio alongside CT addresses both consistency and coverage and can support the initial and periodic internal model validation required under the FRTB framework; see \cite[MAR~30.17]{Bas2019}, \cite[Art.~325bj]{EU_CRR}, and \cite[Section~4.3]{EGIM}.

\subsection{Capital coverage: analysis of safe confidence thresholds}\label{S:market.ct}
\begin{figure}[ht]
    \centering
    \includegraphics[width=0.95\linewidth]{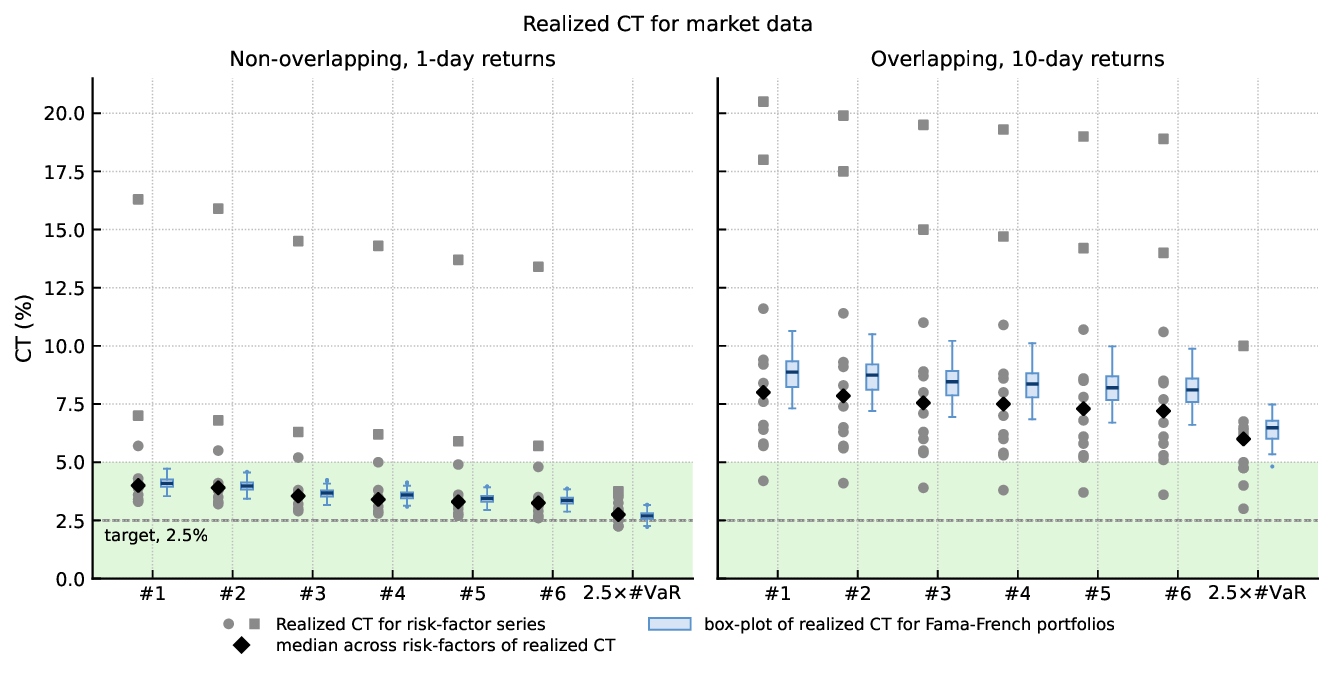}
    \caption{Statistics of realized CT values for the twelve risk-factor series and the Fama--French portfolios; see Table~\ref{T:market.data} for data counts. Each gray marker corresponds to the  realized CT of one risk-factor series; squares for the two short-rate series (SOFR, US 1M Bill) and circles for the remaining risk-factors. Solid black diamonds depict the median of these CT values across the twelve risk factors. Blue box-plots correspond to the set of realized CT values for all Fama--French portfolios. These metrics are displayed for estimators \#1--\#6, together with the CT statistics of \#VaR (multiplied by $2.5$ that aligns it with  ES reference level $\alpha=2.5\%$). The green shaded region depicts the CT safe region below $5\%$. The target CT value of 2.5\% is marked with a dashed horizontal line. The left panel corresponds to non-overlapping, 1-day returns, the right panel to overlapping, 10-day returns. 
    }
    \label{fig:CT_rf_dots}
\end{figure}

We first analyze capital coverage through the realized CT values, focusing solely on market data. For each estimator, each series and each business day, we compute the capital reserve using the previous $n=250$ returns and confront it with the subsequently realized return (the next 1-day return in the non-overlapping scheme or the realized future 10-day return in the overlapping scheme) to construct the secured position. For each series and estimator, we compute a single CT from the full sequence of secured returns, which contains roughly $1\,700$ to $3\,900$ observations, depending on the series. This CT is the empirical counterpart of the safe confidence threshold; see Table~\ref{T:performance.metrics}. We recall that the CT metric can be viewed as a conservativeness-focused ES backtest for internal model validation; see \cite[Proposition~3.1]{MolPit2017}, and cf.~\cite[Article~325bj(3)(b)]{EU_CRR} and \cite[Section~4.3.2]{EGIM}.
We summarize the obtained results in Figure~\ref{fig:CT_rf_dots}, grouped by estimator (x-axis), which depicts the  realized CT for each risk factor (gray markers) and box plots (blue) for the CT of the Fama-French portfolios. Grey squares correspond to the two short-rate series (SOFR, US 1M Bill) and the gray circles for the remaining risk-factors. Black diamond markers correspond to the median across risk factors of the realized CT. See also Table~\ref{T:CT_market} for granular risk-factor-level output. In this section we continue to use the CT traffic-light narrative, with 5\% and 10\% corresponding to amber and red zone entry points, and accordingly we shade the safe zone CT$\,\leq5\%$ in light green. The target CT value of 2.5\% is marked with a dashed horizontal line. See also Section~\ref{S:assessment} for details on CT metric.

\textbf{Under the overlapping scheme, the realized CT doubles.} Putting aside the two short-rate series, the realized daily thresholds range from $2.6\%$ to $5.7\%$ across risk-factors series and estimators. The median across risk-factors of the realized CT (black diamond) ranges  from $4.0\%$ (\#1) to $3.2\%$ (\#6) against the $2.5\%$ target, sitting in the safe CT zone. In contrast, under overlapping, 10-day construction the realized CT  roughly doubles, to $3.6$--$11.6\%$, with medians of $8.0\%$ (\#1) to $7.2\%$ (\#6). This moves the median across risk-factors  of every estimator into the CT amber zone. The per-series deterioration ranges from $1$ to $6$ percentage points and is smallest for the strongly mean-reverting VIX, see Table~\ref{T:CT_market}.  We also note that the \#VaR benchmark moves into the same direction, with its median increasing from $2.8\%$ to $6.0\%$. Since \#VaR shares none of the ES weighting structure, the CT \textbf{deterioration is due to the overlapping scheme, not the estimators' form}. Nevertheless, ES ends deeper in the warning zones, at $3.2$ times its reference level against $2.4$ for \#VaR. This is the empirical counterpart of the regime shift documented in Section~\ref{S:numerical.study}. The market data CT for overlapping scheme matches the simulated heavy-tailed and GARCH ranges, which supports the inclusion of serial dependent and heavy tails models in the simulation design.

\textbf{The short-rate series lead to insufficient capital coverage.} For SOFR and the US 1M Bill the realized CT reaches  $5.8$--$16.3\%$ already on daily returns and $14.0$--$20.5\%$ under the overlapping scheme, compounding the ordinary distributional misspecification rather than replacing it. The output is usually in the red CT zone due to the 10\% breach point exceedance, indicating insufficient capital coverage.

\textbf{The diversified Fama-French portfolios show the same pattern}, with CT values (blue box plots) clustering at $3$--$5\%$ on daily returns and shifting to medians of $8$--$9\%$ under the overlapping, 10-day construction, with wider interquartile ranges and outliers approaching $11\%$. The wider dispersion is consistent with the reduced effective sample size. Diversification across a large cross-section of equities therefore offers no protection.

\subsection{Simulated versus realized estimation risk: consequences for validation and monitoring}\label{S:sim.vs.market}

Once capital requirements are computed using ES rather than VaR, two practical questions arise: what a simulation study must include for its conclusions to carry over to production P\&Ls, and what must be monitored once the capital figure is an ES estimate. In this section, we address the first question, and the second question is postponed to the next section. 

\begin{figure}[ht]
    \centering
    \includegraphics[width=0.95\linewidth]{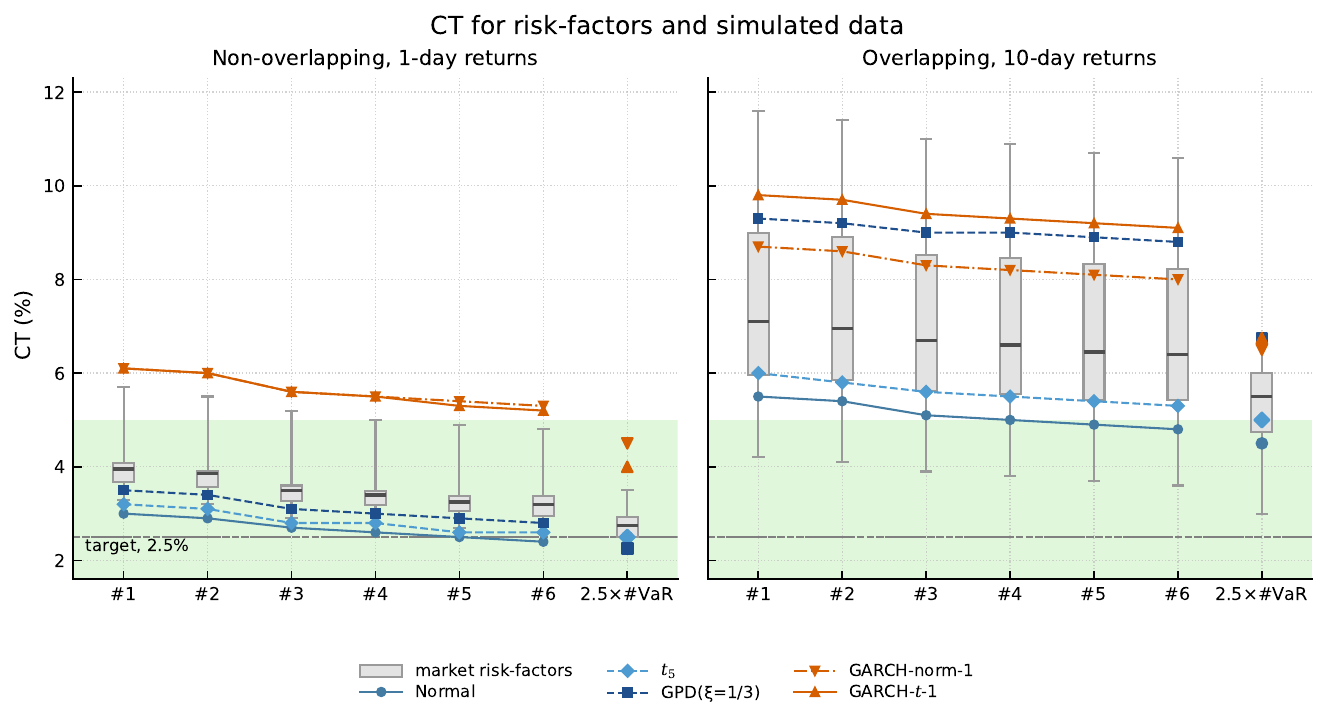}
    \caption{CT values for the realized risk-factor series and for the simulated data, shown for the ES estimators \#1--\#6 and for the \#VaR benchmark (multiplied by $2.5$ to align it to the reference ES target level $\alpha=2.5\%$).  Box plots depict the distribution of the realized CT across the ten regular risk-factor series of Table~\ref{T:market.data} (the two short-rate series are excluded; whiskers span the minimum--maximum range). Colored markers and curves correspond to CT values for selected models from Section~\ref{S:numerical.study}: Gaussian, Student-$t_5$, GPD with $\xi=1/3$, GARCH-norm-1 and GARCH-$t$-1. The left panel shows the non-overlapping scheme, 1-day returns, and the right panel the overlapping scheme, 10-day returns. The green shading marks the safe zone CT$\,\leq5\%$, and the dashed horizontal line the target CT value of $2.5\%$.  
  }
    \label{fig:sim_vs_market}
\end{figure}

In Figure~\ref{fig:sim_vs_market} we display  CT  values for both, market risk-factors  and their simulated counterparts of Section~\ref{S:numerical.study}, estimator by estimator. 

\textbf{Static i.i.d. models understate the realized market CT values.} 
For non-overlapping, 1-day, returns, the market boxes lie above every i.i.d. curve, including the heavier-tailed Student-$t_5$ and GPD($\xi=1/3$) distributions, and below the two curves corresponding to GARCH models. Tail heaviness alone, even at the level of $\xi=1/3$, is therefore not enough to reproduce the CT observed on market data, while including serially dependent volatility in the models brings the CT values above those of the market. Moreover, for overlapping scheme, the Gaussian and Student-$t_5$ models still remain below the lower end of the market boxes, and thus models based on i.i.d. innovations would give a systematically optimistic picture of the ES estimation accuracy.

\textbf{Overlapping widens the CT range, and both heavy tails and serial dependence matter.}
Moving from the left to the right panel, the market CT shifts upward and widens. The boxes lie entirely inside the safe zone for daily returns but entirely above it under the overlapping, 10-day construction, and the whisker span roughly triples. The simulated models react to the rolling sum to different degrees. The GPD($\xi=1/3$) curve, which falls below the market boxes on daily data, now lies above them together with GARCH-$t$-1, while GARCH-norm-1 moves inside the boxes, close to their upper quartile. Likewise, the two GARCH curves, nearly indistinguishable on daily data, separate by roughly one percentage point, so the tail of the innovation distribution becomes relevant once overlapping returns are used. A simulation study meant to justify an ES estimation approach under the overlapping scheme should therefore combine both features: heavy left tails, as in Student-$t_5$ or GPD($\xi=1/3$), and serially dependent volatility, as in GARCH.

\subsection{Capital ratio: analysis of realized ES/VaR quotient}\label{S:market.ratio}

\begin{figure}[ht]
    \centering
    \includegraphics[width=0.95\linewidth]{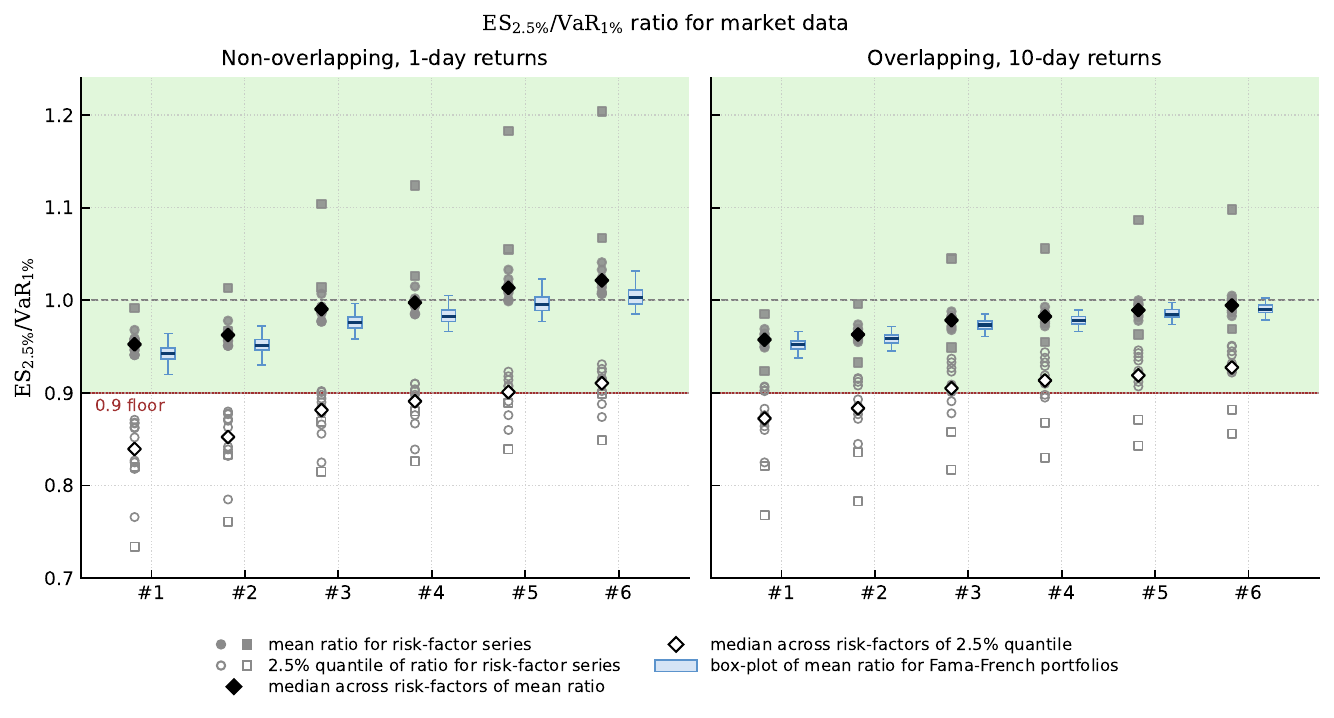}
    \caption{Statistics of realized $\mathrm{ES}_{2.5\%}/\var_{1\%}$ ratio for the twelve risk-factor series and the Fama-French portfolios; see Table~\ref{T:market.data} for data counts. Each filled gray marker corresponds to the mean realized ratio of one risk-factor series, while hollow markers correspond to 2.5\% quantiles of the ratios. Square markers are reserved for the two short-rate series (SOFR, US 1M Bill) and circles for the remaining risk-factors. Black diamonds depict the medians of the  mean (solid diamond) and 2.5\% quantile (hollow diamond) ratios values across the twelve risk factors. Blue box-plots correspond to the set of realized $\mathrm{ES}_{2.5\%}/\var_{1\%}$ values for all Fama--French portfolios. The green shaded region depicts the 'safe region' above the 0.9 floor, and  the gray horizontal dashed line marks the centering level $1$. The left panel corresponds to non-overlapping, 1-day returns, the right panel to overlapping, 10-day returns. }
    \label{fig:ratio_rf_dots}
\end{figure}

The second diagnostic that we consider is the ratio between the $\mathrm{ES}_{2.5\%}$ and $\var_{1\%}$ estimates. For each series and each business day, both quantities are estimated from the same preceding $n=250$ returns and the ratio is recorded. Key statistics of the ratios for each risk factor and each estimator are presented in Table~\ref{T:ratios:combined}. The results are summarized in Figure~\ref{fig:ratio_rf_dots}. The green shaded area depicts the `safe region', above the $0.9$ floor, and the gray horizontal dashed line marks the centering level $1$. As above, the left panel is devoted to non-overlapping, 1-day returns, while the right panel is focused on overlapping, 10-day returns. For each estimator, we display the mean of the realized ratios for each risk factor (filled markers) and the 2.5\% quantiles of the ratios (hollow markers), along with their medians across risk factors (diamonds). The blue box plots, to the right of each estimator, correspond to the Fama-French portfolios.

\textbf{The floor is met for the mean ratio values, but not in the lower ratio distribution tail}. The sample means of the ratios (filled markers) never breach the $0.9$ floor: across the twelve risk factors their median runs from $0.953$ (\#1) to $1.022$ (\#6) on daily returns and from $0.958$ to $0.995$ under overlapping, and no individual series falls below $0.9$ in either scheme. Consequently, no structural cross risk estimation problem is observed for the market data -- if the mean of the ratio would fall below, this would indicate a modeling consistency issue at the estimator design level, recall \cite[Section~4.3.2]{EGIM} for the supervisory context. Note that this is expected behavior, as the chosen VaR and ES estimation methodologies are based on similar non-parametric estimation schemes. That said, the $2.5\%$ quantiles (hollow markers) tell a different story. Their median runs from $0.840$ (\#1) to $0.911$ (\#6) daily and from $0.873$ to $0.928$ under overlapping, clearing the floor only from \#5 onwards in the daily scheme and from \#3 onwards under overlapping.  At the series level the number of breaches falls monotonically with the degree of conservativeness, from twelve out of twelve for \#1 to four for \#6 on daily returns, and from eight to two (the two short-rate series) under overlapping. The floor applies to the ratio averaged over the $250$-day window rather than to individual daily ratios, so reading it off the $2.5\%$ quantile is deliberately stricter than the test itself. The point for estimator selection is nonetheless clear: the empirical plug-in estimator sits at the floor with no margin, while conservative weighting buys material headroom.

\textbf{The ratio is insensitive to the sampling scheme.} Between the two schemes the cross-series medians move by at most three percentage points; the medians of the means are slightly down, and the medians of the $2.5\%$ quantiles are slightly up, which is in sharp contrast to the CT values that roughly double. The dispersion tightens rather than widens: for \#1 the median standard deviation of the ratio falls from $0.065$ on daily returns to $0.036$ under overlapping. The reason is that overlapping distorts the $\mathrm{ES}_{2.5\%}$ and $\var_{1\%}$ estimates through the same channel, and thus the ratio nets the shared degradation out.  A ratio above $0.9$ test on overlapping data therefore says little about the absolute ES estimation error, and should be combined with other diagnostics, such as the CT values.

\textbf{The ratio has no universally optimal target.} Even if the true $\mathrm{ES}_{2.5\%}$ and $\var_{1\%}$ were known exactly, their ratio would depend on the underlying law of returns: close to one under the Gaussian benchmark, but above one for heavy-tailed laws. In fact, recall that the EGIM-induced 0.9 ratio floor is a Student-$t_5$ reference based on a confidence bound for the ratio averaged over 250 observations, see \cite[Section~4.3.2, footnote~19]{EGIM}, rather than a level the data must universally attain, and the realized ratio confounds the tail of the data with the estimation error of both quantities. Since all six estimators are applied to the same sample, the spread across \#1--\#6 for a fixed series isolates the weighting effect, and it is monotone for every series and in both schemes.

The overall message of this section is one of caution: \textbf{no single diagnostic is sufficient to validate ES-based capital}. The $\mathrm{ES}_{2.5\%}/\var_{1\%}$ ratio is the more robust and less illuminating of the two diagnostics investigated: it is reassuring by construction since it  nets the effect that the overlapping scheme inflicts on both of its arguments. The ratio should therefore be read alongside other coverage diagnostics such as the realized CT. Conversely, if the mean of the ratio falls below the threshold, it might indicate a modelling consistency issue.

\section{Concluding remarks}\label{sec:concluding-rems}

In this study, working within the framework of coherent risk estimators, we compare the overlapping and non-overlapping sampling schemes in the estimation of $\mathrm{ES}$ and $\var$ for regulatory capital purposes. Our main findings can be summarized as follows.

First, \textit{the overlapping scheme induces a regime shift, not merely a variance penalty.} Overlapping samples, standard in FRTB computations of ES, do more than shrink the effective sample size for the i.i.d.-based estimators, they are associated with a directional bias in the adverse direction for capital coverage. Namely, they simultaneously add a systematic, directional negative bias when estimating ES, degrade fit quality, and lead to less conservative safety metrics. Our conclusions are consistent with \cite[Annex I]{EBA2023RTS05} and \cite{AicCroReh2021}, which likewise stress the importance of estimator selection in the overlapping regime. Our analysis, however, is considerably more granular: we investigate a substantially broader set of data and a wider range of performance metrics.

Second, \textit{within the considered CRE-based estimators, there is no uniformly best choice}, and the choice of weights fixes a position on a fit--bias--safety trade-off rather than locating an optimum. Moving mass toward the smallest order statistic improves bias, downside error and safety while degrading symmetric fit. The trade-off is sharp under non-overlapping sampling and flattens into a single conservatism axis under overlapping and GARCH data, where symmetric fit is poor for every estimator. \textit{Conservative weighting helps, but does not always suffice}, and when heavy tails, serial dependence and overlapping sampling are jointly present, even the most conservative estimators \#5-\#6 remain outside the safe zone, while under non-overlapping sampling they stay within the green zone for the static models.

Third, \textit{the market data confirm the simulated conclusions}, and show that a single diagnostic is not enough. Across twelve risk-factor series and 99 Fama--French portfolios the median realized CT roughly doubles under the overlapping scheme, moving every estimator from the safe zone into the amber zone. On daily data the observed levels lie between the static i.i.d.\ and the GARCH simulated regimes, while under overlapping sampling the Gaussian and Student-$t_5$ models fall below them and only the GPD and GARCH models reach or exceed them. Over the same data the $\mathrm{ES}_{2.5\%}/\var_{1\%}$ ratio moves less, because the overlapping scheme impacts numerator and denominator through the same channel; a ratio above the prescribed 0.9 threshold is therefore insufficient evidence of adequate capital, and must be read alongside a coverage diagnostic such as the realized CT.

Fourth, \textit{ES estimation adequacy requires enhanced monitoring}. ES's greater tail sensitivity increases the need for monitoring 
and some ES supervisory mitigants are already in place, such as the $\mathrm{ES}_{2.5\%}/\var_{1\%}$ ratio and other tests~\cite[Section~4.3.2]{EGIM} and direct ES backtesting, without specifying a particular approach ~\cite[Art. 21(1)(b)]{EU2024_1085}. However, there is currently no standard practice to our knowledge that would sufficiently quantify the aforementioned risks. This underscores the clear need for further research in this area.

These findings matter for financial stability beyond the validation of a single model. Capital formulas add some generic safeguards on top of the ES estimate through averaging, stress calibration, and a multiplication factor, but the estimate itself remains the base input to the capital requirement and to risk-weighted assets, see \citet{Bas1996b}, \citet[MAR~33]{Bas2019}, and \citet[Article~92]{EU_CRR}. In particular, our results indicate that the regulatory 1-day backtest, which links the multiplication factor to model performance, may not detect the bias documented here, see \citet[MAR~32]{Bas2019}. On our market data, the 1-day \#VaR estimator has a median exception rate of $1.1\%$, close to its nominal level of $1\%$, while the 10-day overlapping ES estimate for estimator \#1 secures the position only at a median confidence level of $8.0\%$, against a target of $2.5\%$, see Table~\ref{T:CT_market}. Direct ES backtests, required by \citet[Article~21(1)(b)]{EU2024_1085} and \citet[Section~4.3.2]{EGIM}, and coverage diagnostics, such as the realized CT, should therefore account for the 10-day overlapping horizon used for capital. Moreover, many banks apply historical simulation with order-statistic weights taken from a small set of standard formulas, such as those in \citet[Annex~I]{EBA2023RTS05}, see \citet[para.~136]{EBA2025}. The resulting understatement is therefore likely to be common for many  institutions and hard to detect through supervisory benchmarking exercises, which compare banks with one another. This makes estimator-level validation relevant for the resilience of the banking sector as a whole, cf.~\citet{DanJamVal2016}.

Overall, our results make clear that, under overlapping scenarios, ES might be easily and often substantially underestimated. Closing this gap calls for a dedicated theory of estimator construction for overlapping returns, and re-weighting estimators inherited from the i.i.d. setting is unlikely to suffice.

\section*{Disclaimer}
\noindent The views and opinions expressed in this paper are the authors’ own and do not necessarily reflect the views and opinions of their current or past employers. In particular, they cannot be taken to represent those of the European Central Bank (ECB) or to state the ECB’s policy. Neither the ECB nor any person acting on its behalf may be held responsible for the use which may be made of the information contained in this publication, or for any errors which, despite careful preparation and checking, may appear therein.

\section*{Declaration of generative AI and AI-assisted technologies}
\noindent During the preparation of this work, the authors used large language model assistants (Anthropic Claude; OpenAI ChatGPT) to support code generation, refine the figures' layout, and to improve the clarity and flow of the exposition. All AI-assisted output was used in a supervised manner: the authors verified every derivation, numerical result, and figure, and edited the text as needed. The authors take full responsibility for the content of the publication.

\section*{Funding}
\noindent Igor Cialenco acknowledges support from the US National Science Foundation grant DMS-2407549.  Damian Jelito acknowledges support from the National Science Centre, Poland, via project 2024/53/B/ST1/00703. Marcin Pitera acknowledges support from the National Science Centre, Poland, via project 2024/53/B/HS4/00433.

\bibliography{bibliography}

\clearpage
\appendix
\addtocontents{toc}{\protect\setcounter{tocdepth}{1}}

\section{Data tables}
\subsection{Simulated data}\label{S:tables}

\begin{table}[htp!]
\centering
\small
\setlength{\tabcolsep}{4pt}

\scalebox{1}{
\begin{minipage}[t]{0.48\textwidth}
\centering
\caption*{\textbf{Normal}}
\scalebox{0.7}{
\begin{tabular}{l|ccc|ccc|ccc}
\toprule
Est. & $ME$ & $SE$ & $DE$ & $SB$ & $MB$ & $RB$ & $UP_{5}$ & $UP_{10}$ & $CT$ \\
\midrule
\#1 & \textbf{6.9\%} & \textbf{8.6\%} & 6.7\% & -1.4\% & -1.7\% & -3.2\% & 34.9\% & 15.7\% & 3.0\% \\
\#2 & 7.0\% & 8.7\% & 6.4\% & \textbf{-0.8\%} & -1.1\% & -2.6\% & 32.4\% & 14.2\% & 2.9\% \\
\#3 & 7.2\% & 9.0\% & 5.7\% & \textbf{0.8\%} & \textbf{0.5\%} & -1.2\% & 26.5\% & 11.0\% & 2.7\% \\
\#4 & 7.3\% & 9.2\% & 5.5\% & 1.4\% & 1.0\% & -0.7\% & 24.9\% & 10.2\% & 2.6\% \\
\#5 & 7.6\% & 9.6\% & 5.2\% & 2.3\% & 1.9\% & \textbf{0.1\%} & 22.4\% & 8.9\% & \textbf{2.5\%} \\
\#6 & 7.8\% & 9.9\% & \textbf{5.0\%} & 2.8\% & 2.4\% & 0.6\% & \textbf{20.9\%} & \textbf{8.3\%} & 2.4\% \\
\midrule
\gray{\#VaR} & 8.5\% & 10.9\% & 5.1\% & 3.4\% & 2.8\% & 0.7\% & 21.1\% & 8.8\% & 1.0\% \\
\bottomrule
\end{tabular}}
\end{minipage}
\hspace{0.02\textwidth}
\begin{minipage}[t]{0.48\textwidth}
\centering
\caption*{\textbf{Normal}}
\scalebox{0.7}{
\begin{tabular}{l|ccc|ccc|ccc}
\toprule
Est. & $ME$ & $SE$ & $DE$ & $SB$ & $MB$ & $RB$ & $UP_{5}$ & $UP_{10}$ & $CT$ \\
\midrule
\#1 & 15.3\% & 18.9\% & 16.0\% & -6.7\% & -8.0\% & -14.4\% & 57.1\% & 45.4\% & 5.5\% \\
\#2 & 15.2\% & 18.8\% & 15.7\% & -6.2\% & -7.6\% & -14.0\% & 55.9\% & 44.3\% & 5.4\% \\
\#3 & 15.0\% & 18.7\% & 15.1\% & -5.2\% & -6.5\% & -13.1\% & 53.6\% & 42.0\% & 5.1\% \\
\#4 & 15.0\% & 18.7\% & 14.9\% & -4.8\% & -6.2\% & -12.7\% & 52.7\% & 41.1\% & 5.0\% \\
\#5 & \textbf{14.9\%} & \textbf{18.6\%} & 14.7\% & -4.3\% & -5.7\% & -12.3\% & 51.5\% & 40.0\% & 4.9\% \\
\#6 & \textbf{14.9\%} & \textbf{18.6\%} & \textbf{14.5\%} & \textbf{-3.9\%} & \textbf{-5.3\%} & \textbf{-12.0\%} & \textbf{50.7\%} & \textbf{39.2\%} & \textbf{4.8\%} \\
\midrule
\gray{\#VaR} & 15.3\% & 19.2\% & 14.0\% & -2.5\% & -4.1\% & -11.1\% & 48.0\% & 37.0\% & 1.8\% \\
\bottomrule
\end{tabular}}
\end{minipage}
} 

\par\medskip

\scalebox{1}{
\begin{minipage}[t]{0.48\textwidth}
\centering
\caption*{\textbf{Student's $t_5$}}
\scalebox{0.7}{
\begin{tabular}{l|ccc|ccc|ccc}
\toprule
Est. & $ME$ & $SE$ & $DE$ & $SB$ & $MB$ & $RB$ & $UP_{5}$ & $UP_{10}$ & $CT$ \\
\midrule
\#1 & \textbf{13.0\%} & \textbf{16.7\%} & 11.4\% & -1.8\% & -3.9\% & -6.6\% & 47.1\% & 33.6\% & 3.2\% \\
\#2 & 13.2\% & 17.1\% & 10.9\% & \textbf{-0.7\%} & -2.9\% & -5.8\% & 44.6\% & 31.4\% & 3.1\% \\
\#3 & 14.3\% & 19.4\% & 9.8\% & 2.8\% & \textbf{-0.0\%} & -3.3\% & 37.9\% & 25.8\% & 2.8\% \\
\#4 & 14.6\% & 20.0\% & 9.5\% & 3.7\% & 0.8\% & -2.6\% & 36.1\% & 24.4\% & 2.8\% \\
\#5 & 15.6\% & 21.9\% & 9.0\% & 5.8\% & 2.4\% & -1.3\% & 33.0\% & 22.0\% & \textbf{2.6\%} \\
\#6 & 16.0\% & 22.7\% & \textbf{8.7\%} & 6.7\% & 3.3\% & \textbf{-0.6\%} & \textbf{31.4\%} & \textbf{20.8\%} & \textbf{2.6\%} \\
\midrule
\gray{\#VaR} & 16.0\% & 21.8\% & 7.7\% & 8.2\% & 5.4\% & 1.3\% & 26.9\% & 17.0\% & 1.0\% \\
\bottomrule
\end{tabular}}
\end{minipage}
\hspace{0.02\textwidth}
\begin{minipage}[t]{0.48\textwidth}
\centering
\caption*{\textbf{Student's $t_5$}}
\scalebox{0.7}{
\begin{tabular}{l|ccc|ccc|ccc}
\toprule
Est. & $ME$ & $SE$ & $DE$ & $SB$ & $MB$ & $RB$ & $UP_{5}$ & $UP_{10}$ & $CT$ \\
\midrule
\#1 & 18.9\% & 23.8\% & 19.0\% & -8.5\% & -11.4\% & -17.4\% & 62.7\% & 53.0\% & 6.0\% \\
\#2 & 18.8\% & 23.7\% & 18.7\% & -8.0\% & -10.9\% & -17.0\% & 61.7\% & 52.0\% & 5.8\% \\
\#3 & 18.5\% & 23.6\% & 18.1\% & -7.0\% & -9.9\% & -16.2\% & 59.8\% & 49.9\% & 5.6\% \\
\#4 & 18.5\% & 23.6\% & 17.9\% & -6.6\% & -9.6\% & -15.9\% & 59.0\% & 49.2\% & 5.5\% \\
\#5 & \textbf{18.3\%} & \textbf{23.5\%} & 17.6\% & -6.1\% & -9.1\% & -15.4\% & 58.1\% & 48.2\% & 5.4\% \\
\#6 & \textbf{18.3\%} & \textbf{23.5\%} & \textbf{17.4\%} & \textbf{-5.7\%} & \textbf{-8.7\%} & \textbf{-15.1\%} & \textbf{57.4\%} & \textbf{47.4\%} & \textbf{5.3\%} \\
\midrule
\gray{\#VaR} &  18.5\% & 24.2\% & 16.4\% & -3.1\% & -6.4\% & -13.4\% & 52.7\% & 43.0\% & 2.0\% \\
\bottomrule
\end{tabular}}
\end{minipage}
} 

\par\medskip

\scalebox{1}{
\begin{minipage}[t]{0.48\textwidth}
\centering
\caption*{\textbf{GPD ($\xi=1/3$)}}
\scalebox{0.7}{
\begin{tabular}{l|ccc|ccc|ccc}
\toprule
Est. & $ME$ & $SE$ & $DE$ & $SB$ & $MB$ & $RB$ & $UP_{5}$ & $UP_{10}$ & $CT$ \\
\midrule
\#1 & \textbf{25.8\%} & \textbf{37.2\%} & 20.6\% & -2.0\% & -9.2\% & -14.1\% & 55.8\% & 48.8\% & 3.5\% \\
\#2 & 26.2\% & 38.4\% & 19.9\% & \textbf{-0.1\%} & -7.6\% & -12.9\% & 53.6\% & 46.5\% & 3.4\% \\
\#3 & 29.3\% & 47.2\% & 18.0\% & 7.2\% & -2.4\% & -8.9\% & 46.6\% & 39.9\% & 3.1\% \\
\#4 & 30.1\% & 48.9\% & 17.5\% & 8.9\% & \textbf{-1.0\%} & -7.8\% & 44.9\% & 38.3\% & 3.0\% \\
\#5 & 32.7\% & 55.8\% & 16.6\% & 13.5\% & 2.0\% & -5.7\% & 41.5\% & 35.1\% & 2.9\% \\
\#6 & 33.7\% & 57.8\% & \textbf{16.1\%} & 15.3\% & 3.4\% & \textbf{-4.7\%} & \textbf{40.0\%} & \textbf{33.7\%} & \textbf{2.8\%} \\
\midrule
\gray{\#VaR} & 31.7\% & 46.1\% & 12.9\% & 18.6\% & 10.4\% & 2.3\% & 31.4\% & 25.3\% & 0.9\% \\
\bottomrule
\end{tabular}}
\end{minipage}
\hspace{0.02\textwidth}
\begin{minipage}[t]{0.48\textwidth}
\centering
\caption*{\textbf{GPD ($\xi=1/3$)}}
\scalebox{0.7}{
\begin{tabular}{l|ccc|ccc|ccc}
\toprule
Est. & $ME$ & $SE$ & $DE$ & $SB$ & $MB$ & $RB$ & $UP_{5}$ & $UP_{10}$ & $CT$ \\
\midrule
\#1 & 42.4\% & 61.2\% & 39.0\% & -20.6\% & -33.6\% & -40.1\% & 77.1\% & 73.8\% & 9.3\% \\
\#2 & 42.3\% & 61.2\% & 38.8\% & -20.2\% & -33.2\% & -39.8\% & 76.8\% & 73.5\% & 9.2\% \\
\#3 & 42.0\% & 61.1\% & 38.3\% & -19.5\% & -32.6\% & -39.3\% & 76.3\% & 72.9\% & 9.0\% \\
\#4 & 41.9\% & \textbf{61.0\%} & 38.1\% & -19.2\% & -32.3\% & -39.1\% & 76.1\% & 72.7\% & 9.0\% \\
\#5 & 41.7\% & \textbf{61.0\%} & 37.9\% & -18.9\% & -32.0\% & -38.8\% & 75.9\% & 72.5\% & 8.9\% \\
\#6 & \textbf{41.6\%} & \textbf{61.0\%} & \textbf{37.8\%} & \textbf{-18.7\%} & \textbf{-31.8\%} & \textbf{-38.6\%} & \textbf{75.7\%} & \textbf{72.2\%} & \textbf{8.8\%} \\
\midrule
\gray{\#VaR} & 40.6\% & 66.0\% & 32.7\% & -8.0\% & -22.7\% & -31.7\% & 67.1\% & 62.9\% & 2.7\% \\
\bottomrule
\end{tabular}}
\end{minipage}
} 

\par\medskip

\scalebox{1}{
\begin{minipage}[t]{0.48\textwidth}
\centering
\caption*{\textbf{GPD ($\xi=1/4$)}}
\scalebox{0.7}{
\begin{tabular}{l|ccc|ccc|ccc}
\toprule
Est. & $ME$ & $SE$ & $DE$ & $SB$ & $MB$ & $RB$ & $UP_{5}$ & $UP_{10}$ & $CT$ \\
\midrule
\#1 & \textbf{21.3\%} & \textbf{28.4\%} & 17.7\% & -2.0\% & -6.8\% & -11.2\% & 52.9\% & 44.5\% & 3.4\% \\
\#2 & 21.6\% & 29.2\% & 17.0\% & \textbf{-0.4\%} & -5.3\% & -10.0\% & 50.5\% & 42.2\% & 3.3\% \\
\#3 & 23.9\% & 34.4\% & 15.3\% & 5.5\% & -0.7\% & -6.3\% & 43.5\% & 35.7\% & 2.9\% \\
\#4 & 24.4\% & 35.6\% & 14.9\% & 7.0\% & \textbf{0.5\%} & -5.3\% & 41.8\% & 34.1\% & 2.9\% \\
\#5 & 26.4\% & 39.8\% & 14.1\% & 10.6\% & 3.1\% & -3.3\% & 38.4\% & 31.1\% & \textbf{2.7\%} \\
\#6 & 27.2\% & 41.2\% & \textbf{13.7\%} & 12.1\% & 4.4\% & \textbf{-2.3\%} & \textbf{36.9\%} & \textbf{29.7\%} & \textbf{2.7\%} \\
\midrule
\gray{\#VaR} & 26.6\% & 37.4\% & 11.7\% & 14.8\% & 8.8\% & 1.9\% & 30.7\% & 23.9\% & 0.9\% \\
\bottomrule
\end{tabular}}
\end{minipage}
\hspace{0.02\textwidth}
\begin{minipage}[t]{0.48\textwidth}
\centering
\caption*{\textbf{GPD ($\xi=1/4$)}}
\scalebox{0.7}{
\begin{tabular}{l|ccc|ccc|ccc}
\toprule
Est. & $ME$ & $SE$ & $DE$ & $SB$ & $MB$ & $RB$ & $UP_{5}$ & $UP_{10}$ & $CT$ \\
\midrule
\#1 & 34.4\% & 45.2\% & 32.8\% & -16.7\% & -25.7\% & -33.3\% & 73.3\% & 68.8\% & 8.2\% \\
\#2 & 34.3\% & 45.2\% & 32.5\% & -16.2\% & -25.2\% & -32.9\% & 72.8\% & 68.2\% & 8.1\% \\
\#3 & 34.0\% & 45.1\% & 32.0\% & -15.4\% & -24.5\% & -32.3\% & 72.0\% & 67.4\% & 7.9\% \\
\#4 & 33.9\% & 45.0\% & 31.8\% & -15.1\% & -24.2\% & -32.0\% & 71.7\% & 67.0\% & 7.8\% \\
\#5 & \textbf{33.7\%} & 45.0\% & 31.6\% & -14.8\% & -23.8\% & -31.7\% & 71.4\% & 66.6\% & \textbf{7.7\%} \\
\#6 & \textbf{33.7\%} & \textbf{44.9\%} & \textbf{31.4\%} & \textbf{-14.4\%} & \textbf{-23.5\%} & \textbf{-31.5\%} & \textbf{71.0\%} & \textbf{66.3\%} & \textbf{7.7\%} \\
\midrule
\gray{\#VaR} & 33.1\% & 46.7\% & 28.2\% & -7.6\% & -17.4\% & -26.8\% & 63.9\% & 58.8\% & 2.6\% \\
\bottomrule
\end{tabular}}
\end{minipage}
} 

\par\medskip

\scalebox{1}{
\begin{minipage}[t]{0.48\textwidth}
\centering
\caption*{\textbf{NIG ($\alpha=0.4,\ \beta=0.14$)}}
\scalebox{0.7}{
\begin{tabular}{l|ccc|ccc|ccc}
\toprule
Est. & $ME$ & $SE$ & $DE$ & $SB$ & $MB$ & $RB$ & $UP_{5}$ & $UP_{10}$ & $CT$ \\
\midrule
\#1 & \textbf{17.4\%} & \textbf{21.9\%} & 15.2\% & -2.3\% & -5.0\% & -8.7\% & 50.0\% & 40.2\% & 3.2\% \\
\#2 & 17.6\% & 22.4\% & 14.7\% & \textbf{-0.9\%} & -3.6\% & -7.6\% & 47.5\% & 37.8\% & 3.1\% \\
\#3 & 19.0\% & 24.9\% & 13.1\% & 3.8\% & \textbf{0.5\%} & -4.2\% & 40.4\% & 31.5\% & 2.8\% \\
\#4 & 19.5\% & 25.7\% & 12.7\% & 5.0\% & 1.6\% & -3.3\% & 38.6\% & 29.9\% & 2.8\% \\
\#5 & 20.8\% & 27.8\% & 12.0\% & 7.7\% & 4.0\% & -1.4\% & 35.2\% & 27.1\% & 2.6\% \\
\#6 & 21.4\% & 28.8\% & \textbf{11.7\%} & 9.0\% & 5.1\% & \textbf{-0.5\%} & \textbf{33.6\%} & \textbf{25.7\%} & \textbf{2.5\%} \\
\midrule
\gray{\#VaR} & 22.4\% & 30.4\% & 11.0\% & 11.2\% & 7.0\% & 2.4\% & 31.1\% & 23.5\% & 0.9\% \\
\bottomrule
\end{tabular}}
\end{minipage}
\hspace{0.02\textwidth}
\begin{minipage}[t]{0.48\textwidth}
\centering
\caption*{\textbf{NIG ($\alpha=0.4,\ \beta=0.14$)}}
\scalebox{0.7}{
\begin{tabular}{l|ccc|ccc|ccc}
\toprule
Est. & $ME$ & $SE$ & $DE$ & $SB$ & $MB$ & $RB$ & $UP_{5}$ & $UP_{10}$ & $CT$ \\
\midrule
\#1 & 22.1\% & 27.0\% & 21.8\% & -10.1\% & -14.5\% & -20.9\% & 65.7\% & 57.9\% & 6.6\% \\
\#2 & 22.0\% & 27.0\% & 21.5\% & -9.6\% & -14.0\% & -20.6\% & 65.0\% & 57.1\% & 6.5\% \\
\#3 & 21.7\% & 26.9\% & 21.0\% & -8.7\% & -13.1\% & -19.8\% & 63.5\% & 55.5\% & 6.3\% \\
\#4 & 21.7\% & 26.9\% & 20.8\% & -8.4\% & -12.8\% & -19.5\% & 63.0\% & 54.9\% & 6.2\% \\
\#5 & \textbf{21.5\%} & \textbf{26.8\%} & 20.5\% & -8.0\% & -12.4\% & -19.2\% & 62.3\% & 54.2\% & 6.1\% \\
\#6 & \textbf{21.5\%} & \textbf{26.8\%} & \textbf{20.3\%} & \textbf{-7.6\%} & \textbf{-12.0\%} & \textbf{-18.9\%} & \textbf{61.8\%} & \textbf{53.6\%} & \textbf{6.0\%} \\
\midrule
\gray{\#VaR} & 21.6\% & 27.6\% & 19.0\% & -4.5\% & -9.3\% & -16.8\% & 56.9\% & 48.7\% & 2.2\% \\
\bottomrule
\end{tabular}}
\end{minipage}
} 

\caption{Performance output for ES estimators \#1--\#6 under Normal, Student's $t$ ($\nu=5$), Negative GPD ($\xi\in\{1/3,1/4\}$), and NIG ($\alpha=0.4,\ \beta=0.14$) distributions. Left sub-panel: \textbf{non-overlapping} data; right sub-panel: \textbf{overlapping} data. Sample size $n=250$, confidence threshold $\alpha=2.5\%$. See Table~\ref{T:estimators} for estimator definitions and Table~\ref{T:performance.metrics} for metric definitions. The best results for each metric are highlighted in bold.}
\label{tab:biases:table1}
\end{table}

\begin{table}[htp!]
\centering
\small
\setlength{\tabcolsep}{4pt}

\scalebox{1}{
\begin{minipage}[t]{0.48\textwidth}
\centering
\caption*{\textbf{NIG ($\alpha=0.4,\ \beta=-0.14$)}}
\scalebox{0.7}{
\begin{tabular}{l|ccc|ccc|ccc}
\toprule
Est. & $ME$ & $SE$ & $DE$ & $SB$ & $MB$ & $RB$ & $UP_{5}$ & $UP_{10}$ & $CT$ \\
\midrule
\#1 & \textbf{21.3\%} & \textbf{26.9\%} & 18.4\% & -2.6\% & -6.1\% & -11.8\% & 51.7\% & 43.8\% & 3.3\% \\
\#2 & 21.6\% & 27.5\% & 17.7\% & \textbf{-0.8\%} & -4.5\% & -10.5\% & 49.2\% & 41.4\% & 3.2\% \\
\#3 & 23.5\% & 30.8\% & 15.9\% & 4.9\% & \textbf{0.6\%} & -6.4\% & 42.0\% & 34.8\% & 2.9\% \\
\#4 & 24.1\% & 31.8\% & 15.4\% & 6.5\% & 1.9\% & -5.4\% & 40.3\% & 33.2\% & 2.9\% \\
\#5 & 25.8\% & 34.7\% & 14.5\% & 9.8\% & 4.8\% & -3.2\% & 36.9\% & 30.1\% & 2.7\% \\
\#6 & 26.5\% & 35.9\% & \textbf{14.1\%} & 11.4\% & 6.2\% & \textbf{-2.1\%} & \textbf{35.3\%} & \textbf{28.7\%} & \textbf{2.6\%} \\
\midrule
\gray{\#VaR} & 28.3\% & 38.7\% & 13.1\% & 15.0\% & 9.4\% & 2.3\% & 31.8\% & 25.5\% & 0.9\% \\
\bottomrule
\end{tabular}}
\end{minipage}
\hspace{0.02\textwidth}
\begin{minipage}[t]{0.48\textwidth}
\centering
\caption*{\textbf{NIG ($\alpha=0.4,\ \beta=-0.14$)}}
\scalebox{0.7}{
\begin{tabular}{l|ccc|ccc|ccc}
\toprule
Est. & $ME$ & $SE$ & $DE$ & $SB$ & $MB$ & $RB$ & $UP_{5}$ & $UP_{10}$ & $CT$ \\
\midrule
\#1 & 30.7\% & 37.2\% & 30.5\% & -15.6\% & -22.2\% & -29.6\% & 70.1\% & 65.1\% & 7.7\% \\
\#2 & 30.6\% & 37.2\% & 30.3\% & -15.1\% & -21.7\% & -29.3\% & 69.6\% & 64.5\% & 7.6\% \\
\#3 & 30.4\% & 37.0\% & 29.8\% & -14.3\% & -20.9\% & -28.6\% & 68.7\% & 63.5\% & 7.4\% \\
\#4 & 30.3\% & 37.0\% & 29.6\% & -14.0\% & -20.6\% & -28.3\% & 68.4\% & 63.1\% & 7.4\% \\
\#5 & \textbf{30.1\%} & 36.9\% & 29.3\% & -13.6\% & -20.2\% & -28.0\% & 68.0\% & 62.7\% & 7.3\% \\
\#6 & \textbf{30.1\%} & \textbf{36.8\%} & \textbf{29.1\%} & \textbf{-13.3\%} & \textbf{-19.9\%} & \textbf{-27.8\%} & \textbf{67.6\%} & \textbf{62.3\%} & \textbf{7.2\%} \\
\midrule
\gray{\#VaR} & 29.9\% & 37.5\% & 27.3\% & -9.1\% & -16.2\% & -25.1\% & 63.0\% & 57.5\% & 2.5\% \\
\bottomrule
\end{tabular}}
\end{minipage}
} 

\par\medskip

\scalebox{1}{
\begin{minipage}[t]{0.48\textwidth}
\centering
\caption*{\textbf{NIG ($\alpha=0.55,\ \beta=0.3025$)}}
\scalebox{0.7}{
\begin{tabular}{l|ccc|ccc|ccc}
\toprule
Est. & $ME$ & $SE$ & $DE$ & $SB$ & $MB$ & $RB$ & $UP_{5}$ & $UP_{10}$ & $CT$ \\
\midrule
\#1 & \textbf{15.0\%} & \textbf{18.9\%} & 13.3\% & -2.1\% & -4.3\% & -7.4\% & 48.4\% & 37.0\% & 3.2\% \\
\#2 & 15.2\% & 19.2\% & 12.8\% & \textbf{-0.9\%} & -3.1\% & -6.4\% & 45.9\% & 34.8\% & 3.1\% \\
\#3 & 16.3\% & 21.3\% & 11.4\% & 3.1\% & \textbf{0.5\%} & -3.4\% & 38.8\% & 28.7\% & 2.8\% \\
\#4 & 16.7\% & 21.9\% & 11.1\% & 4.2\% & 1.5\% & -2.6\% & 37.0\% & 27.2\% & 2.7\% \\
\#5 & 17.8\% & 23.7\% & 10.4\% & 6.5\% & 3.5\% & -1.0\% & 33.8\% & 24.5\% & 2.6\% \\
\#6 & 18.3\% & 24.5\% & \textbf{10.1\%} & 7.6\% & 4.5\% & \textbf{-0.2\%} & \textbf{32.2\%} & \textbf{23.2\%} & \textbf{2.5\%} \\
\midrule
\gray{\#VaR} & 19.2\% & 25.8\% & 9.6\% & 9.4\% & 6.1\% & 2.2\% & 29.8\% & 21.2\% & 0.9\% \\
\bottomrule
\end{tabular}}
\end{minipage}
\hspace{0.02\textwidth}
\begin{minipage}[t]{0.48\textwidth}
\centering
\caption*{\textbf{NIG ($\alpha=0.55,\ \beta=0.3025$)}}
\scalebox{0.7}{
\begin{tabular}{l|ccc|ccc|ccc}
\toprule
Est. & $ME$ & $SE$ & $DE$ & $SB$ & $MB$ & $RB$ & $UP_{5}$ & $UP_{10}$ & $CT$ \\
\midrule
\#1 & 17.6\% & 21.7\% & 17.6\% & -7.8\% & -10.9\% & -16.5\% & 62.3\% & 52.0\% & 6.1\% \\
\#2 & 17.5\% & 21.7\% & 17.3\% & -7.4\% & -10.5\% & -16.1\% & 61.4\% & 51.0\% & 6.0\% \\
\#3 & 17.3\% & \textbf{21.5\%} & 16.8\% & -6.5\% & -9.6\% & -15.3\% & 59.6\% & 49.1\% & 5.8\% \\
\#4 & 17.3\% & \textbf{21.5\%} & 16.6\% & -6.2\% & -9.3\% & -15.1\% & 59.0\% & 48.4\% & 5.7\% \\
\#5 & 17.2\% & \textbf{21.5\%} & 16.4\% & -5.7\% & -8.9\% & -14.7\% & 58.1\% & 47.5\% & 5.6\% \\
\#6 & \textbf{17.1\%} & \textbf{21.5\%} & \textbf{16.2\%} & \textbf{-5.4\%} & \textbf{-8.5\%} & \textbf{-14.4\%} & \textbf{57.4\%} & \textbf{46.8\%} & \textbf{5.5\%} \\
\midrule
\gray{\#VaR} & 17.4\% & 22.2\% & 15.3\% & -3.0\% & -6.4\% & -12.7\% & 52.9\% & 42.5\% & 2.0\% \\
\bottomrule
\end{tabular}}
\end{minipage}
} 

\par\medskip

\scalebox{1}{
\begin{minipage}[t]{0.48\textwidth}
\centering
\caption*{\textbf{NIG ($\alpha=0.55,\ \beta=-0.3025$)}}
\scalebox{0.7}{
\begin{tabular}{l|ccc|ccc|ccc}
\toprule
Est. & $ME$ & $SE$ & $DE$ & $SB$ & $MB$ & $RB$ & $UP_{5}$ & $UP_{10}$ & $CT$ \\
\midrule
\#1 & \textbf{21.4\%} & \textbf{26.9\%} & 18.5\% & -2.5\% & -6.0\% & -11.8\% & 51.5\% & 43.6\% & 3.3\% \\
\#2 & 21.7\% & 27.6\% & 17.8\% & \textbf{-0.7\%} & -4.3\% & -10.4\% & 49.0\% & 41.2\% & 3.2\% \\
\#3 & 23.5\% & 30.9\% & 15.9\% & 5.0\% & \textbf{0.8\%} & -6.3\% & 41.8\% & 34.6\% & 2.9\% \\
\#4 & 24.1\% & 31.8\% & 15.4\% & 6.5\% & 2.1\% & -5.3\% & 40.0\% & 32.9\% & 2.8\% \\
\#5 & 25.8\% & 34.6\% & 14.6\% & 9.9\% & 5.0\% & -3.0\% & 36.6\% & 29.9\% & 2.7\% \\
\#6 & 26.6\% & 35.9\% & \textbf{14.1\%} & 11.5\% & 6.4\% & \textbf{-1.9\%} & \textbf{35.0\%} & \textbf{28.5\%} & \textbf{2.6\%} \\
\midrule
\gray{\#VaR} & 28.6\% & 39.0\% & 13.1\% & 15.4\% & 9.8\% & 2.4\% & 31.4\% & 25.3\% & 1.0\% \\
\bottomrule
\end{tabular}}
\end{minipage}
\hspace{0.02\textwidth}
\begin{minipage}[t]{0.48\textwidth}
\centering
\caption*{\textbf{NIG ($\alpha=0.55,\ \beta=-0.3025$)}}
\scalebox{0.7}{
\begin{tabular}{l|ccc|ccc|ccc}
\toprule
Est. & $ME$ & $SE$ & $DE$ & $SB$ & $MB$ & $RB$ & $UP_{5}$ & $UP_{10}$ & $CT$ \\
\midrule
\#1 & 31.5\% & 38.1\% & 31.5\% & -16.3\% & -22.9\% & -30.6\% & 70.5\% & 65.5\% & 7.9\% \\
\#2 & 31.4\% & 38.0\% & 31.2\% & -15.8\% & -22.4\% & -30.3\% & 70.0\% & 65.0\% & 7.8\% \\
\#3 & 31.1\% & 37.9\% & 30.7\% & -15.0\% & -21.6\% & -29.7\% & 69.1\% & 64.1\% & 7.6\% \\
\#4 & 31.1\% & 37.8\% & 30.5\% & -14.7\% & -21.3\% & -29.4\% & 68.8\% & 63.7\% & 7.5\% \\
\#5 & \textbf{30.9\%} & \textbf{37.7\%} & 30.3\% & -14.4\% & -21.0\% & -29.1\% & 68.4\% & 63.3\% & \textbf{7.4\%} \\
\#6 & \textbf{30.9\%} & \textbf{37.7\%} & \textbf{30.1\%} & \textbf{-14.0\%} & \textbf{-20.7\%} & \textbf{-28.9\%} & \textbf{68.1\%} & \textbf{62.9\%} & \textbf{7.4\%} \\
\midrule
\gray{\#VaR} & 30.7\% & 38.3\% & 28.4\% & -10.1\% & -17.1\% & -26.2\% & 63.7\% & 58.4\% & 2.6\% \\
\bottomrule
\end{tabular}}
\end{minipage}
} 

\par\medskip

\scalebox{1}{
\begin{minipage}[t]{0.48\textwidth}
\centering
\caption*{\textbf{NIG ($\alpha=0.4,\ \beta=0.22$)}}
\scalebox{0.7}{
\begin{tabular}{l|ccc|ccc|ccc}
\toprule
Est. & $ME$ & $SE$ & $DE$ & $SB$ & $MB$ & $RB$ & $UP_{5}$ & $UP_{10}$ & $CT$ \\
\midrule
\#1 & \textbf{17.9\%} & \textbf{22.5\%} & 15.6\% & -2.3\% & -5.2\% & -9.2\% & 50.4\% & 40.8\% & 3.3\% \\
\#2 & 18.1\% & 23.1\% & 15.0\% & \textbf{-0.8\%} & -3.8\% & -8.0\% & 47.9\% & 38.5\% & 3.2\% \\
\#3 & 19.6\% & 25.8\% & 13.4\% & 4.0\% & \textbf{0.4\%} & -4.6\% & 40.8\% & 32.1\% & 2.9\% \\
\#4 & 20.1\% & 26.6\% & 13.0\% & 5.2\% & 1.6\% & -3.7\% & 39.0\% & 30.5\% & 2.8\% \\
\#5 & 21.5\% & 28.9\% & 12.3\% & 8.1\% & 4.0\% & -1.8\% & 35.6\% & 27.7\% & \textbf{2.6\%} \\
\#6 & 22.2\% & 29.9\% & \textbf{11.9\%} & 9.4\% & 5.2\% & \textbf{-0.9\%} & \textbf{34.0\%} & \textbf{26.3\%} & \textbf{2.6\%} \\
\midrule
\gray{\#VaR} & 23.1\% & 31.5\% & 11.2\% & 11.7\% & 7.2\% & 2.2\% & 31.3\% & 23.8\% & 0.9\% \\
\bottomrule
\end{tabular}}
\end{minipage}
\hspace{0.02\textwidth}
\begin{minipage}[t]{0.48\textwidth}
\centering
\caption*{\textbf{NIG ($\alpha=0.4,\ \beta=0.22$)}}
\scalebox{0.7}{
\begin{tabular}{l|ccc|ccc|ccc}
\toprule
Est. & $ME$ & $SE$ & $DE$ & $SB$ & $MB$ & $RB$ & $UP_{5}$ & $UP_{10}$ & $CT$ \\
\midrule
\#1 & 21.3\% & 26.2\% & 21.0\% & -9.8\% & -14.3\% & -20.0\% & 66.0\% & 57.9\% & 6.7\% \\
\#2 & 21.3\% & 26.2\% & 20.7\% & -9.4\% & -13.8\% & -19.7\% & 65.3\% & 57.1\% & 6.5\% \\
\#3 & 21.0\% & 26.0\% & 20.2\% & -8.5\% & -13.0\% & -19.0\% & 63.9\% & 55.6\% & 6.3\% \\
\#4 & 20.9\% & 26.0\% & 20.0\% & -8.2\% & -12.7\% & -18.7\% & 63.4\% & 55.0\% & 6.2\% \\
\#5 & \textbf{20.8\%} & \textbf{25.9\%} & 19.8\% & -7.8\% & -12.3\% & -18.4\% & 62.7\% & 54.2\% & \textbf{6.1\%} \\
\#6 & \textbf{20.8\%} & \textbf{25.9\%} & \textbf{19.6\%} & \textbf{-7.5\%} & \textbf{-12.0\%} & \textbf{-18.1\%} & \textbf{62.1\%} & \textbf{53.6\%} & \textbf{6.1\%} \\
\midrule
\gray{\#VaR} & 20.9\% & 26.7\% & 18.3\% & -4.3\% & -9.2\% & -15.9\% & 57.1\% & 48.5\% & 2.2\% \\
\bottomrule
\end{tabular}}
\end{minipage}
} 

\par\medskip

\scalebox{1}{
\begin{minipage}[t]{0.48\textwidth}
\centering
\caption*{\textbf{NIG ($\alpha=0.4,\ \beta=-0.22$)}}
\scalebox{0.7}{
\begin{tabular}{l|ccc|ccc|ccc}
\toprule
Est. & $ME$ & $SE$ & $DE$ & $SB$ & $MB$ & $RB$ & $UP_{5}$ & $UP_{10}$ & $CT$ \\
\midrule
\#1 & \textbf{25.1\%} & \textbf{31.7\%} & 21.4\% & -3.0\% & -7.7\% & -14.5\% & 53.6\% & 47.0\% & 3.5\% \\
\#2 & 25.5\% & 32.5\% & 20.7\% & \textbf{-1.0\%} & -5.9\% & -13.1\% & 51.2\% & 44.6\% & 3.4\% \\
\#3 & 27.8\% & 36.9\% & 18.6\% & 5.8\% & \textbf{0.0\%} & -8.7\% & 44.0\% & 37.8\% & 3.1\% \\
\#4 & 28.5\% & 38.1\% & 18.0\% & 7.6\% & 1.6\% & -7.5\% & 42.2\% & 36.1\% & 3.0\% \\
\#5 & 30.7\% & 41.7\% & 17.0\% & 11.7\% & 5.0\% & -5.1\% & 38.8\% & 33.0\% & 2.8\% \\
\#6 & 31.6\% & 43.2\% & \textbf{16.6\%} & 13.5\% & 6.6\% & \textbf{-4.0\%} & \textbf{37.2\%} & \textbf{31.5\%} & \textbf{2.7\%} \\
\midrule
\gray{\#VaR} & 33.9\% & 47.0\% & 15.1\% & 18.4\% & 10.9\% & 2.3\% & 33.0\% & 27.6\% & 1.0\% \\
\bottomrule
\end{tabular}}
\end{minipage}
\hspace{0.02\textwidth}
\begin{minipage}[t]{0.48\textwidth}
\centering
\caption*{\textbf{NIG ($\alpha=0.4,\ \beta=-0.22$)}}
\scalebox{0.7}{
\begin{tabular}{l|ccc|ccc|ccc}
\toprule
Est. & $ME$ & $SE$ & $DE$ & $SB$ & $MB$ & $RB$ & $UP_{5}$ & $UP_{10}$ & $CT$ \\
\midrule
\#1 & 36.3\% & 43.6\% & 35.9\% & -19.4\% & -27.8\% & -35.5\% & 72.6\% & 68.6\% & 8.6\% \\
\#2 & 36.1\% & 43.5\% & 35.7\% & -19.0\% & -27.4\% & -35.2\% & 72.3\% & 68.2\% & 8.6\% \\
\#3 & 35.9\% & 43.4\% & 35.2\% & -18.3\% & -26.7\% & -34.6\% & 71.7\% & 67.5\% & 8.4\% \\
\#4 & 35.8\% & 43.3\% & 35.1\% & -18.0\% & -26.4\% & -34.4\% & 71.4\% & 67.2\% & 8.3\% \\
\#5 & 35.7\% & \textbf{43.2\%} & 34.9\% & -17.7\% & -26.1\% & -34.2\% & 71.1\% & 66.9\% & \textbf{8.2\%} \\
\#6 & \textbf{35.6\%} & \textbf{43.2\%} & \textbf{34.7\%} & \textbf{-17.4\%} & \textbf{-25.8\%} & \textbf{-34.0\%} & \textbf{70.9\%} & \textbf{66.6\%} & \textbf{8.2\%} \\
\midrule
\gray{\#VaR} & 35.2\% & 43.9\% & 32.5\% & -12.6\% & -21.5\% & -31.1\% & 66.3\% & 61.9\% & 2.8\% \\
\bottomrule
\end{tabular}}
\end{minipage}
} 

\caption{Performance output for ES estimators \#1--\#6 under remaining NIG distributions. Left sub-panel: \textbf{non-overlapping} data; right sub-panel: \textbf{overlapping} data. Sample size $n=250$, confidence threshold $\alpha=2.5\%$. See Table~\ref{T:estimators} for estimator definitions and Table~\ref{T:performance.metrics} for metric definitions. The best results for each metric are highlighted in bold.}
\label{tab:biases:table2}
\end{table}

\begin{table}[htp!]
\centering
\small
\setlength{\tabcolsep}{4pt}

\scalebox{1}{
\begin{minipage}[t]{0.48\textwidth}
\centering
\caption*{\textbf{GARCH-norm ($\omega = 0.000004$, $\alpha=0.17$, $\beta=0.8$)}}
\scalebox{0.7}{
\begin{tabular}{l|ccc|ccc|ccc}
\toprule
Est. & $ME$ & $SE$ & $DE$ & $SB$ & $MB$ & $RB$ & $UP_{5}$ & $UP_{10}$ & $CT$ \\
\midrule
\#1 & 27.7\% & \textbf{38.4\%} & 24.8\% & -10.7\% & -19.4\% & -23.5\% & 70.5\% & 64.6\% & 6.1\% \\
\#2 & 27.6\% & 38.7\% & 24.4\% & -9.9\% & -18.8\% & -22.9\% & 69.6\% & 63.5\% & 6.0\% \\
\#3 & \textbf{27.4\%} & 39.4\% & 23.3\% & -7.8\% & -17.0\% & -21.5\% & 67.0\% & 60.7\% & 5.6\% \\
\#4 & \textbf{27.4\%} & 39.7\% & 23.0\% & -7.1\% & -16.4\% & -21.1\% & 66.3\% & 59.9\% & 5.5\% \\
\#5 & \textbf{27.4\%} & 40.1\% & 22.5\% & -6.0\% & -15.4\% & -20.3\% & 64.9\% & 58.3\% & 5.4\% \\
\#6 & \textbf{27.4\%} & 40.5\% & \textbf{22.2\%} & \textbf{-5.3\%} & \textbf{-14.9\%} & \textbf{-19.9\%} & \textbf{64.1\%} & \textbf{57.5\%} & \textbf{5.3\%} \\
\midrule
\gray{\#VaR} & 27.5\% & 43.1\% & 19.7\% & 0.0\% & -10.3\% & -15.5\% & 57.6\% & 50.5\% & 1.8\% \\
\bottomrule
\end{tabular}}
\end{minipage}
\hspace{0.02\textwidth}
\begin{minipage}[t]{0.48\textwidth}
\centering
\caption*{\textbf{GARCH-norm ($\omega = 0.000004$, $\alpha=0.17$, $\beta=0.8$)}}
\scalebox{0.7}{
\begin{tabular}{l|ccc|ccc|ccc}
\toprule
Est. & $ME$ & $SE$ & $DE$ & $SB$ & $MB$ & $RB$ & $UP_{5}$ & $UP_{10}$ & $CT$ \\
\midrule
\#1 & 34.5\% & \textbf{45.1\%} & 32.7\% & -17.7\% & -27.4\% & -32.8\% & 75.3\% & 71.1\% & 8.7\% \\
\#2 & 34.4\% & 45.2\% & 32.4\% & -17.1\% & -26.9\% & -32.4\% & 74.8\% & 70.5\% & 8.6\% \\
\#3 & 34.1\% & 45.3\% & 31.7\% & -15.9\% & -25.9\% & -31.6\% & 73.7\% & 69.3\% & 8.3\% \\
\#4 & 34.1\% & 45.4\% & 31.5\% & -15.5\% & -25.5\% & -31.3\% & 73.3\% & 68.9\% & 8.2\% \\
\#5 & \textbf{33.9\%} & 45.5\% & 31.2\% & -14.9\% & -25.0\% & -30.9\% & 72.8\% & 68.3\% & 8.1\% \\
\#6 & \textbf{33.9\%} & 45.6\% & \textbf{31.0\%} & \textbf{-14.4\%} & \textbf{-24.6\%} & \textbf{-30.6\%} & \textbf{72.3\%} & \textbf{67.8\%} & \textbf{8.0\%} \\
\midrule
\gray{\#VaR} & 33.3\% & 48.1\% & 27.6\% & -7.4\% & -18.6\% & -26.2\% & 65.6\% & 60.5\% & 2.6\% \\
\bottomrule
\end{tabular}}
\end{minipage}
} 

\par\medskip

\scalebox{1}{
\begin{minipage}[t]{0.48\textwidth}
\centering
\caption*{\textbf{GARCH-norm ($\omega = 0.00001$, $\alpha=0.05$, $\beta=0.94$)}}
\scalebox{0.7}{
\begin{tabular}{l|ccc|ccc|ccc}
\toprule
Est. & $ME$ & $SE$ & $DE$ & $SB$ & $MB$ & $RB$ & $UP_{5}$ & $UP_{10}$ & $CT$ \\
\midrule
\#1 & 19.1\% & \textbf{23.7\%} & 18.6\% & -8.4\% & -12.5\% & -16.8\% & 64.5\% & 55.2\% & 5.5\% \\
\#2 & 19.0\% & \textbf{23.7\%} & 18.3\% & -7.8\% & -11.9\% & -16.3\% & 63.4\% & 53.9\% & 5.4\% \\
\#3 & 18.7\% & \textbf{23.7\%} & 17.4\% & -6.2\% & -10.3\% & -15.0\% & 60.4\% & 50.7\% & 5.1\% \\
\#4 & 18.7\% & 23.8\% & 17.1\% & -5.6\% & -9.8\% & -14.6\% & 59.4\% & 49.6\% & 5.0\% \\
\#5 & 18.6\% & 23.9\% & 16.7\% & -4.8\% & -9.0\% & -13.9\% & 57.8\% & 48.0\% & 4.8\% \\
\#6 & \textbf{18.5\%} & 24.0\% & \textbf{16.4\%} & \textbf{-4.2\%} & \textbf{-8.5\%} & \textbf{-13.5\%} & \textbf{56.9\%} & \textbf{47.0\%} & \textbf{4.7\%} \\
\midrule
\gray{\#VaR} & 18.6\% & 24.6\% & 15.4\% & -2.1\% & -6.6\% & -12.0\% & 53.0\% & 43.2\% & 1.8\% \\
\bottomrule
\end{tabular}}
\end{minipage}
\hspace{0.02\textwidth}
\begin{minipage}[t]{0.48\textwidth}
\centering
\caption*{\textbf{GARCH-norm ($\omega = 0.00001$, $\alpha=0.05$, $\beta=0.94$)}}
\scalebox{0.7}{
\begin{tabular}{l|ccc|ccc|ccc}
\toprule
Est. & $ME$ & $SE$ & $DE$ & $SB$ & $MB$ & $RB$ & $UP_{5}$ & $UP_{10}$ & $CT$ \\
\midrule
\#1 & 25.3\% & 30.7\% & 25.7\% & -14.0\% & -18.8\% & -25.7\% & 70.4\% & 63.9\% & 7.9\% \\
\#2 & 25.2\% & 30.7\% & 25.4\% & -13.4\% & -18.4\% & -25.3\% & 69.7\% & 63.1\% & 7.8\% \\
\#3 & 24.9\% & 30.5\% & 24.8\% & -12.4\% & -17.4\% & -24.5\% & 68.3\% & 61.6\% & 7.5\% \\
\#4 & 24.8\% & 30.5\% & 24.6\% & -12.0\% & -17.0\% & -24.2\% & 67.7\% & 61.0\% & 7.4\% \\
\#5 & 24.7\% & \textbf{30.4\%} & 24.3\% & -11.5\% & -16.5\% & -23.8\% & 67.1\% & 60.3\% & 7.3\% \\
\#6 & \textbf{24.6\%} & \textbf{30.4\%} & \textbf{24.1\%} & \textbf{-11.1\%} & \textbf{-16.2\%} & \textbf{-23.5\%} & \textbf{66.5\%} & \textbf{59.7\%} & \textbf{7.2\%} \\
\midrule
\gray{\#VaR} &24.4\% & 30.8\% & 22.7\% & -8.1\% & -13.5\% & -21.5\% & 62.4\% & 55.4\% & 2.6\% \\
\bottomrule
\end{tabular}}
\end{minipage}
} 

\par\medskip

\scalebox{1}{
\begin{minipage}[t]{0.48\textwidth}
\centering
\caption*{\textbf{GARCH-$t$ ($\omega = 0.000004$, $\alpha=0.17$, $\beta=0.8$, $\nu=5$)}}
\scalebox{0.7}{
\begin{tabular}{l|ccc|ccc|ccc}
\toprule
Est. & $ME$ & $SE$ & $DE$ & $SB$ & $MB$ & $RB$ & $UP_{5}$ & $UP_{10}$ & $CT$ \\
\midrule
\#1 & \textbf{35.1\%} & \textbf{57.7\%} & 29.3\% & -10.8\% & -23.9\% & -28.8\% & 72.0\% & 67.5\% & 6.1\% \\
\#2 & \textbf{35.1\%} & 58.5\% & 28.8\% & -9.7\% & -23.1\% & -28.1\% & 71.0\% & 66.4\% & 6.0\% \\
\#3 & 35.2\% & 61.0\% & 27.5\% & -6.4\% & -20.6\% & -26.2\% & 68.0\% & 63.1\% & 5.6\% \\
\#4 & 35.3\% & 61.7\% & 27.1\% & -5.5\% & -19.8\% & -25.7\% & 67.2\% & 62.2\% & 5.5\% \\
\#5 & 35.5\% & 63.3\% & 26.4\% & -3.7\% & -18.4\% & -24.6\% & 65.5\% & 60.4\% & 5.3\% \\
\#6 & 35.6\% & 64.2\% & \textbf{26.0\%} & \textbf{-2.7\%} & \textbf{-17.7\%} & \textbf{-24.1\%} & \textbf{64.6\%} & \textbf{59.5\%} & \textbf{5.2\%} \\
\midrule
\gray{\#VaR} & 36.2\% & 71.0\% & 21.8\% & 6.5\% & -10.1\% & -16.2\% & 55.9\% & 50.1\% & 1.6\% \\
\bottomrule
\end{tabular}}
\end{minipage}
\hspace{0.02\textwidth}
\begin{minipage}[t]{0.48\textwidth}
\centering
\caption*{\textbf{GARCH-$t$ ($\omega = 0.000004$, $\alpha=0.17$, $\beta=0.8$, $\nu=5$)}}
\scalebox{0.7}{
\begin{tabular}{l|ccc|ccc|ccc}
\toprule
Est. & $ME$ & $SE$ & $DE$ & $SB$ & $MB$ & $RB$ & $UP_{5}$ & $UP_{10}$ & $CT$ \\
\midrule
\#1 & 42.5\% & \textbf{61.8\%} & 38.6\% & -21.1\% & -35.1\% & -39.7\% & 78.7\% & 75.6\% & 9.8\% \\
\#2 & 42.4\% & 62.1\% & 38.3\% & -20.5\% & -34.6\% & -39.3\% & 78.2\% & 75.1\% & 9.7\% \\
\#3 & 42.2\% & 62.6\% & 37.7\% & -19.3\% & -33.6\% & -38.5\% & 77.4\% & 74.2\% & 9.4\% \\
\#4 & 42.1\% & 62.9\% & 37.5\% & -18.9\% & -33.3\% & -38.3\% & 77.1\% & 73.9\% & 9.3\% \\
\#5 & 42.0\% & 63.1\% & 37.2\% & -18.3\% & -32.8\% & -37.9\% & 76.7\% & 73.4\% & 9.2\% \\
\#6 & \textbf{41.9\%} & 63.4\% & \textbf{36.9\%} & \textbf{-17.8\%} & \textbf{-32.4\%} & \textbf{-37.6\%} & \textbf{76.4\%} & \textbf{73.0\%} & \textbf{9.1\%} \\
\midrule
\gray{\#VaR} &41.2\% & 69.7\% & 32.0\% & -7.4\% & -24.2\% & -29.7\% & 68.6\% & 64.6\% & 2.7\% \\
\bottomrule
\end{tabular}}
\end{minipage}
} 

\par\medskip

\scalebox{1}{
\begin{minipage}[t]{0.48\textwidth}
\centering
\caption*{\textbf{GARCH-$t$ ($\omega = 0.00001$, $\alpha=0.05$, $\beta=0.94$, $\nu=5$)}}
\scalebox{0.7}{
\begin{tabular}{l|ccc|ccc|ccc}
\toprule
Est. & $ME$ & $SE$ & $DE$ & $SB$ & $MB$ & $RB$ & $UP_{5}$ & $UP_{10}$ & $CT$ \\
\midrule
\#1 & 29.3\% & \textbf{40.2\%} & 26.4\% & -11.3\% & -20.3\% & -25.0\% & 70.2\% & 64.7\% & 5.9\% \\
\#2 & 29.2\% & 40.5\% & 25.9\% & -10.4\% & -19.5\% & -24.3\% & 69.1\% & 63.5\% & 5.8\% \\
\#3 & \textbf{29.1\%} & 41.8\% & 24.6\% & -7.5\% & -17.1\% & -22.5\% & 65.9\% & 59.9\% & 5.4\% \\
\#4 & \textbf{29.1\%} & 42.2\% & 24.2\% & -6.6\% & -16.4\% & -21.9\% & 64.9\% & 59.0\% & 5.3\% \\
\#5 & 29.3\% & 43.2\% & 23.6\% & -5.0\% & -15.0\% & -20.9\% & 63.1\% & 57.0\% & 5.1\% \\
\#6 & 29.3\% & 43.6\% & \textbf{23.2\%} & \textbf{-4.1\%} & \textbf{-14.3\%} & \textbf{-20.4\%} & \textbf{62.1\%} & \textbf{56.0\%} & \textbf{5.0\%} \\
\midrule
\gray{\#VaR} & 29.1\% & 45.1\% & 20.8\% & 0.5\% & -10.0\% & -15.9\% & 56.6\% & 50.0\% & 1.7\% \\
\bottomrule
\end{tabular}}
\end{minipage}
\hspace{0.02\textwidth}
\begin{minipage}[t]{0.48\textwidth}
\centering
\caption*{\textbf{GARCH-$t$ ($\omega = 0.00001$, $\alpha=0.05$, $\beta=0.94$, $\nu=5$)}}
\scalebox{0.7}{
\begin{tabular}{l|ccc|ccc|ccc}
\toprule
Est. & $ME$ & $SE$ & $DE$ & $SB$ & $MB$ & $RB$ & $UP_{5}$ & $UP_{10}$ & $CT$ \\
\midrule
\#1 & 34.9\% & \textbf{44.0\%} & 34.1\% & -20.5\% & -29.6\% & -34.8\% & 77.8\% & 73.8\% & 9.7\% \\
\#2 & 34.8\% & \textbf{44.0\%} & 33.8\% & -20.0\% & -29.1\% & -34.5\% & 77.3\% & 73.3\% & 9.6\% \\
\#3 & 34.5\% & \textbf{44.0\%} & 33.2\% & -19.0\% & -28.3\% & -33.8\% & 76.4\% & 72.2\% & 9.3\% \\
\#4 & 34.4\% & 44.1\% & 33.0\% & -18.6\% & -27.9\% & -33.5\% & 76.1\% & 71.8\% & 9.2\% \\
\#5 & 34.2\% & 44.1\% & 32.7\% & -18.1\% & -27.5\% & -33.1\% & 75.6\% & 71.3\% & 9.1\% \\
\#6 & \textbf{34.1\%} & 44.1\% & \textbf{32.5\%} & \textbf{-17.8\%} & \textbf{-27.2\%} & \textbf{-32.9\%} & \textbf{75.3\%} & \textbf{71.0\%} & \textbf{9.0\%} \\
\midrule
\gray{\#VaR} & 33.2\% & 45.4\% & 29.6\% & -12.0\% & -22.2\% & -28.5\% & 69.7\% & 64.9\% & 3.0\% \\
\bottomrule
\end{tabular}}
\end{minipage}
} 

\caption{Performance output for ES estimators \#1--\#6 under GARCH-norm and GARCH-$t$ specifications. Left sub-panel: \textbf{non-overlapping} data; right sub-panel: \textbf{overlapping} data. Sample size $n=250$, confidence threshold $\alpha=2.5\%$. See Table~\ref{T:estimators} for estimator definitions and Table~\ref{T:performance.metrics} for metric definitions. The best results for each metric are highlighted in bold.}
\label{tab:biases:garch}
\end{table}

\clearpage

\subsection{Market data}\label{S:tables.market}


\begin{table}[htp!]

\scalebox{0.92}{
\begin{minipage}[t]{0.15\textwidth}
\centering
\caption*{\textbf{SOFR}}
\scalebox{0.7}{
\begin{tabular}{l|cc}
\toprule
Est. & Daily & 10-day \\
\midrule
\#1 & 16.3\% & 18.0\%  \\
\#2 & 15.9\%  & 17.5\% \\
\#3 & 14.5\%  & 15.0\% \\
\#4 & 14.3\%  & 14.7\%\\
\#5 & 13.7\%  & 14.2\% \\
\#6 & 13.4\% & 14.0\% \\ \midrule
\gray{\#VaR}  & 1.1\% & 2.5\%\\
\bottomrule
\end{tabular}}
\end{minipage}

\hspace{0.02\textwidth}

\begin{minipage}[t]{0.15\textwidth}
\centering
\caption*{\textbf{US 1M Bill}}
\scalebox{0.7}{
\begin{tabular}{l|cc}
\toprule
Est. & Daily & 10-day \\
\midrule
\#1 & 7.0\% & 20.5\%  \\
\#2 & 6.8\%  & 19.9\% \\
\#3 & 6.3\%  & 19.5\% \\
\#4 & 6.2\%  & 19.3\%\\
\#5 & 5.9\%  & 19.0\% \\
\#6 & 5.8\% & 18.9\% \\ \midrule
\gray{\#VaR}  & 1.5\% & 4.0\%\\
\bottomrule
\end{tabular}}
\end{minipage}

\hspace{0.02\textwidth}

\begin{minipage}[t]{0.15\textwidth}
\centering
\caption*{\textbf{US 2Y Note}}
\scalebox{0.7}{
\begin{tabular}{l|cc}
\toprule
Est. & Daily & 10-day \\
\midrule
\#1 & 5.7\% & 11.6\%  \\
\#2 & 5.5\%  & 11.4\% \\
\#3 & 5.2\%  & 11.0\% \\
\#4 & 5.0\%  & 10.9\%\\
\#5 & 4.9\%  &10.7\% \\
\#6 & 4.8\% &10.6\% \\ \midrule
\gray{\#VaR}  & 1.3\% & 2.6\%\\
\bottomrule
\end{tabular}}
\end{minipage}

\hspace{0.02\textwidth}

\begin{minipage}[t]{0.15\textwidth}
\centering
\caption*{\textbf{US 10 Y Note}}
\scalebox{0.7}{
\begin{tabular}{l|cc}
\toprule
Est. & Daily & 10-day \\
\midrule
\#1 & 3.9\% & 7.6\%  \\
\#2 & 3.8\%  & 7.4\% \\
\#3 & 3.5\%  & 7.1\% \\
\#4 & 3.4\%  & 7.0\%\\
\#5 & 3.2\%  &6.8\% \\
\#6 & 3.1\% &6.7\% \\ \midrule
\gray{\#VaR}  & 1.1\% & 2.4\%\\
\bottomrule
\end{tabular}}
\end{minipage}

\hspace{0.02\textwidth}

\begin{minipage}[t]{0.15\textwidth}
\centering
\caption*{\textbf{DE 10Y Bund}}
\scalebox{0.7}{
\begin{tabular}{l|cc}
\toprule
Est. & Daily & 10-day \\
\midrule
\#1 & 4.1\% & 5.7\%  \\
\#2 & 3.9\%  & 5.6\% \\
\#3 & 3.6\%  & 5.4\% \\
\#4 & 3.4\%  & 5.3\%\\
\#5 & 3.4\%  & 5.2\% \\
\#6 & 3.4\% &5.1\% \\ \midrule
\gray{\#VaR}  & 1.2\% & 2.0\%\\
\bottomrule
\end{tabular}}
\end{minipage}

\hspace{0.02\textwidth}

\begin{minipage}[t]{0.15\textwidth}
\centering
\caption*{\textbf{Xtrackers}}
\scalebox{0.7}{
\begin{tabular}{l|cc}
\toprule
Est. & Daily & 10-day \\
\midrule
\#1 & 4.0\% & 9.4\%  \\
\#2 & 3.9\%  & 9.3\% \\
\#3 & 3.6\%  & 8.9\% \\
\#4 & 3.5\%  & 8.8\%\\
\#5 & 3.3\%  &8.6\% \\
\#6 & 3.3\% &8.5\% \\ \midrule
\gray{\#VaR}  & 1.0\% & 1.9\%\\
\bottomrule
\end{tabular}}
\end{minipage}

}

\par\medskip

\scalebox{0.92}{
\begin{minipage}[t]{0.15\textwidth}
\centering
\caption*{\textbf{S\&P500}}
\scalebox{0.7}{
\begin{tabular}{l|cc}
\toprule
Est. & Daily & 10-day \\
\midrule
\#1 & 4.3\% & 9.2\%  \\
\#2 & 4.1\%  & 9.1\% \\
\#3 & 3.8\%  & 8.7\% \\
\#4 & 3.8\%  & 8.6\%\\
\#5 & 3.6\%  &8.5\% \\
\#6 & 3.5\% &8.4\% \\ \midrule
\gray{\#VaR}  & 1.4\% & 2.7\%\\
\bottomrule
\end{tabular}}
\end{minipage}

\hspace{0.02\textwidth}

\begin{minipage}[t]{0.15\textwidth}
\centering
\caption*{\textbf{VIX}}
\scalebox{0.7}{
\begin{tabular}{l|cc}
\toprule
Est. & Daily & 10-day \\
\midrule
\#1 & 3.3\% & 4.2\%  \\
\#2 & 3.2\%  & 4.1\% \\
\#3 & 2.9\%  & 3.9\% \\
\#4 & 2.8\%  & 3.8\%\\
\#5 & 2.7\%  &3.7\% \\
\#6 & 2.6\% &3.6\% \\ \midrule
\gray{\#VaR}  & 0.9\% & 1.2\%\\
\bottomrule
\end{tabular}}
\end{minipage}

\hspace{0.02\textwidth}

\begin{minipage}[t]{0.15\textwidth}
\centering
\caption*{\textbf{EUR-USD}}
\scalebox{0.7}{
\begin{tabular}{l|cc}
\toprule
Est. & Daily & 10-day \\
\midrule
\#1 & 3.4\% & 6.6\%  \\
\#2 & 3.3\%  & 6.5\% \\
\#3 & 3.0\%  & 6.3\% \\
\#4 & 2.9\%  & 6.2\%\\
\#5 & 2.8\%  &6.1\% \\
\#6 & 2.7\% &6.1\% \\ \midrule
\gray{\#VaR}  & 1.1\% & 2.4\%\\
\bottomrule
\end{tabular}}
\end{minipage}

\hspace{0.02\textwidth}

\begin{minipage}[t]{0.15\textwidth}
\centering
\caption*{\textbf{USD-JPY}}
\scalebox{0.7}{
\begin{tabular}{l|cc}
\toprule
Est. & Daily & 10-day \\
\midrule
\#1 & 3.9\% & 6.4\%  \\
\#2 & 3.8\%  & 6.3\% \\
\#3 & 3.5\%  & 6.0\% \\
\#4 & 3.4\%  & 6.0\%\\
\#5 & 3.2\%  &5.8\% \\
\#6 & 3.2\% &5.8\% \\ \midrule
\gray{\#VaR}  & 1.1\% & 1.9\%\\
\bottomrule
\end{tabular}}
\end{minipage}

\hspace{0.02\textwidth}

\begin{minipage}[t]{0.15\textwidth}
\centering
\caption*{\textbf{Brent}}
\scalebox{0.7}{
\begin{tabular}{l|cc}
\toprule
Est. & Daily & 10-day \\
\midrule
\#1 & 4.0\% & 8.4\%  \\
\#2 & 3.9\%  & 8.3\% \\
\#3 & 3.5\%  & 8.0\% \\
\#4 & 3.4\%  & 8.0\%\\
\#5 & 3.3\%  &7.8\% \\
\#6 & 3.2\% &7.7\% \\ \midrule
\gray{\#VaR}  & 0.9\% & 2.4\%\\
\bottomrule
\end{tabular}}
\end{minipage}

\hspace{0.02\textwidth}

\begin{minipage}[t]{0.15\textwidth}
\centering
\caption*{\textbf{Gold}}
\scalebox{0.7}{
\begin{tabular}{l|cc}
\toprule
Est. & Daily & 10-day \\
\midrule
\#1 & 3.6\% & 5.8\%  \\
\#2 & 3.5\%  & 5.7\% \\
\#3 & 3.2\%  & 5.5\% \\
\#4 & 3.1\%  & 5.4\%\\
\#5 & 3.0\%  &5.3\% \\
\#6 & 2.9\% &5.3\% \\ \midrule
\gray{\#VaR}  & 1.0\% & 1.6\%\\
\bottomrule
\end{tabular}}
\end{minipage}

}

\par\medskip

    \caption{Realized  CT for market data. The capital reserves are estimated from the previous $n=250$ returns, daily (left) and 10-day overlapping (right). The metrics are computed on the secured sample constructed for each business day, by confronting the estimated reserve with the subsequently realized return (the next 1-day return in the daily scheme, the realized future 10-day return in the overlapping scheme). The top row uses absolute returns, the bottom row relative returns. We can see that the 1-day \#VaR estimator has a median exception rate of $1.1\%$ across the market data timeseries analysed, close to its nominal level of $1\%$, while the 10-day overlapping ES estimate secures the position only at a median confidence level of $8.0\%$ for estimator \#1, against a target of $2.5\%$. The median confidence level slightly improves to $7.2\%$ for the most conservative estimator \#6 analysed.
    }
    \label{T:CT_market}
\end{table}

\providecommand{\cthr}{2.5\% CI}
\providecommand{\ratHdrJoint}{%
 & \multicolumn{3}{c|}{Daily} & \multicolumn{3}{c}{10-day} \\
\cmidrule(lr){2-4}\cmidrule(lr){5-7}
Est. & mean & std & \cthr & mean & std & \cthr \\}

\begin{table}[htbp!]
\centering
\renewcommand{\arraystretch}{0.9}
\setlength{\tabcolsep}{4pt}

\begin{minipage}[t]{0.45\textwidth}\centering
\caption*{\textbf{SOFR}}
\resizebox{\linewidth}{!}{%
\begin{tabular}{l|ccc|ccc}
\toprule
\ratHdrJoint
\midrule
\#1 & 99.2\%  & 21.2\% & [73.4\%, 146.4\%] & 98.5\%  & 18.0\% & [76.8\%, 152.8\%] \\
\#2 & 101.3\% & 22.0\% & [76.1\%, 150.8\%] & 99.6\%  & 18.5\% & [78.3\%, 156.8\%] \\
\#3 & 110.4\% & 29.6\% & [81.5\%, 178.1\%] & 104.5\% & 24.8\% & [81.7\%, 187.0\%] \\
\#4 & 112.4\% & 30.6\% & [82.6\%, 182.8\%] & 105.6\% & 25.6\% & [83.0\%, 191.5\%] \\
\#5 & 118.3\% & 36.4\% & [83.9\%, 201.6\%] & 108.7\% & 30.4\% & [84.3\%, 212.9\%] \\
\#6 & 120.4\% & 37.6\% & [84.9\%, 206.7\%] & 109.8\% & 31.4\% & [85.6\%, 217.9\%] \\
\bottomrule
\end{tabular}}
\end{minipage}\hfill
\begin{minipage}[t]{0.45\textwidth}\centering
\caption*{\textbf{US 1M Bill}}
\resizebox{\linewidth}{!}{%
\begin{tabular}{l|ccc|ccc}
\toprule
\ratHdrJoint
\midrule
\#1 & 95.4\%  & 9.7\%  & [82.0\%, 117.3\%] & 92.4\% & 6.3\% & [82.1\%, 103.3\%] \\
\#2 & 96.7\%  & 9.8\%  & [83.3\%, 119.7\%] & 93.3\% & 6.0\% & [83.6\%, 104.2\%] \\
\#3 & 101.4\% & 12.2\% & [86.9\%, 134.1\%] & 94.9\% & 5.8\% & [85.8\%, 107.7\%] \\
\#4 & 102.6\% & 12.4\% & [88.0\%, 136.3\%] & 95.5\% & 5.6\% & [86.8\%, 108.0\%] \\
\#5 & 105.5\% & 14.6\% & [88.9\%, 146.4\%] & 96.3\% & 5.8\% & [87.1\%, 109.5\%] \\
\#6 & 106.7\% & 15.0\% & [89.9\%, 148.8\%] & 96.9\% & 5.7\% & [88.2\%, 109.8\%] \\
\bottomrule
\end{tabular}}
\end{minipage}

\par\smallskip

\begin{minipage}[t]{0.45\textwidth}\centering
\caption*{\textbf{US 2Y Note}}
\resizebox{\linewidth}{!}{%
\begin{tabular}{l|ccc|ccc}
\toprule
\ratHdrJoint
\midrule
\#1 & 94.6\%  & 6.8\% & [81.8\%, 106.2\%] & 95.5\%  & 4.6\% & [86.9\%, 104.3\%] \\
\#2 & 95.6\%  & 6.8\% & [83.2\%, 107.1\%] & 96.2\%  & 4.5\% & [88.1\%, 105.1\%] \\
\#3 & 98.6\%  & 8.0\% & [85.6\%, 112.3\%] & 98.1\%  & 4.7\% & [90.1\%, 108.3\%] \\
\#4 & 99.4\%  & 8.0\% & [86.7\%, 113.2\%] & 98.6\%  & 4.6\% & [91.3\%, 108.8\%] \\
\#5 & 101.1\% & 9.1\% & [87.6\%, 117.6\%] & 99.7\%  & 5.0\% & [91.8\%, 111.0\%] \\
\#6 & 102.0\% & 9.2\% & [88.8\%, 118.9\%] & 100.2\% & 5.0\% & [92.9\%, 111.6\%] \\
\bottomrule
\end{tabular}}
\end{minipage}\hfill
\begin{minipage}[t]{0.45\textwidth}\centering
\caption*{\textbf{US 10Y Note}}
\resizebox{\linewidth}{!}{%
\begin{tabular}{l|ccc|ccc}
\toprule
\ratHdrJoint
\midrule
\#1 & 95.8\%  & 4.8\% & [87.1\%, 104.7\%] & 96.3\%  & 2.9\% & [90.3\%, 101.7\%] \\
\#2 & 96.6\%  & 4.8\% & [88.0\%, 106.1\%] & 96.9\%  & 2.8\% & [91.6\%, 102.4\%] \\
\#3 & 98.8\%  & 5.5\% & [89.8\%, 109.8\%] & 98.3\%  & 2.9\% & [93.7\%, 104.9\%] \\
\#4 & 99.5\%  & 5.6\% & [90.4\%, 110.4\%] & 98.7\%  & 2.9\% & [94.4\%, 105.5\%] \\
\#5 & 100.8\% & 6.3\% & [90.8\%, 113.0\%] & 99.5\%  & 3.2\% & [94.6\%, 107.1\%] \\
\#6 & 101.4\% & 6.4\% & [91.6\%, 113.7\%] & 100.0\% & 3.2\% & [95.0\%, 107.7\%] \\
\bottomrule
\end{tabular}}
\end{minipage}

\par\smallskip

\begin{minipage}[t]{0.45\textwidth}\centering
\caption*{\textbf{DE 10Y Bund}}
\resizebox{\linewidth}{!}{%
\begin{tabular}{l|ccc|ccc}
\toprule
\ratHdrJoint
\midrule
\#1 & 95.7\%  & 6.3\% & [86.2\%, 113.1\%] & 96.9\%  & 2.4\% & [90.7\%, 100.6\%] \\
\#2 & 96.5\%  & 6.4\% & [87.0\%, 114.3\%] & 97.4\%  & 2.4\% & [91.2\%, 100.9\%] \\
\#3 & 99.3\%  & 8.0\% & [89.4\%, 122.0\%] & 98.8\%  & 2.6\% & [92.7\%, 102.9\%] \\
\#4 & 100.0\% & 8.1\% & [90.2\%, 123.3\%] & 99.3\%  & 2.6\% & [93.3\%, 103.6\%] \\
\#5 & 101.7\% & 9.5\% & [91.4\%, 128.6\%] & 100.0\% & 3.0\% & [93.9\%, 105.3\%] \\
\#6 & 102.4\% & 9.7\% & [92.3\%, 130.0\%] & 100.5\% & 3.1\% & [94.5\%, 105.7\%] \\
\bottomrule
\end{tabular}}
\end{minipage}\hfill
\begin{minipage}[t]{0.45\textwidth}\centering
\caption*{\textbf{Xtrackers}}
\resizebox{\linewidth}{!}{%
\begin{tabular}{l|ccc|ccc}
\toprule
\ratHdrJoint
\midrule
\#1 & 94.7\%  & 7.6\% & [76.6\%, 111.2\%] & 95.6\% & 3.9\% & [86.0\%, 101.7\%] \\
\#2 & 95.7\%  & 7.5\% & [78.5\%, 112.7\%] & 96.2\% & 3.7\% & [87.7\%, 102.1\%] \\
\#3 & 99.3\%  & 8.5\% & [82.5\%, 120.2\%] & 97.7\% & 3.6\% & [90.7\%, 104.3\%] \\
\#4 & 100.2\% & 8.6\% & [83.9\%, 121.5\%] & 98.2\% & 3.6\% & [91.4\%, 104.9\%] \\
\#5 & 102.3\% & 9.7\% & [86.0\%, 126.9\%] & 99.0\% & 3.9\% & [92.4\%, 106.8\%] \\
\#6 & 103.3\% & 9.8\% & [87.4\%, 128.3\%] & 99.5\% & 3.9\% & [93.2\%, 107.5\%] \\
\bottomrule
\end{tabular}}
\end{minipage}

\par\smallskip

\begin{minipage}[t]{0.45\textwidth}\centering
\caption*{\textbf{S\&P500}}
\resizebox{\linewidth}{!}{%
\begin{tabular}{l|ccc|ccc}
\toprule
\ratHdrJoint
\midrule
\#1 & 94.1\%  & 5.6\% & [86.3\%, 102.5\%] & 95.0\% & 3.3\% & [88.3\%, 101.3\%] \\
\#2 & 95.1\%  & 5.8\% & [87.2\%, 103.7\%] & 95.7\% & 3.1\% & [89.3\%, 101.7\%] \\
\#3 & 97.7\%  & 7.7\% & [89.0\%, 109.6\%] & 97.0\% & 3.1\% & [90.9\%, 104.5\%] \\
\#4 & 98.5\%  & 7.9\% & [89.8\%, 110.8\%] & 97.5\% & 3.1\% & [91.9\%, 105.0\%] \\
\#5 & 100.0\% & 9.5\% & [90.7\%, 117.2\%] & 98.2\% & 3.3\% & [92.0\%, 107.0\%] \\
\#6 & 100.7\% & 9.8\% & [91.4\%, 119.4\%] & 98.7\% & 3.2\% & [92.6\%, 107.5\%] \\
\bottomrule
\end{tabular}}
\end{minipage}\hfill
\begin{minipage}[t]{0.45\textwidth}\centering
\caption*{\textbf{VIX}}
\resizebox{\linewidth}{!}{%
\begin{tabular}{l|ccc|ccc}
\toprule
\ratHdrJoint
\midrule
\#1 & 96.8\%  & 6.8\%  & [86.8\%, 111.9\%] & 95.9\% & 3.5\% & [87.6\%, 102.4\%] \\
\#2 & 97.8\%  & 7.0\%  & [87.7\%, 113.2\%] & 96.4\% & 3.4\% & [88.6\%, 103.0\%] \\
\#3 & 100.7\% & 9.0\%  & [90.2\%, 119.8\%] & 97.8\% & 3.7\% & [90.3\%, 105.5\%] \\
\#4 & 101.5\% & 9.2\%  & [91.0\%, 120.8\%] & 98.2\% & 3.6\% & [91.0\%, 106.0\%] \\
\#5 & 103.3\% & 10.8\% & [92.3\%, 125.5\%] & 98.9\% & 4.0\% & [91.7\%, 107.7\%] \\
\#6 & 104.1\% & 11.1\% & [93.1\%, 126.7\%] & 99.4\% & 4.0\% & [92.5\%, 108.2\%] \\
\bottomrule
\end{tabular}}
\end{minipage}

\par\smallskip

\begin{minipage}[t]{0.45\textwidth}\centering
\caption*{\textbf{EUR-USD}}
\resizebox{\linewidth}{!}{%
\begin{tabular}{l|ccc|ccc}
\toprule
\ratHdrJoint
\midrule
\#1 & 95.1\%  & 6.6\% & [82.5\%, 112.4\%] & 94.9\% & 3.7\% & [86.4\%, 100.2\%] \\
\#2 & 96.0\%  & 6.5\% & [84.2\%, 113.3\%] & 95.5\% & 3.5\% & [87.2\%, 100.4\%] \\
\#3 & 98.7\%  & 7.6\% & [87.8\%, 119.6\%] & 96.8\% & 3.2\% & [89.1\%, 101.5\%] \\
\#4 & 99.4\%  & 7.6\% & [88.9\%, 120.4\%] & 97.2\% & 3.0\% & [89.8\%, 102.0\%] \\
\#5 & 101.0\% & 8.7\% & [90.0\%, 124.9\%] & 97.8\% & 3.1\% & [91.2\%, 103.1\%] \\
\#6 & 101.8\% & 8.8\% & [90.9\%, 125.9\%] & 98.3\% & 3.0\% & [92.2\%, 103.6\%] \\
\bottomrule
\end{tabular}}
\end{minipage}\hfill
\begin{minipage}[t]{0.45\textwidth}\centering
\caption*{\textbf{USD-JPY}}
\resizebox{\linewidth}{!}{%
\begin{tabular}{l|ccc|ccc}
\toprule
\ratHdrJoint
\midrule
\#1 & 94.1\%  & 4.6\% & [86.7\%, 103.5\%] & 96.0\%  & 3.0\% & [90.2\%, 103.0\%] \\
\#2 & 95.1\%  & 4.5\% & [87.8\%, 104.4\%] & 96.6\%  & 2.8\% & [90.8\%, 103.5\%] \\
\#3 & 97.7\%  & 5.0\% & [90.1\%, 108.3\%] & 98.2\%  & 2.9\% & [92.3\%, 106.1\%] \\
\#4 & 98.5\%  & 5.0\% & [91.0\%, 109.0\%] & 98.6\%  & 2.9\% & [92.9\%, 106.6\%] \\
\#5 & 99.9\%  & 5.6\% & [91.8\%, 111.6\%] & 99.5\%  & 3.2\% & [93.5\%, 108.4\%] \\
\#6 & 100.7\% & 5.7\% & [92.6\%, 112.7\%] & 100.0\% & 3.2\% & [94.1\%, 108.9\%] \\
\bottomrule
\end{tabular}}
\end{minipage}

\par\smallskip

\begin{minipage}[t]{0.45\textwidth}\centering
\caption*{\textbf{Brent}}
\resizebox{\linewidth}{!}{%
\begin{tabular}{l|ccc|ccc}
\toprule
\ratHdrJoint
\midrule
\#1 & 95.9\%  & 6.4\%  & [85.2\%, 114.4\%] & 96.4\% & 3.0\% & [90.6\%, 102.0\%] \\
\#2 & 96.8\%  & 6.5\%  & [86.3\%, 116.1\%] & 96.9\% & 2.8\% & [91.5\%, 102.2\%] \\
\#3 & 99.3\%  & 8.4\%  & [88.5\%, 127.8\%] & 97.9\% & 2.7\% & [93.3\%, 103.6\%] \\
\#4 & 100.1\% & 8.6\%  & [89.3\%, 129.7\%] & 98.3\% & 2.6\% & [93.8\%, 103.8\%] \\
\#5 & 101.6\% & 10.3\% & [90.2\%, 137.8\%] & 98.8\% & 2.8\% & [94.3\%, 104.7\%] \\
\#6 & 102.3\% & 10.6\% & [91.2\%, 139.8\%] & 99.2\% & 2.7\% & [95.1\%, 105.0\%] \\
\bottomrule
\end{tabular}}
\end{minipage}\hfill
\begin{minipage}[t]{0.45\textwidth}\centering
\caption*{\textbf{Gold}}
\resizebox{\linewidth}{!}{%
\begin{tabular}{l|ccc|ccc}
\toprule
\ratHdrJoint
\midrule
\#1 & 94.1\%  & 6.0\% & [82.7\%, 111.8\%] & 95.3\% & 4.2\% & [82.5\%, 101.9\%] \\
\#2 & 95.1\%  & 5.9\% & [83.9\%, 113.2\%] & 95.9\% & 3.9\% & [84.5\%, 102.4\%] \\
\#3 & 97.8\%  & 7.0\% & [86.6\%, 121.9\%] & 97.2\% & 3.7\% & [87.8\%, 103.7\%] \\
\#4 & 98.6\%  & 7.1\% & [87.6\%, 123.2\%] & 97.7\% & 3.5\% & [89.5\%, 103.8\%] \\
\#5 & 100.1\% & 8.1\% & [89.1\%, 128.4\%] & 98.4\% & 3.6\% & [90.7\%, 105.0\%] \\
\#6 & 100.9\% & 8.3\% & [90.2\%, 129.8\%] & 98.9\% & 3.5\% & [92.4\%, 105.6\%] \\
\bottomrule
\end{tabular}}
\end{minipage}

\caption{Market data: realized ratio of $ES_{2.5\%}$ estimate to $VaR_{1\%}$
estimate for the ES estimators defined in Table~\ref{T:estimators}. For each
instrument, results are shown jointly for 250 daily returns (Daily) and 10-day
overlapping returns (10-day); the last column of each block reports the
$[2.5\%,\,97.5\%]$ quantile interval of the realized ratio. The first six
instruments (SOFR, US 1M Bill, US 2Y/10Y Notes, DE 10Y Bund, Xtrackers) use
absolute returns; the remaining six (S\&P500, VIX, EUR-USD, USD-JPY, Brent,
Gold) use relative returns.}
\label{T:ratios:combined}
\end{table}

\clearpage

\section{Supplementary plots}

\begin{figure}[htp!]
\centering
\begin{minipage}[t]{0.32\textwidth}
    \centering
    \includegraphics[width=\linewidth]{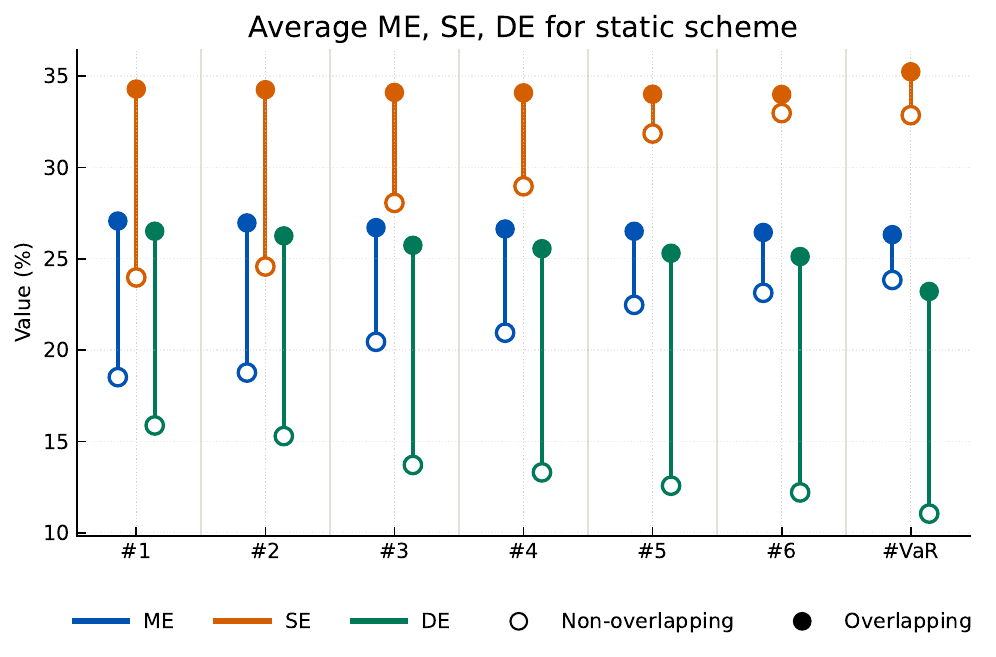}
\end{minipage}
\hfill
\begin{minipage}[t]{0.32\textwidth}
    \centering
    \includegraphics[width=\linewidth]{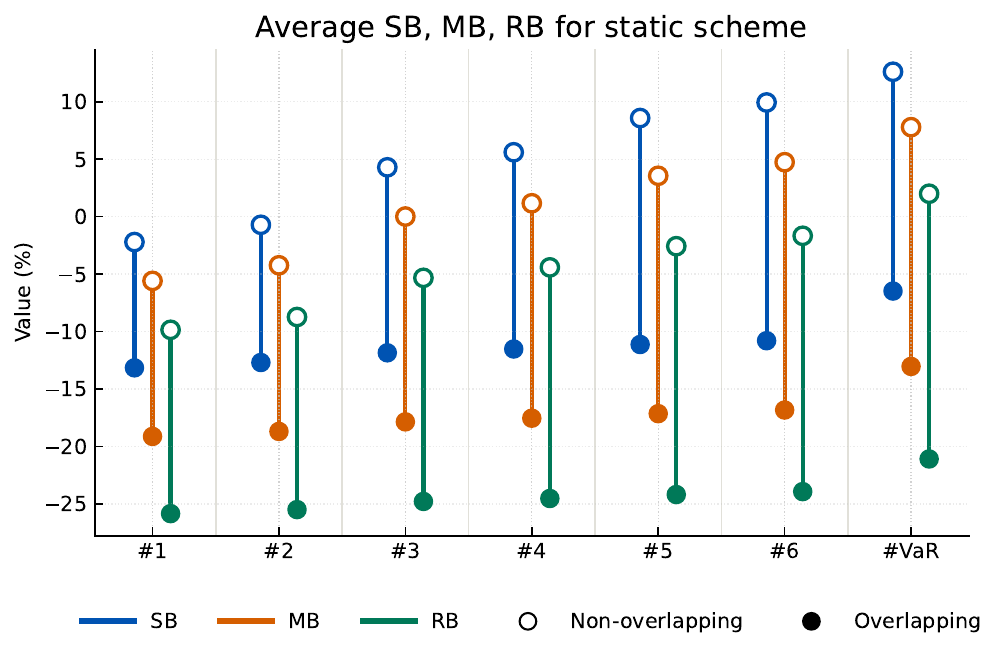}
\end{minipage}
\hfill
\begin{minipage}[t]{0.32\textwidth}
    \centering
    \includegraphics[width=\linewidth]{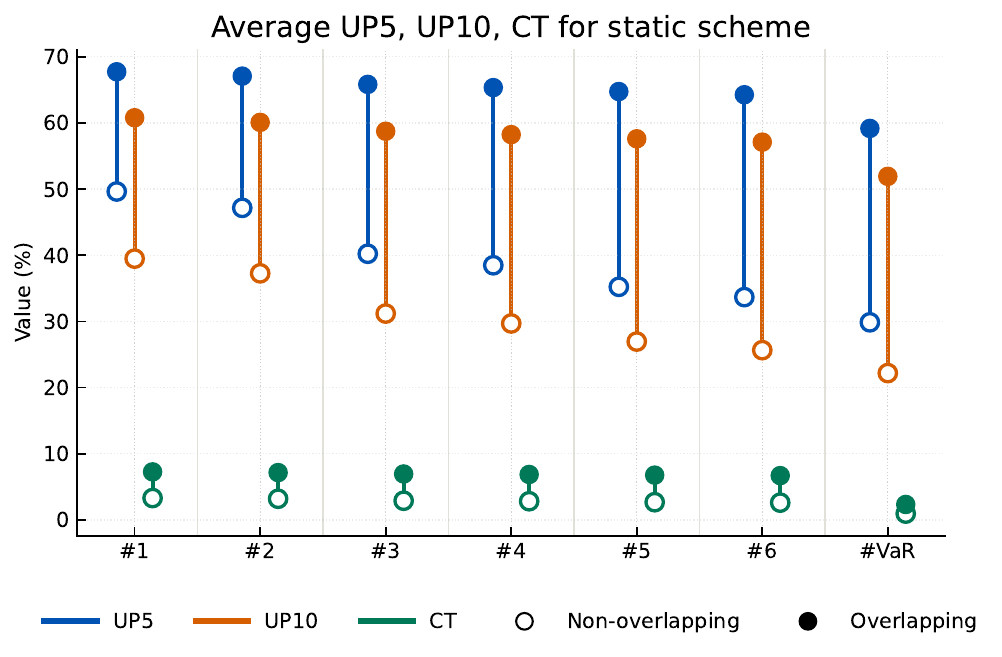}
\end{minipage}

\begin{minipage}[t]{0.32\textwidth}
    \centering
    \includegraphics[width=\linewidth]{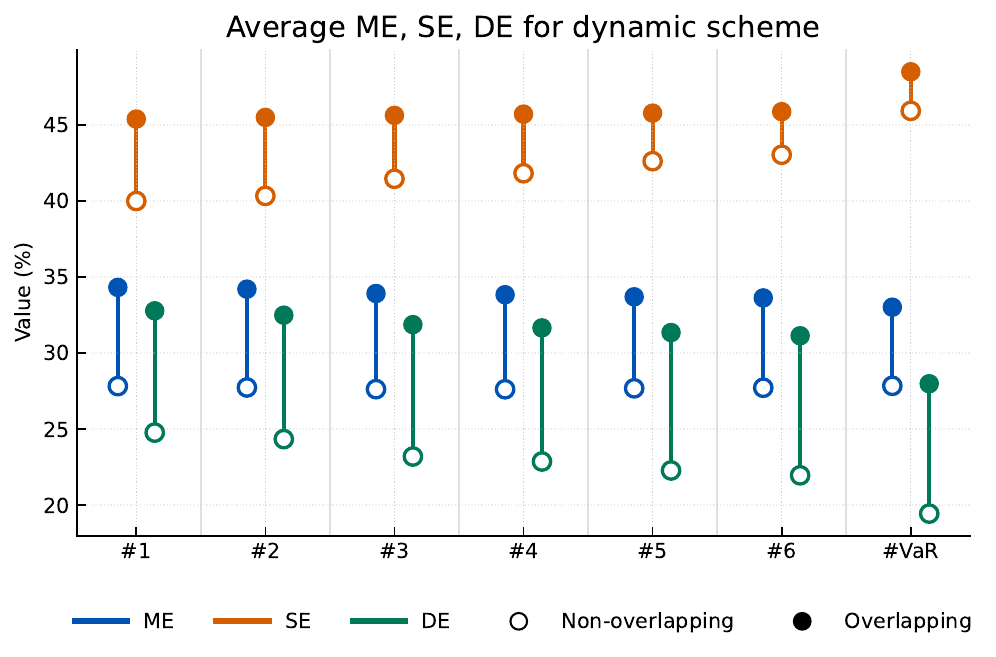}
\end{minipage}
\hfill
\begin{minipage}[t]{0.32\textwidth}
    \centering
    \includegraphics[width=\linewidth]{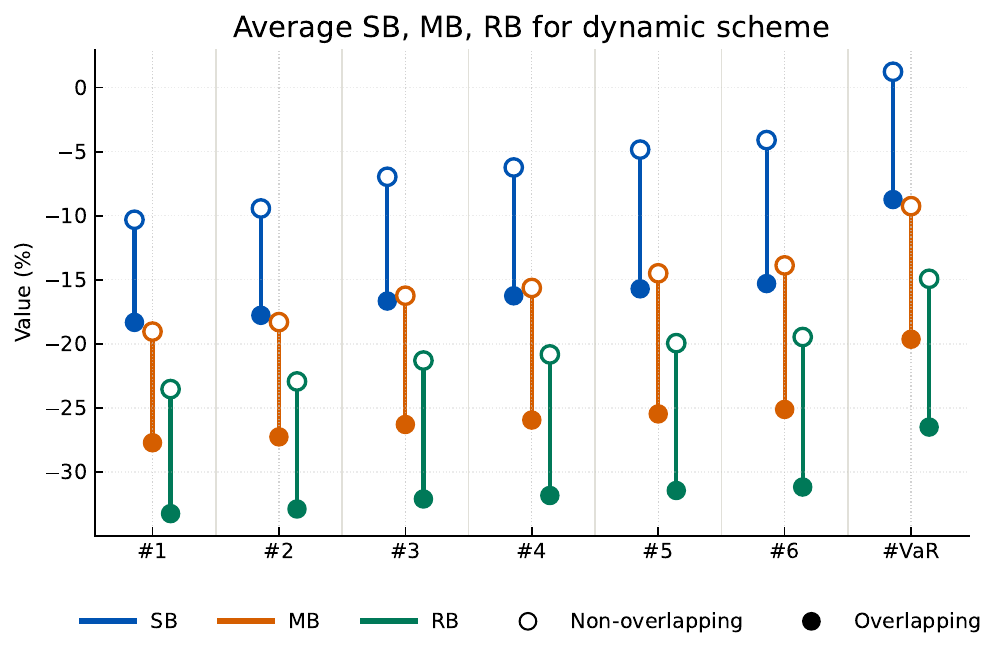}
\end{minipage}
\hfill
\begin{minipage}[t]{0.32\textwidth}
    \centering
    \includegraphics[width=\linewidth]{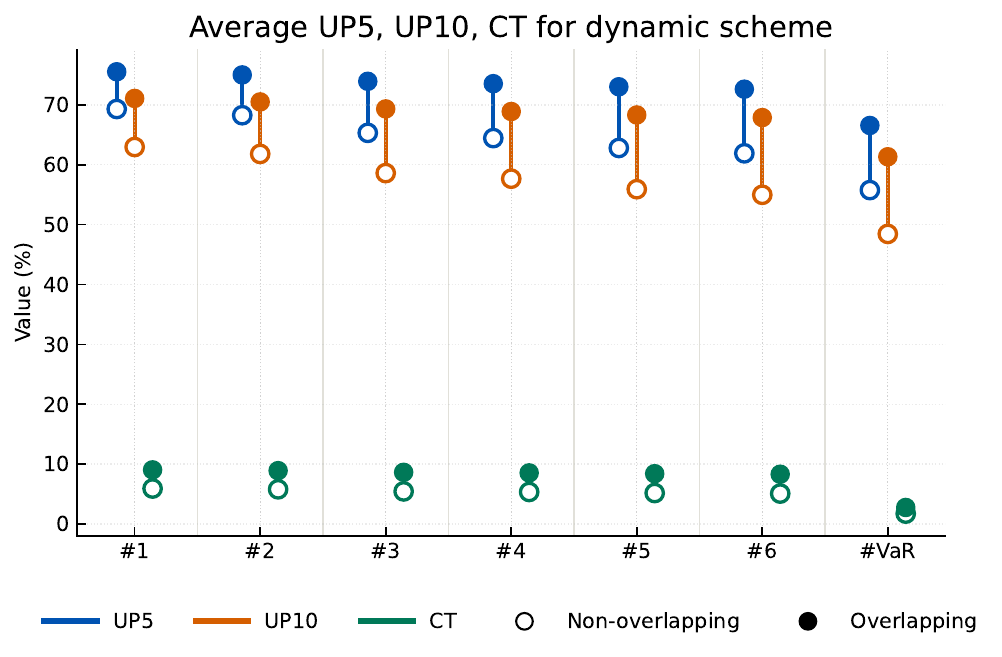}
\end{minipage}
\caption{Values of metrics averaged across static (top row) and dynamic (bottom row)
models, separately for each estimator and sampling scheme.}
\label{fig:dumbell_grouped_all}
\end{figure}

\end{document}

%% file: headings.tex
\usepackage{amsthm, amsmath, amssymb, amsfonts, graphicx, epsfig}
\usepackage{accents}
\usepackage{ulem} 

\usepackage{bbm}
\usepackage{mathrsfs}

\usepackage{epstopdf}

\usepackage{graphicx}
\usepackage[font=sl,labelfont=bf]{caption}
\usepackage{subcaption}

\usepackage{enumerate}

\usepackage[usenames,dvipsnames]{color}

\definecolor{labelkey}{rgb}{0,0,1}

\usepackage[colorlinks=true, pdfstartview=FitV, linkcolor=blue,
            citecolor=blue, urlcolor=blue]{hyperref}
\usepackage[usenames]{color}
\usepackage{multirow}

\usepackage{url}
\makeatletter\def\url@leostyle{%
 \@ifundefined{selectfont}{\def\UrlFont{\sf}}{\def\UrlFont{\scriptsize\ttfamily}}} \makeatother

\newtheorem{theorem}{Theorem}

\theoremstyle{definition}
\newtheorem{definition}[theorem]{Definition}

\theoremstyle{remark}

\usepackage{mdframed}

\theoremstyle{definition}
\newmdtheoremenv{definition2}[theorem]{Definition}
\newmdtheoremenv{proposition2}[theorem]{Proposition}
\newmdtheoremenv{theorem2}[theorem]{Theorem}
\newmdtheoremenv{corollary2}[theorem]{Corollary}

\numberwithin{equation}{section}
\numberwithin{theorem}{section}

\definecolor{Red}{rgb}{0.9,0,0.0}
\definecolor{Blue}{rgb}{0,0.0,1.0}

\newcommand{\gray}[1]{\begin{color}[rgb]{0.5, 0.5, 0.5}{#1}\end{color}}

\def\cS{\mathcal{S}}

\def\cX{\mathcal{X}}

\def\bE{\mathbb{E}}

\def\bN{\mathbb{N}}

\def\bP{\mathbb{P}}

\def\bR{\mathbb{R}}

\def\bfX{\mathbf{X}}

\def\bfx{\mathbf{x}}
\def\bfy{\mathbf{y}}

\newcommand{\1}{\mathbbm{1}}            
\newcommand{\set}[1]{\{#1\}}            
\renewcommand{\mid}{\;|\;}              
\newcommand{\dif}{\operatorname{d}\!}   

\DeclareMathOperator{\var}{\mathrm{VaR}}           
\DeclareMathOperator{\ES}{\mathrm{ES}}                    
     
\makeatletter
\def\namedlabel#1#2{\begingroup
    #2%
    \def\@currentlabel{#2}%
    \phantomsection\label{#1}\endgroup
}
\makeatother